\documentclass[final,3p,times,fleqn]{elsarticle}
\usepackage{amssymb}
\usepackage{amsbsy}
\usepackage{amsmath}
\usepackage{color}
\usepackage{mymacros}
\usepackage{pdflscape}
\usepackage{afterpage}
\usepackage{capt-of}

 \usepackage{tikz}
 \usetikzlibrary{positioning}
 \usepackage{graphicx}
 \usepackage[outdir=./]{epstopdf}
\usepackage[normal,tight,center]{subfigure}
\usepackage{array}
\usepackage{calc}
\usepackage[thinlines]{easytable}
\usepackage{comment}
\usepackage{tabu}
\usepackage{hyperref}
\makeatletter
\newcommand{\mathleft}{\@fleqntrue\@mathmargin2em}
\newcommand{\mathcenter}{\@fleqnfalse}
\makeatother
\newcommand\rev[1]{\textcolor{black}{#1}}

\def\uhe    {u_h^e}

\def\omge   {\Omega_e}

\journal{CMAME}

\begin{document}

\begin{frontmatter}



\title{An asynchronous discontinuous Galerkin method for massively parallel PDE solvers}

\author[label1]{Shubham K.\ Goswami}
\ead{shubhamkg@iisc.ac.in}
\author[label1]{Konduri Aditya\corref{cor1}\fnref{fnt1}}
\ead{konduriadi@iisc.ac.in}
\cortext[cor1]{Corresponding author.}
\address[label1]{Department of Computational and Data Sciences, Indian Institute of Science, Bengaluru, India}

\begin{abstract}
The discontinuous Galerkin (DG) method is widely being used to solve hyperbolic partial differential equations (PDEs) due to its ability to provide high-order accurate solutions in complex geometries, capture discontinuities, and exhibit high arithmetic intensity. However, the scalability of DG-based solvers is impeded by communication bottlenecks arising from the data movement and synchronization requirements at extreme scales. To address these challenges, recent studies have focused on the development of asynchronous computing approaches for PDE solvers. Herein, we introduce the asynchronous DG (ADG) method, which combines the benefits of the DG method with asynchronous computing to overcome communication bottlenecks. The ADG method relaxes the need for data communication and synchronization at a mathematical level, allowing processing elements to operate independently regardless of the communication status, thus potentially improving the scalability of solvers. The proposed ADG method ensures flux conservation and effectively addresses challenges arising from asynchrony. To assess its stability, Fourier-mode analysis is employed to examine the dissipation and dispersion behavior of fully-discrete equations that use the DG and ADG schemes along with the Runge-Kutta (RK) time integration scheme. Furthermore, an error analysis within a statistical framework is presented, which demonstrates that the ADG method with standard numerical fluxes achieves at most first-order accuracy. To recover accuracy, we introduce asynchrony-tolerant (AT) fluxes that utilize data from multiple time levels. Extensive numerical experiments were conducted to validate the performance of the ADG-AT scheme for both linear and nonlinear problems. Overall, the proposed ADG-AT method demonstrates the potential to achieve accurate and scalable DG-based PDE solvers, paving the way for simulations of complex physical systems on massively parallel supercomputers.
\end{abstract}

\begin{keyword}
Asynchronous schemes \sep Partial differential equations \sep Parallel computing \sep Massive computations \sep Discontinuous Galerkin method



\end{keyword}

\end{frontmatter}

%\begin{comment}
%%domainΩintoNE
% \linenumbers


\section{Introduction}
\label{sec:introduction}

% Introduction to the DG method
The discontinuous Galerkin (DG) method, a class of finite-element methods, utilizes piecewise polynomials as basis functions to approximate the solutions of partial differential equations. These basis functions can be completely discontinuous, which allows adaptability to functions across various elements. The DG method has several advantages, including its ability to handle complex geometries, capture discontinuities/shocks in solutions, achieve high-order accuracy, and maintain local conservation \cite{hesthaven2007}. These attributes make the DG method particularly well-suited for simulating hyperbolic problems that are frequently encountered in fluid dynamics.

% DG HPC
The computations resulting from the DG method possess a compact structure that provides high arithmetic intensity and is suitable for developing massively parallel solvers. However, at extreme scales, data movement costs are prohibitively expensive and degrade the parallel performance. Therefore, significant advancements have been made to enhance the scalability of DG-based solvers. One such development is the hybridized discontinuous Galerkin (HDG) method, which employs element and hybrid unknowns to reduce the number of globally coupled degrees of freedom \cite{cockburn2009unified,roca2013scalable}.
\rev{The \textit{parareal} approach has also shown great promise in improving the scalability of parallel time-dependent PDE solvers \cite{LIONS2001661, burrage1997parallel, lirias1119269}. 
The method, in addition to spatial domain decomposition, also decomposes the computations in time integration among different processing elements and uses a predictor-corrector approach to maintain the desired accuracy.
}
Another notable progress is the utilization of GPUs, where asynchronous data transfer techniques have been incorporated into DG solvers to minimize the parallelization overhead and improve parallel efficiency \cite{xia2015openacc,kirby2020gpu}. For example, an OpenACC directive-based GPU parallel scheme was proposed in \cite{xia2015openacc} to solve compressible Navier-Stokes equations on hybrid unstructured grids, thereby demonstrating its effectiveness and extensibility for high-order CFD solvers on GPUs. In addition, the use of OCCA, a thread-programming abstraction model, has enabled the solution of compressible Euler equations using a discontinuous Galerkin method on CPUs and GPUs \cite{kirby2020gpu}, offering a unified strategy for achieving performance portability across multi-threaded hardware architectures. In recent years, the matrix-free approach has gained popularity for the development of highly scalable DG-based solvers. These solvers are particularly suitable for solving large-scale problems on high-performance computing platforms \cite{nguyen2023implicit,kronbichler2019fast}. Matrix-free DG schemes have demonstrated superior performance compared to other methodologies, including hybridized DG methods \cite{kronbichler2018performance}. An example of utilizing a matrix-free approach for DG in fluid dynamics simulations on exascale computers is the \textit{ExaDG} project \cite{exadg2020}. Building on the finite element library \texttt{deal.II} \cite{dealII94}, \textit{ExaDG} addresses the demand for high-order accuracy in complex geometries and efficient scaling in exascale computing environments. The software leverages optimized data structures and computation-communication overlap to enhance the solver's performance and scalability.
% DUNE: Peter Bastian
\rev{Another software library that provides the necessary framework for the development of the DG method based PDE solvers is the \textit{Distributed Unified Numerical Environment (DUNE)} \cite{bastian2008-dune1, bastian2008-dune2, BASTIAN2021}. It facilitates multiple grid managers for various discretization methods, such as finite element, finite volume, and DG, and provides Python-bindings for several of its modules, thus making it easier to use. Ref.~\cite{klofkorn2012-dune-dg} especially demonstrates the performance improvement of the DG-solver based on the \textit{DUNE} framework with computation-communication overlap using non-blocking communication and matrix-free implementation. The extension towards the exascale architecture has been demonstrated with the \textit{Exa-DUNE} project, which is aimed at the development of hardware-specific software components and scalable high-level algorithms for PDEs \cite{bastian2014exadune, exadune1-bastian2016, exadune2-bastian2016}.}
% FLEXI
\rev{Additionally, the DG method based numerical framework \textit{FLEXI}, which uses explicit schemes for time integration, has shown its capabilities to perform large-scale CFD simulations for solving compressible Navier-Stokes equations with great scalability properties \cite{KRAIS2021186-flexi, blind2023exascale-flexi}.}

Current state-of-the-art simulations using the DG method are performed on hundreds of thousands of PEs \cite{exadg2020,kirby2020gpu,melander2023, blind2023exascale-flexi}. At this extreme scale, it is observed that the communication time owing to data movement dominates the total execution time, significantly reducing the scalability of the solver. As we are moving towards the exascale, where the machines comprise millions of processing elements, data communication and its synchronization overhead pose a major bottleneck in achieving high parallel efficiency. To address this issue, an asynchronous computing approach based on the finite difference method was introduced in \cite{konduri2012async,donzis2014}. This approach relaxes the data communication/synchronization requirements between processing elements (PEs) in parallel solvers at a mathematical level, which enables computations to proceed regardless of the status of communications. Computations at grid points near PE boundaries, which have data dependencies from neighboring PEs, are allowed to advance with older or delayed values in the stencil. However, the standard finite difference schemes implemented using the asynchronous computing approach exhibited at most first-order accurate solutions in the presence of delayed data. To overcome this challenge, asynchrony-tolerant (AT) schemes have been developed \cite{konduri2017at,komal2023reactions-at}. These schemes use extended stencils in space and/or time to recover the accuracy compromised by the asynchrony. Extensive numerical experiments have demonstrated the robustness of AT schemes in accurately capturing the solution dynamics for both linear and nonlinear problems, including simulations of one-dimensional reacting flow problems and three-dimensional direct numerical simulations of compressible turbulent flows \cite{aditya2019arXiv,komal2020dns-at,komal2023reactions-at,goswami2023lserkat-jcp}. A similar concept, where inter-processor communications are enabled owing to event-triggered communication, was explored in \cite{ghosh2018event,ghosh2019parallel}. 
\rev{A different idea that uses asynchrony in the choice of schemes and time steps is the heterogeneous asynchronous time integrators (HATI) approach, where various time schemes with different time steps are utilized to account for multi-scale effects that are typically present in transient structural dynamics \cite{Gravouil2014, mahjoubi2009, fekak2017}.}

The objective of this work is to develop an asynchronous discontinuous Galerkin (ADG)\footnote{\rev{Ader-DG and averaging discontinuous Galerkin methods are also sometimes referred to as ADG methods. However, in this paper, the ADG method always refers to the asynchronous discontinuous Galerkin method.}} method based on the asynchronous computing approach that would significantly improve the scalability of DG solvers. Preliminary results of the proposed method were reported in \cite{goswami2022asyncdg-aviation}. This paper presents a comprehensive analysis of the viability of the ADG method. The key contributions of this study are as follows.
\begin{itemize}
    \item Develop the asynchronous discontinuous Galerkin (ADG) method that preserves local conservation while allowing computations with delayed data.
    \item Investigate the effect of asynchrony on the stability and accuracy of schemes implemented using the ADG method.
    \item Develop asynchrony-tolerant (AT) numerical fluxes that provide high-order accurate solutions.
    \item Validate performance of the ADG method based on numerical simulations of linear and nonlinear PDEs.
\end{itemize}

%Outline
The remainder of this paper is organized as follows. Section~\ref{sec:background} provides a brief background of the discontinuous Galerkin method and its parallel implementation. In Sec.~\ref{sec:asyncDG}, the asynchronous discontinuous Galerkin (ADG) method is introduced. The numerical properties of the ADG method, including conservation, stability, and accuracy, are analyzed in Sec.~\ref{sec:nm}. To improve the accuracy of asynchronous DG schemes, new asynchrony-tolerant (AT) fluxes are proposed in Sec.~\ref{sec:atflux}. In Sec.~\ref{sec:numexp}, we validate the performance of the ADG and ADG-AT schemes through numerical experiments involving both linear and nonlinear problems. Finally, conclusions of this study are presented in Sec.~\ref{sec:conclusions}.

%====================================
\section{Standard discontinuous Galerkin (DG) method}
\label{sec:background}

To illustrate the standard discontinuous Galerkin (DG) method, consider the one-dimensional linear advection equation,
\begin{equation}
    \frac{\partial u}{\partial t} + \frac{\partial f(u)}{\partial x} = 0, 
    \label{eq:wave}
\end{equation}
where $u(x,t)$ is a scalar defined over the spatial domain $\Omega = [0,L]$, $f(u) = au$ is the linear advective flux, and $a$ is the constant advection speed. The initial condition is $u(x,0)=u_0$, and the boundary condition is periodic. 
% approximate solution and discretization
An approximate solution, $u_h$, to the above equation can be obtained by discretization of the domain into $N_E$ non-overlapping elements, $\Omega \equiv \Omega_h = \bigcup_{e=1}^{N_E} \omge$, where the $e$th element $\Omega_e$ spans $[x_e, x_{e+1}]$. In general, these elements can have different sizes. However, uniform-size elements with $\Delta x = x_{e+1} - x_e$ are considered for simplicity.
% solution space
Let $V_h = \bigoplus_{e=1}^{N_E} V_h^e$ be the solution space defined as a collection of piecewise smooth functions $v_h$ defined on $\Omega_h$ that can be discontinuous across the boundaries of the elements. The element-wise space $V_h^e$ is spanned by basis functions $\phi_j^e(x), 0 \leq j \leq N_p$, which are considered to be polynomials of degree at most $N_p$, defined on $\Omega_e$.
Using these basis functions, we can approximate the local element-wise solution for the element $\Omega_e$ as
\begin{equation}
    \uhe = \sum_{j=0}^{N_p} \hat{u}_j^e(t) \phi_j^e(x),
    \label{eq:DGsolution}
\end{equation}
where $\hat{u}_j^e(t), 0 \leq j \leq N_p$ are unknown local degrees of freedom (DoFs). {We employ Lagrange polynomials as basis functions, which ensures that the function value at the $j$th node $(0 \leq j \leq N_p)$ of an element $\Omega_e$ corresponds to the $j$th DoF of the $e$th element.}
By combining local solutions, the global solution over the discretized spatial domain $\Omega_h$ can be expressed as $u_h = \bigoplus_{e=1}^{N_E} u_h^e$.

\begin{figure}[h!]
\begin{center}
    \includegraphics[width=0.67\linewidth]{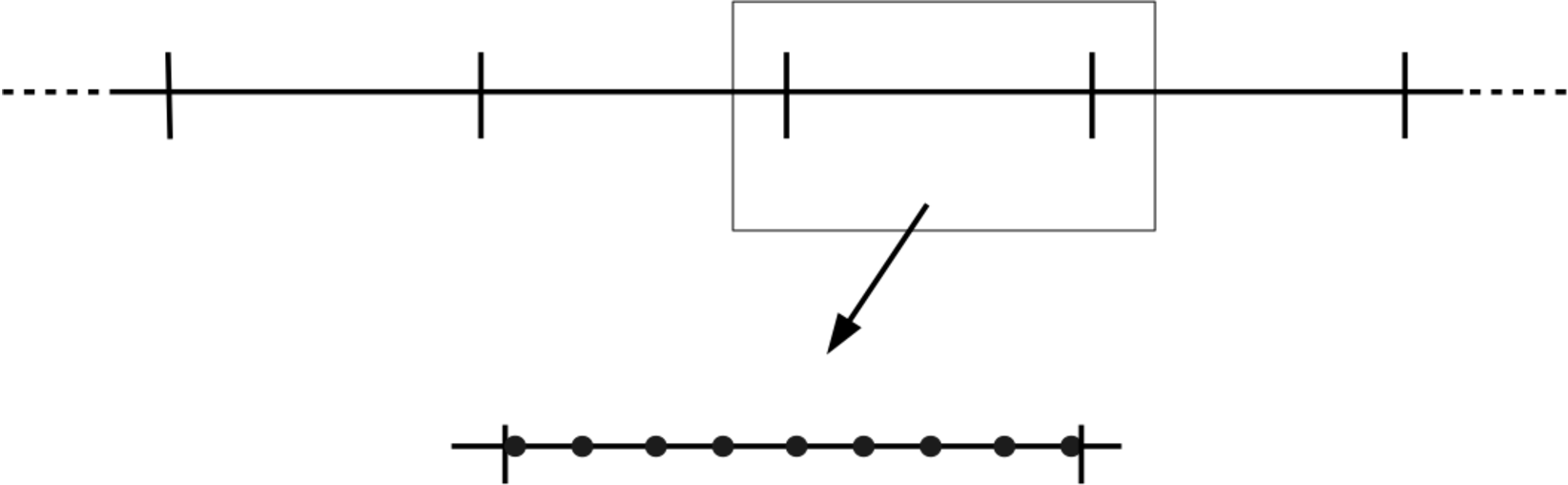}
    \begin{picture}(0,0)
    \put(-162,-5){$\Omega_e$}
    \put(-216,-5){$x_{e}$}
    \put(-106,-5){$x_{e+1}$}
    \put(-218,15){$\hat{u}_0$}
    \put(-204,15){$\hat{u}_1$}
    \put(-188,15){$\hat{u}_2$}
    \put(-158,15){$\cdots$}
    \put(-108,15){$\hat{u}_{N_p}$}
    \put(-162,62){$x_e$}
    \put(-225,62){$x_{e-1}$}
    \put(-288,62){$x_{e-2}$}
    \put(-106,62){$x_{e+1}$}
    \put(-42,62){$x_{e+2}$}
    \put(-130,68){$\Omega_{e}$}
    \put(-194,68){$\Omega_{e-1}$}
    \put(-254,68){$\Omega_{e-2}$}
    \put(-75,68){$\Omega_{e+1}$}

    \put(-170,85){\textcolor{red}{$u_e^-$}}
    \put(-155,85){\textcolor{red}{$u_e^+$}}
    \put(-117,85){\textcolor{red}{$u_{e+1}^-$}}
    \put(-96,85){\textcolor{red}{$u_{e+1}^+$}}
    \put(-178,100){\textcolor{blue}{$\hat{f}(u_{e}^-, u_e^+)$}}
    \put(-120,100){\textcolor{blue}{$\hat{f}(u_{e+1}^-, u_{e+1}^+)$}}
    \end{picture}
    \vspace{0.2cm}
\caption{\small {A schematic of a typical computational element $\Omega_e$ in a discretized domain $\Omega_h = \bigcup_{e+1}^{N_E} \Omega_e $ with numerical fluxes shown in blue color at the boundaries $x_e$  and $x_{e+1}$ of $\Omega_e$.}}
\label{fig:flux}
\end{center}
\end{figure}

The approximate solution $u_h$, in general, does not exactly satisfy the PDE and results in a residual, which is given by
\begin{equation}
    \mathcal{R}_h(x,t) = \frac{\partial u_h}{\partial t} + a \frac{\partial u_h}{\partial x}.
\end{equation}
The unknown DoFs in the local solution can be computed by minimizing this residual. This is achieved by making the residual orthogonal to some test functions. In the Galerkin approach, test functions $v_h = \phi_i(x), i \leq 0 \leq N_p$  are drawn from the solution space $V_h$. The orthogonality condition is imposed by using an inner product based on the $L2$-norm.
This results in $N_p+1$ equations, $\int_{\Omega_e} ({\partial u_h^e}/{\partial t})\phi_i^e(x)dx + \int_{\Omega_e}a(\partial u_h^e/\partial x) \phi_i^e(x)dx = 0, i = 0, 1, \dots , N_p,$ for computing $N_p+1$ unknown DoFs ($\hat{u}_j^e, j = 0, \dots , N_p$) per element.
% This results in $N_p+1$ equations to compute $N_p+1$ unknown DoFs per element.
Further, by performing integration by parts once on the advection term, we can express the integrals in a locally defined weak form as
\begin{align}
    \int_{\Omega_e}\frac{\partial u_h^e}{\partial t}\phi_i^e(x)dx - \int_{\Omega_e}au_h^e\frac{d\phi_i^e(x)}{dx}dx
    &= -\left[\left(au_h\right)^*\phi_i^e(x)\right]_{x_e}^{x_{e+1}}, \quad i = 0, 1, \dots , N_p,
    \label{eq:intbyparts}
\end{align}
where the term $\left[\left(au_h\right)^*\phi_i^e(x)\right]_{x_e}^{x_{e+1}}$ which arises from the integration by parts, represents the flux at the element boundaries. Substituting the expression for $u_h^e$ from Eq.~\eqref{eq:DGsolution} into the above equation, the semi-discrete weak form can be rewritten as
\begin{equation}
    \sum_{j=0}^{N_p}\frac{d\hat{u}_j^e}{dt}\int_{\Omega_e}\phi_j^e(x)\phi_i^e(x)dx - \sum_{j=0}^{N_p}a\hat{u}_j^e\int_{\Omega_e}\phi_j^e(x)\frac{d\phi_i^e(x)}{dx}dx = -\left[\left(au_h\right)^*\phi_i^e(x)\right]_{x_e}^{x_{e+1}}, \quad i = 0, 1, \dots , N_p.
    \label{eq:DGweakform}
\end{equation}
% 
%  Numerical flux
% 
It is important to note that the approximate solution $u_h$ can be discontinuous at the element boundaries, which often leads to a non-unique function value at the boundary nodes. As illustrated in Fig.~\ref{fig:flux}, for example, at the $x_e$ boundary, the function values $u_{e}^-$ and $u_e^+$ arise from the computations at elements $\Omega_{e-1}$ and $\Omega_e$, respectively. To handle such discontinuities, a numerical flux function, $\hat{f}(u_e^-,u_e^+) = (au_e)^*$ is introduced at the boundaries. This flux function is a linear combination of $au_e^-$ and $au_e^+$ and is designed to ensure a single-valued representation. We define $(au_e)^* = a(B^+u_e^- + B^-u_e^+)$, where $B^+ = B^- = 1/2$ corresponds to the central flux and $B^+ = 1$ and $B^- = 0$ for $a>0$ represent the upwind flux.

The integrals in Eq.~\eqref{eq:DGweakform} are often evaluated by mapping each element $\Omega_e = [x_e, x_{e+1}]$ to a reference element $\Omega_R$ in DG solvers. Such a mapping does not require computations of the integrals for every element, reducing the computation cost. Choosing the reference element $\Omega_R$ to be $[-1, 1]$ with a transformation $\xi(x) = (2x-x_e-x_{e+1})/{\Delta x}$, a generalized expression for the system of equations of the element-wise semi-discrete weak form can be expressed as
\begin{equation}
    \boldsymbol{\mathcal{M}}\frac{d\boldsymbol{u}_h^e}{dt} = \frac{2a}{\Delta x}\left( \boldsymbol{\mathcal{K}}^-\boldsymbol{u}_h^{e-1} + (\boldsymbol{\mathcal{K}}_l + \boldsymbol{\mathcal{S}} + \boldsymbol{\mathcal{K}}_r)\boldsymbol{u}_h^e + \boldsymbol{\mathcal{K}}^+\boldsymbol{u}_h^{e+1} \right).\label{eq:DGscheme}
\end{equation}
Here $\boldsymbol{\mathcal{M}}$ is the mass matrix, $\boldsymbol{\mathcal{S}}$ is the stiffness matrix, and $\boldsymbol{\mathcal{K}}$s are the flux matrices.
The element-wise entries of these matrices are
\begin{align}
    [\boldsymbol{\mathcal{M}}]_{ij} &= \int_{-1}^{1}\phi_j(\xi)\phi_i(\xi)d\xi, \quad [\boldsymbol{\mathcal{S}}]_{ij} = \int_{-1}^{1}\phi_j(\xi)\frac{d\phi_i(\xi)}{d\xi}d\xi, \quad [\boldsymbol{\mathcal{K}}^-]_{ij} = B^+\phi_j(1)\phi_i(-1), \nonumber \\
    [\boldsymbol{\mathcal{K}}^+]_{ij} &= -B^-\phi_j(-1)\phi_i(1), \quad
    [\boldsymbol{\mathcal{K}}_l]_{ij} = B^-\phi_j(-1)\phi_i(-1), \quad [\boldsymbol{\mathcal{K}}_r]_{ij} = -B^+\phi_j(1)\phi_i(1),
    \label{eq:matrices-reference}
\end{align}
where $\phi(\xi)$ represents the basis function in the reference element $\Omega_R$, and $\boldsymbol{\mathcal{K}}^- \boldsymbol{u}_h^{e-1} + \boldsymbol{\mathcal{K}}_l \boldsymbol{u}_h^e$ and $\boldsymbol{\mathcal{K}}_r \boldsymbol{u}_h^e + \boldsymbol{\mathcal{K}}^+ \boldsymbol{u}_h^{e+1}$ provide the numerical fluxes at $x_e$ and $x_{e+1}$, respectively.
%  Time integration
Equation~(\ref{eq:DGscheme}) represents a system of $N_p + 1$ linear ODEs with a given initial condition. To perform the time integration, a simple explicit Euler scheme can be used, which results in the fully-discrete matrix equation,
\begin{equation}
    \boldsymbol{u}_h^{e,n+1} = \boldsymbol{u}_h^{e,n} + 2\sigma\boldsymbol{\mathcal{M}}^{-1}\left( \boldsymbol{\mathcal{K}}^-\boldsymbol{u}_h^{e-1,n} + (\boldsymbol{\mathcal{K}}_l + \boldsymbol{\mathcal{S}} + \boldsymbol{\mathcal{K}}_r)\boldsymbol{u}_h^{e,n} + \boldsymbol{\mathcal{K}}^+\boldsymbol{u}_h^{e+1,n} \right).
    \label{eq:EulerDG}
\end{equation}
Here the time advancement is from a level $n$ to $n+1$ that are $\Delta t$ apart. The solution vector at $n$th time level is $\boldsymbol{u}_h^{e,n}$ such that $[\boldsymbol{u}_h^{e,n}]_j = \hat{u}_j^{e,n} = u_h(x_j^e,t^n) = u_h(x_j^e,n\Delta t)$. The Courant number, $\sigma = {a\Delta t}/{\Delta x}$, is a parameter that is commonly bounded to provide a stable solution.  

% serial implementation
An implementation of such a fully discrete scheme into a serial solver involves a time advancement loop within which a linear solve is performed for each element, iterating over all the elements in the domain. In solving the matrix equation (Eq.~\eqref{eq:EulerDG}) at each element $\Omega_e$, some of the right-hand-side computations require DoF values from the neighboring elements. This dependency arises from the definition of the numerical flux function $\hat{f}((u_e^-)^n,(u_e^+)^n)$ or simply $\hat{f}^n(u_e^-,u_e^+)$.
It is important to note that time integration is explicit in this serial implementation. Therefore, there is no need for a global linear solve, and the equations for each element can be solved independently. This property makes the scheme computationally efficient and well-suited for parallelization, where computations at multiple elements can be executed concurrently to enhance the performance of the solver.

\vspace{0.3cm}
\begin{figure}[h!]
\begin{center}
     \includegraphics[scale=0.6]{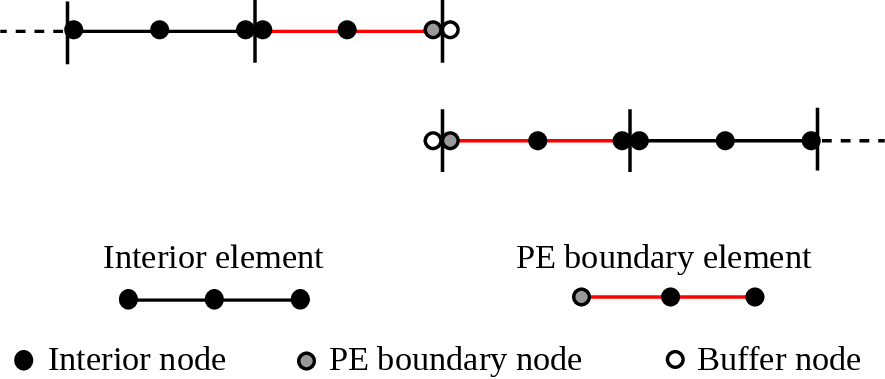}
    \begin{picture}(0,0)
    \put(-212,72){\textcolor{blue}{Communication}}
    \put(-110,84){\textcolor{blue}{Synchronization}}
    \put(-152,110){\textcolor{black}{$(u_e^-)^n$}}
    \put(-152,56){\textcolor{blue}{$(u_e^-)^n$}}
    \put(-130,110){\textcolor{blue}{$(u_e^+)^n$}}
    \put(-130,56){\textcolor{black}{$(u_e^+)^n$}}
    \put(-170,88){$\Omega_{e-1}$}
    \put(-224,88){$\Omega_{e-2}$}
    \put(-107,50){$\Omega_{e}$}
    \put(-58,50){$\Omega_{e+1}$}
    \put(-248,84){$x_{e-2}$}
    \put(-193,84){$x_{e-1}$}
    \put(-134,84){$x_{e}$}
    \put(-134,48){$x_{e}$}
    \put(-82,48){$x_{e+1}$}
    \put(-30,48){$x_{e+2}$}
    \put(-200,120){PE-0}
    \put(-80,105){PE-1}
    \put(-128,76){\rotatebox{90}{\textcolor{blue}{\Huge{$\rightarrow$}}}}
    \put(-144,96){\rotatebox{-90}{\textcolor{blue}{\Huge{$\rightarrow$}}}}
    \end{picture}
\caption{\small{An illustration of communication and synchronization at a PE boundary with two sub-domains.}}
\label{fig:domain-decomposition-sync}
\end{center}
\end{figure}

% Parallel implementation
In a parallel implementation, the computational domain $\Omega_h$ is typically decomposed into smaller subdomains, which are then mapped to different processing elements (PEs). An illustration of this domain decomposition is shown in Fig.~\ref{fig:domain-decomposition-sync}, where the elements in the domain are divided into two processing elements, PE-0 and PE-1, such that the elements to the left of $\Omega_{e}$ belong to PE-0 and the remaining to PE-1, with the PE boundary at $x_e$. For the explicit-in-time formulation presented here, nearest-neighbor communication is required to advance the solution in time. This communication requirement arises from flux computations at the element boundaries near the PE boundaries (at $x_e$ in Fig.~\ref{fig:domain-decomposition-sync}). For example, when the upwind flux is used (assuming $a>0$), the flux at $x_e$ is $au_e^-$. This is trivial to compute at PE-0, because $u_e^-$ is available in its memory. However, at PE-1, $u_e^-$ is unavailable. Therefore, the flux is computed by communicating $u_e^-$ from PE-0 to a buffer node at PE-1. Note that for the upwind flux, only one communication is required at the PE boundary. However, when the central flux is used, the flux at $x_e$ is given by $a(u_e^- +u_e^+)/2$, which requires bidirectional communication. These communications typically involve two steps: a communication initiation step followed by a synchronization call to ensure that the data between the source and destination PEs are the same.  

The elements in parallel DG solvers can be divided into two sets, as shown in Fig.~\ref{fig:domain-decomposition-sync}. The first is the set of interior elements, whose computations are independent of the communication between PEs. Second, PE boundary elements, whose computations depend on communication. At each time step, the solution at the PE boundary elements cannot advance unless communication between the PEs is complete. This is ensured by imposing synchronization after communication, as mentioned previously. Indeed, such synchronization of data across PEs poses a major bottleneck in the scalability of massively parallel solvers. Methods for reducing communication costs leverage the idea of overlapping data movements between PEs with computations at interior points \cite{exadg2020}. In some instances \cite{komal2023reactions-at}, the over-decomposition of subdomains is performed at the cost of redundant computations to avoid communication. Despite such optimizations, communication between PEs and data synchronization continues to pose a challenge in the scalability of state-of-the-art DG solvers at extreme scales \rev{\cite{brus2017dgperformance, KRAIS2021186-flexi, blind2023exascale-flexi, klofkorn2012-dune-dg}}. We refer to such implementations of the DG method, where synchronizations are imposed on communications at each time step advancement, as the \textit{synchronous approach}.

%====================================
\section{Asynchronous DG method}
\label{sec:asyncDG}

In the asynchronous computing approach, the communication cost of parallel solvers is reduced by either avoiding the synchronization of data between processing elements (synchronization-avoiding algorithm, SAA) or by avoiding communication altogether for a few time steps (communication-avoiding algorithm, CAA). The implementation of these algorithms modifies the numerical schemes, and their effects must be investigated. To illustrate the asynchronous discontinuous Galerkin (ADG) method, we utilize the synchronization-avoiding algorithm (SAA).

\vspace{0.25cm}
\begin{figure}[ht!]
\begin{center}
    \includegraphics[scale=0.6]{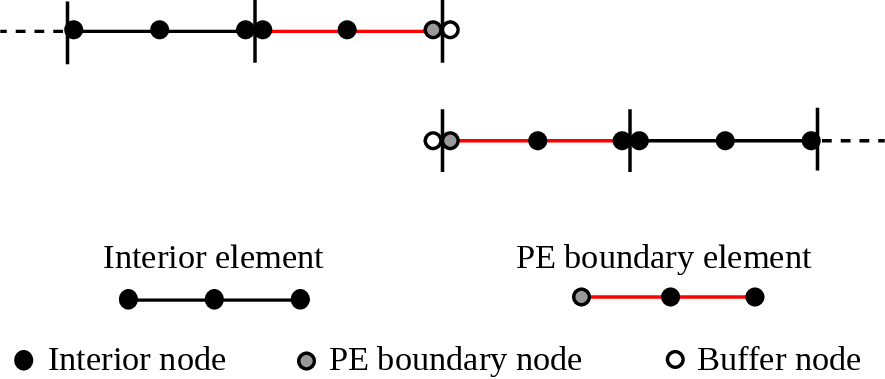}
    \begin{picture}(0,0)
    \put(-212,72){\textcolor{blue}{Communication}}
    \put(-110,84){\textcolor{red}{No synchronization}}
    \put(-152,110){\textcolor{black}{$(u_e^-)^n$}}
    \put(-155,56){\textcolor{red}{$(u_e^-)^{\tilde{n}^-}$}}
    \put(-130,110){\textcolor{red}{$(u_e^+)^{\tilde{n}^+}$}}
    \put(-130,56){\textcolor{black}{$(u_e^+)^n$}}
    \put(-170,88){$\Omega_{e-1}$}
    \put(-224,88){$\Omega_{e-2}$}
    \put(-107,50){$\Omega_{e}$}
    \put(-58,50){$\Omega_{e+1}$}
    \put(-248,84){$x_{e-2}$}
    \put(-193,84){$x_{e-1}$}
    \put(-134,84){$x_{e}$}
    \put(-134,48){$x_{e}$}
    \put(-82,48){$x_{e+1}$}
    \put(-30,48){$x_{e+2}$}
    \put(-200,120){PE-0}
    \put(-80,105){PE-1}
    \put(-128,76){\rotatebox{90}{\textcolor{blue}{\Huge{$\rightarrow$}}}}
    \put(-144,96){\rotatebox{-90}{\textcolor{blue}{\Huge{$\rightarrow$}}}}
    \end{picture}
\caption{\small{An illustration of communication with relaxed synchronization at a PE boundary with two sub-domains.}}
\label{fig:domain-decomposition-async}
\end{center}
\end{figure}

Consider the solution of the linear advection equation, Eq.~(\ref{eq:wave}), as discussed in the previous section. When advancing from time level $n$ to $n+1$, communication between PEs is performed to exchange $u^-$ and $u^+$ from time level $n$, which is required for flux computations at the PE boundary nodes. In the asynchronous approach based on SAA, the communication of these values is initiated; however, the synchronization call to ensure their availability at the destination PE is not enforced. This means that the buffer nodes at the PE boundaries may or may not have $u^-$ or $u^+$ from time level $n$. The time levels of these values depend on the communication speed, which can vary based on hardware and software factors, including network topology, latency, and communication library. As the arrival of messages at PEs can be modeled as a random process \cite{hoefler2010}, we can treat the available time level of $u$ at the buffer nodes, $\tilde{n}$\footnote{$\tilde{n}$ represents the random nature of the variable $n$}, as a random variable. Note that, with the relaxation of synchronization, $u^-$ and $u^+$ at the respective buffer nodes can be at different time levels, as shown in Fig.~\ref{fig:domain-decomposition-async}. Clearly, the delay in data cannot be unbounded. Therefore, we restrict the maximum allowable delay to $L$ time levels. Let $\tilde{k}$ represent the delay at the buffer nodes, such that $\tilde{n}=n-\tilde{k}$ and $\tilde{k} \in \{ 0, 1, 2, \dots , L-1 \}$. When $\tilde{k}=0$, the flux computations are \textit{synchronous}; otherwise, they are \textit{asynchronous}. Furthermore, if $p_k$ represents the probability of delay of $\tilde{k} = k$ time levels at buffer nodes in a simulation, then $\sum_{k=0}^{L-1}p_k=1$. The relaxation of synchronization modifies the fully-discrete matrix equation in Eq.~\eqref{eq:EulerDG} for the PE boundary elements $\Omega_{e-1}$ and $\Omega_e$ (see Fig.~\ref{fig:domain-decomposition-async}) to
\begin{align}
    \boldsymbol{u}_h^{e-1,n+1} &= \boldsymbol{u}_h^{e-1,n} + 2\sigma\boldsymbol{\mathcal{M}}^{-1}\left( \boldsymbol{\mathcal{K}}^-\boldsymbol{u}_h^{e-2,n} + (\boldsymbol{\mathcal{K}}_l + \boldsymbol{\mathcal{S}} + \boldsymbol{\mathcal{K}}_r)\boldsymbol{u}_h^{e-1,n} + \boldsymbol{\mathcal{K}}^+\boldsymbol{u}_h^{e,\tilde{n}^+} \right)  \quad \text{for left PE boundary element}\nonumber\\
    \boldsymbol{u}_h^{e,n+1} &= \boldsymbol{u}_h^{e,n} + 2\sigma\boldsymbol{\mathcal{M}}^{-1}\left( \boldsymbol{\mathcal{K}}^-\boldsymbol{u}_h^{e-1,\tilde{n}^-} + (\boldsymbol{\mathcal{K}}_l + \boldsymbol{\mathcal{S}} + \boldsymbol{\mathcal{K}}_r)\boldsymbol{u}_h^{e,n} + \boldsymbol{\mathcal{K}}^+\boldsymbol{u}_h^{e+1,n} \right)
    \quad \text{for right PE boundary element},
    \label{eq:asyncEulerDG}
\end{align}
where $\boldsymbol{u}_h^{e,\tilde{n}^+}$ and $\boldsymbol{u}_h^{e-1,\tilde{n}^-}$ comprise the DoFs at the buffer nodes of PE-0 and PE-1, respectively. Here, $\tilde{n}^+ = n-\tilde{k}^+$ and $\tilde{n}^- = n-\tilde{k}^-$. Note that these modified equations are effective for all PE boundary elements.

The asynchronous DG method can be summarized as follows. After domain decomposition in parallel solvers, the elements in the subdomains are divided into two sets. First, a set of interior elements whose computations require data that are local to the subdomain and are, therefore, independent of the communication between PEs. The second is a set of PE boundary elements, whose computations require data from neighboring PEs. For the interior elements, a solution update is performed using the standard DG schemes. For example, using Eq.~(\ref{eq:EulerDG}). All the computations at these elements are synchronous. On the other hand, Eq.~(\ref{eq:asyncEulerDG}) is used to advance the solution at the PE boundary elements. Here, the buffer nodes may have delayed DoFs; therefore, computations may be asynchronous. The ADG method is only viable if the schemes produce stable and accurate solutions. Next, an investigation of the numerical properties is presented.

%====================================
\section{Numerical properties}
\label{sec:nm}

%==========
\subsection{Conservation}
\label{sec:conservation}

For equations governing the conservation laws, the DG method, like the finite volume method, ensures conservation because the flux is unique at the boundary or cell interface by consistency. Here, we investigate the effect of asynchrony on the conservative properties of the DG schemes. To aid the discussion, we first consider a synchronous case, as illustrated in Fig.~\ref{fig:domain-decomposition-sync}. The boundary node $x_e$ in the figure is shared between elements $\Omega_{e-1}$ and $\Omega_e$, which are at processing elements PE-0 and PE-1, respectively. For the time advancement from $n$ to $n+1$, the function values ($u^-$ and $u^+$) at $x_e$ are at time level $n$ in both the processing elements, facilitated by data communication and synchronization. As a result, a unique numerical flux can be computed; for example, $\hat{f}\left((u_e^-)^n, (u_e^+)^{n}\right) = a\left(B^+(u_e^-)^n + B^-(u_e^+)^{n}\right)$ which is used for solution updates using Eq.~\eqref{eq:EulerDG} at $\Omega_{e-1}$ and $\Omega_e$. The use of such a numerical flux ensures the conservation property of the DG method. 
Now, consider the asynchronous case shown in Fig.~\ref{fig:domain-decomposition-async}. The expressions for the numerical flux at $x_e$ in elements $\Omega_{e-1}$ and $\Omega_e$ are $\hat{f}\left((u_e^-)^n, (u_e^+)^{\tilde{n}^+}\right) = a\left(B^+(u_e^-)^n + B^-(u_e^+)^{\tilde{n}^+}\right)$ and $\hat{f}\left((u_e^-)^{\tilde{n}^-}, (u_e^+)^n\right) = a\left(B^+(u_e^-)^{\tilde{n}^-} + B^-(u_e^+)^n\right)$, respectively. In the presence of delays ($\tilde{n}^+, \tilde{n}^- \neq n$), these two fluxes are unequal, resulting in a violation of conservation. Hence, a naive implementation of the delays makes the asynchronous DG method non-conservative.

% 
% Restoring flux conservation
% 
To maintain conservation under asynchrony, we propose using all function values at $x_e$ from a common time level that is available to both PEs to compute numerical fluxes at PE boundaries. The common time level can be obtained as $\tilde{n} = \min(\tilde{n}^+, \tilde{n}^-)$, which provides a single-valued numerical flux $\hat{f}\left((u_e^-)^{\tilde{n}}, (u_e^+)^{\tilde{n}}\right) = \hat{f}^{\tilde{n}}(u_e^-,u_e^+) = a(B^+(u_e^-)^{\tilde{n}} + B^-(u_e^+)^{\tilde{n}})$ for the elements $\Omega_{e-1}$ and $\Omega_e$ at $x_e$. Based on this choice, the matrix update equations for the ADG scheme at PE boundaries are
\begin{align}
    \boldsymbol{u}_h^{e-1,n+1} &= \boldsymbol{u}_h^{e-1,n} + 2\sigma\boldsymbol{\mathcal{M}}^{-1}\left( \boldsymbol{\mathcal{K}}^-\boldsymbol{u}_h^{e-2,n} + (\boldsymbol{\mathcal{K}}_l + \boldsymbol{\mathcal{S}})\boldsymbol{u}_h^{e-1,n} + \boldsymbol{\mathcal{K}}_r\boldsymbol{u}_h^{e-1,\tilde{n}} + \boldsymbol{\mathcal{K}}^+\boldsymbol{u}_h^{e,\tilde{n}} \right) \nonumber\\
    \boldsymbol{u}_h^{e,n+1} &= \boldsymbol{u}_h^{e,n} + 2\sigma\boldsymbol{\mathcal{M}}^{-1}\left( \boldsymbol{\mathcal{K}}^-\boldsymbol{u}_h^{e-1,\tilde{n}} + \boldsymbol{\mathcal{K}}_l\boldsymbol{u}_h^{e,\tilde{n}} + (\boldsymbol{\mathcal{S}} + \boldsymbol{\mathcal{K}}_r)\boldsymbol{u}_h^{e,n} + \boldsymbol{\mathcal{K}}^+\boldsymbol{u}_h^{e+1,n} \right).
    \label{eq:asyncEulerDGconserved}
\end{align}
Here, $\boldsymbol{\mathcal{K}}_r\boldsymbol{u}_h^{e-1,\tilde{n}} + \boldsymbol{\mathcal{K}}^+\boldsymbol{u}_h^{e,\tilde{n}}$ and $\boldsymbol{\mathcal{K}}^-\boldsymbol{u}_h^{e-1,\tilde{n}} + \boldsymbol{\mathcal{K}}_l\boldsymbol{u}_h^{e,\tilde{n}}$ are the numerical fluxes at $x_e$ for the two elements $\Omega_{e-1}$ and $\Omega_e$, respectively, which balance each other. These equations are used instead of Eq.~(\ref{eq:asyncEulerDG}) at the PE boundaries in the asynchronous discontinuous Galerkin (ADG) method, which preserve the conservation property. 

% \rev{In a parallel setting, it is implemented based on a communication-avoiding algorithm, where communication and synchronization are performed at a fixed interval, such that delays are deterministic \cite{aditya2019arXiv, goswami2023lserkat-jcp}. Although the local conservation is ensured with these modifications, the effect on the accuracy due to asynchrony is significant, as discussed in Sec.~\ref{sec:accuracy}.}

%generalization

%==========
\subsection{Stability}
\label{sec:stability}

% stability
The stability of a numerical method is crucial to ensure that the associated error does not grow unbounded over time. One of the most popular methods for analyzing numerical stability is the von Neumann method \cite{neumann1950,charney1990numerical}, which computes the amplification factor for an update equation. However, this method renders inapplicable when the update equation comprises of multiple time levels \cite{komal2021stability}. A more generalized approach to assess stability is based on Fourier mode analysis, which also provides information on the dispersive and dissipative nature of errors introduced by numerical schemes \cite{vichnevetsky1982fourier, fourierdg-fang1999, ALHAWWARY2018JCP}. Here, we will use the Fourier analysis to investigate the stability limits of the asynchronous discontinuous Galerkin method. For reference, the stability limits for the standard discontinuous Galerkin method are first obtained.

\subsubsection{Fully-discrete synchronous DG-RK schemes}
We now review the Fourier mode analysis for the fully-discrete equation that uses DG schemes with Lagrange polynomials as basis functions and the Runge-Kutta (RK) scheme for time integration. Henceforth, these schemes are referred to as DG($N_p$)-RK$q$, where $N_p$ is the degree of polynomial basis, and $q$ is the order of accuracy of the RK scheme in time. First, let us assume an initial condition
\begin{equation}
    u(x,0) = u_0(x) = e^{\mathrm{i} \kappa x}, \quad x \in (-\infty , \infty),
    \label{eq:initial-condition}
\end{equation}
to solve the linear advection problem in Eq.~\eqref{eq:wave}.
This results in a wave solution of the form
\begin{equation}
    u(x,t) = e^{\mathrm{i}(\kappa x - \omega t)}.
    \label{eq:analytical-fourier}
\end{equation}
Here, $\mathrm{i} = \sqrt{-1}$, $\kappa$ is the wavenumber of the initial condition, and $\omega$ is the frequency satisfying $\omega = \kappa a$, which is the exact dispersion relation for the linear advection equation.
To compute the numerical dispersion relation of the DG-RK scheme, we seek a solution for an element $\Omega_e$ of the form $\boldsymbol{u}_h^e(x,t) = \hat{\boldsymbol{u}}(t) e^{\mathrm{i}\kappa x_e} = \boldsymbol{\mu} e^{\mathrm{i} ( \kappa x_e -\tilde{\omega} t)}$, where  $\boldsymbol{\hat{u}}(t)$ is the vector of Fourier amplitudes, $\tilde{\omega}$ is the numerical frequency, and $\boldsymbol{\mu} \in  \mathbb{R}^{N_p+1}$.

The fully discrete DG-RK scheme is obtained by considering the semi-discrete form of the PDE for an element $\Omega_e$, provided in Eq.~\eqref{eq:DGscheme}, which \rev{is comprised} of $N_p+1$ ordinary differential equations,
\begin{equation}
    \frac{d \boldsymbol{u}_h^e}{dt} = \mathcal{L}(\boldsymbol{u}_h^e) = \dfrac{2a}{\Delta x} \boldsymbol{\mathcal{M}}^{-1} \left( \boldsymbol{\mathcal{K}}^-\boldsymbol{u}_h^{e-1} + (\boldsymbol{\mathcal{K}}_l + \boldsymbol{\mathcal{S}} + \boldsymbol{\mathcal{K}}_r)\boldsymbol{u}_h^e + \boldsymbol{\mathcal{K}}^+\boldsymbol{u}_h^{e+1} \right),
    \label{eq:DGscheme-modified}
\end{equation}
and the second-order Runge-Kutta (RK2) scheme, as an example, for the time integration,
\begin{align}
    &\boldsymbol{k}_1^e = \Delta t \mathcal{L}(\boldsymbol{u}_h^{e,n}), \quad 
    \boldsymbol{k}_2^e = \Delta t \mathcal{L}(\boldsymbol{u}_h^{e,n} + \boldsymbol{k}_1^e) \nonumber \\
    &\boldsymbol{u}_h^{e,n+1} = \boldsymbol{u}_h^{e,n} + \frac{1}{2}(\boldsymbol{k}_1^e + \boldsymbol{k}_2^e).
    \label{eq:dg2-rk2}
\end{align}
Note that in the first RK stage, the computation of $\boldsymbol{k}_1^e$ depends on $\boldsymbol{u}_h^{e-1,n}, \boldsymbol{u}_h^{e,n}$ and $\boldsymbol{u}_h^{e+1,n}$. Similarly, in the subsequent stage, $\boldsymbol{k}_2^e$ is computed based on $\boldsymbol{k}_1^{e-1}, \boldsymbol{k}_1^e$ and $\boldsymbol{k}_1^{e+1}$ along with $\boldsymbol{u}_h^{e-1,n}, \boldsymbol{u}_h^{e,n}$ and $\boldsymbol{u}_h^{e+1,n}$.
Substituting the numerical solution into the fully-discrete equation, we obtain the update equations from time level $n$ to $n+1$ as
\begin{align}
% K1e
    \boldsymbol{k}_1^e &= 2\sigma \boldsymbol{\mathcal{M}}^{-1} \left( e^{-\mathrm{i} \kappa \Delta x} \boldsymbol{\mathcal{K}}^- + (\boldsymbol{\mathcal{K}}_l + \boldsymbol{\mathcal{S}} + \boldsymbol{\mathcal{K}}_r) + e^{\mathrm{i} \kappa \Delta x } \boldsymbol{\mathcal{K}}^+  \right) \boldsymbol{u}_h^{e,n} = \hat{\boldsymbol{K}}_1^e \boldsymbol{u}_h^{e,n} \nonumber \\
% K1e-1
    \boldsymbol{k}_1^{e-1} &= 2\sigma \boldsymbol{\mathcal{M}}^{-1} \left( e^{-2\mathrm{i} \kappa \Delta x} \boldsymbol{\mathcal{K}}^- + e^{-\mathrm{i} \kappa \Delta x }(\boldsymbol{\mathcal{K}}_l + \boldsymbol{\mathcal{S}} + \boldsymbol{\mathcal{K}}_r) + \boldsymbol{\mathcal{K}}^+  \right)\boldsymbol{u}_h^{e,n} = \hat{\boldsymbol{K}}_1^{e-1} \boldsymbol{u}_h^{e,n} 
    \nonumber \\
%  K1e+1
    \boldsymbol{k}_1^{e+1} &= 2 \sigma \boldsymbol{\mathcal{M}}^{-1} \left( \boldsymbol{\mathcal{K}}^- + e^{\mathrm{i} \kappa \Delta x } (\boldsymbol{\mathcal{K}}_l + \boldsymbol{\mathcal{S}} + \boldsymbol{\mathcal{K}}_r) + e^{2\mathrm{i} \kappa \Delta x } \boldsymbol{\mathcal{K}}^+  \right)\boldsymbol{u}_h^{e,n} = \hat{\boldsymbol{K}}_1^{e+1} \boldsymbol{u}_h^{e,n}
    \nonumber \\
% K2e
    \boldsymbol{k}_2^e &= 2 \sigma \boldsymbol{\mathcal{M}}^{-1} \left( \boldsymbol{\mathcal{K}}^- (e^{-\mathrm{i} \kappa \Delta x} \mathbb{I} + \hat{\boldsymbol{K}}_1^{e-1}) + (\boldsymbol{\mathcal{K}}_l + \boldsymbol{\mathcal{S}} + \boldsymbol{\mathcal{K}}_r )(\mathbb{I} + \hat{\boldsymbol{K}}_1^{e}) + \boldsymbol{\mathcal{K}}^+ (e^{\mathrm{i} \kappa \Delta x} \mathbb{I} + \hat{\boldsymbol{K}}_1^{e+1}) \right) \boldsymbol{u}_h^{e,n} = \hat{\boldsymbol{K}}_2^e \boldsymbol{u}_h^{e,n} \nonumber \\
% Un+1
    \boldsymbol{u}_h^{e,n+1} &= \left(\mathbb{I} + \frac{1}{2}(\hat{\boldsymbol{K}}_1^e + \hat{\boldsymbol{K}}_2^e ) \right) \boldsymbol{u}_h^{e,n} = \boldsymbol{\mathcal{G}}^{\text{s}} \boldsymbol{u}_h^{e,n}.
    \label{eq:dg2-rk2-updates}
\end{align}
Here, $\mathbb{I} $ is the identity matrix of size $(N_p+1) \times (N_p+1) $ and $\boldsymbol{\mathcal{G}}^{s}$ is the amplification matrix for the synchronous scheme considered.
The above update equation can further be transformed into an eigenvalue problem by substituting the numerical solution as
\begin{equation}
    e^{-\mathrm{i} \tilde{\omega} \Delta t} \boldsymbol{\mu} = \boldsymbol{\mathcal{G}}^{\text{s}} \boldsymbol{\mu},
    \label{eq:fully-discrete-eigen-eqn}
\end{equation}
where the eigenvalues are $\lambda_j = e^{-\mathrm{i} \tilde{\omega}_j \Delta t}$ and the respective eigenvectors are $\boldsymbol{\mu}_j$ for  $j = 0, \dots , N_p$. Furthermore, the numerical solution can be written as a linear expansion in the eigenvector space as
\begin{equation}
    \boldsymbol{u}_h^{e,n} = \sum_{j=0}^{N_p} \vartheta_j \boldsymbol{\mu}_j e^{\mathrm{i}(\kappa x_e - \tilde{\omega}_j n\Delta t) } = \sum_{j=0}^{N_p} \vartheta_j \lambda_j^n \boldsymbol{\mu}_j e^{\mathrm{i} \kappa x_e },
    \label{eq:DGRKsolution-eigenvectors}
\end{equation}
where the coefficient $\vartheta_j$ can be derived from the initial conditions.

The numerical solution obtained for the $e$th element in Eq.~\eqref{eq:DGRKsolution-eigenvectors} is a linear combination of $N_p+1$ modes. Each of these modes has its own dispersion and dissipation behavior induced by the eigenvalues $\lambda_j = e^{-\mathrm{i}\tilde{\omega}_j \Delta t}, 0 \leq j \leq N_p$, through the numerical frequency $\tilde{\omega}_j $. It is noteworthy that eigenvalues are generally complex. To analyze the numerical dispersion and dissipation, the wavenumber can be non-dimensionalized as $ K = \dfrac{\kappa \Delta x}{N_p+1}$, and the corresponding non-dimensional numerical frequency is $\tilde{\Omega} = \mathrm{i} \dfrac{ln(\lambda)}{\sigma (N_p+1)}$, where $\sigma  = a\Delta t/\Delta x$.
While the non-dimensional wavenumber is a real number, the non-dimensional numerical frequency can be a complex number that can be expressed as $\tilde{\Omega} = \tilde{\Omega}_r + \mathrm{i} \tilde{\Omega}_{\mathrm{i}} $. The exact dispersion relation requires $\tilde{\Omega} = K$, which means $\tilde{\Omega}_r = K$ and $\tilde{\Omega}_{\mathrm{i}} = 0$.
Among the $N_p+1$ eigenmodes, only the physical mode satisfies the exact dispersion relation for a wide range of wavenumbers. The other modes are known as parasite modes.
Additionally, for a stable scheme, all eigenmodes should have a non-positive $\tilde{\Omega}_\mathrm{i}$, i.e., the numerical dissipation should satisfy the relation $\tilde{\Omega}_{\mathrm{i}}(K) \leq 0$.
Clearly, the behavior of the eigenmodes depends on the Courant number $\sigma$, which can be bounded to obtain a stable solution.

\begin{figure}[h!]
    \centering
    \includegraphics[width=6.5cm]{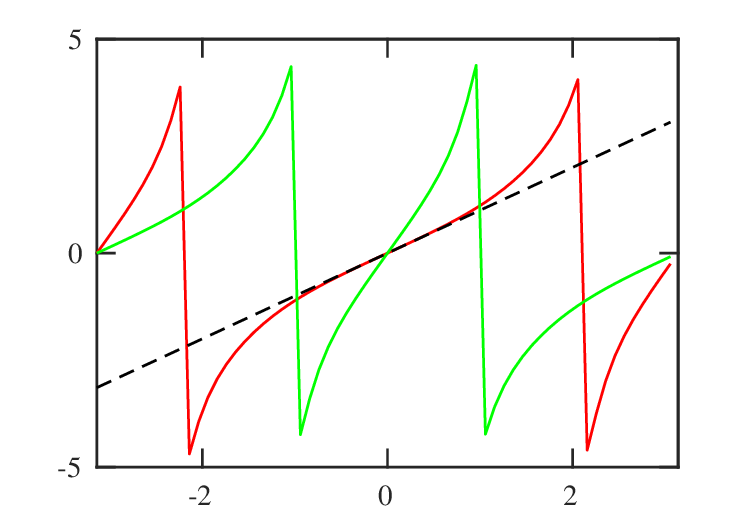}
    \begin{picture}(0,0)
    \put(-155,125){\small (a)}
    \put(-56,125){\small{$\sigma = 0.333$}}
    \put(-190,64){\small{\rotatebox{90}{$\tilde{\Omega}_r$}}}
    \put(-100,-5){\small{$K$}}
    \end{picture}
    \hspace{0.8cm}
    \includegraphics[width=6.5cm]{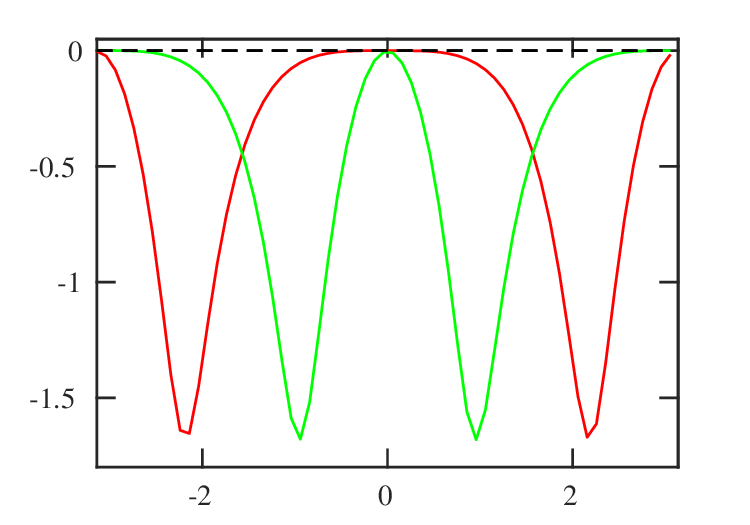}
    \begin{picture}(0,0)
    \put(-155,125){\small (b)}
    \put(-56,125){\small{$\sigma = 0.333$}}
    \put(-190,64){\small{\rotatebox{90}{$\tilde{\Omega}_i$}}}
    \put(-100,-8){\small{$K$}}
    \end{picture}
    
    \vspace{0.4cm}
    \includegraphics[width=6.5cm]{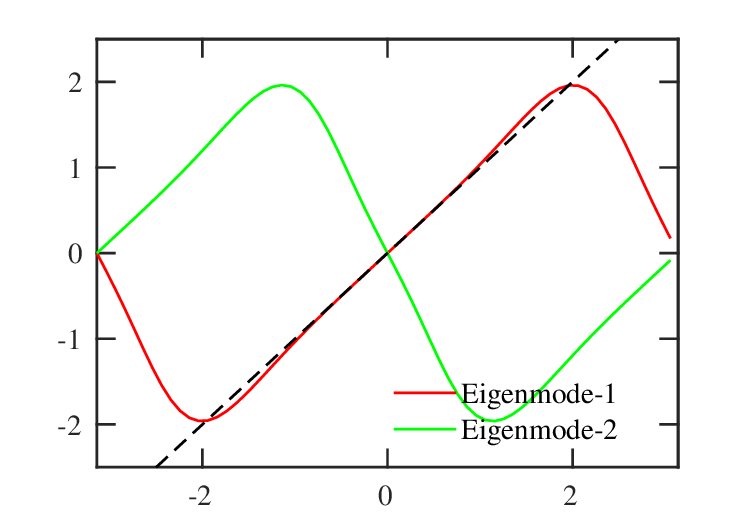}
    \begin{picture}(0,0)
    \put(-155,125){\small (c)}
    \put(-48,125){\small $\sigma = 0.1$}
    \put(-190,64){\small{\rotatebox{90}{$\tilde{\Omega}_r$}}}
    \put(-100,-8){\small{$K$}}
    \end{picture}
    \hspace{0.8cm}
    \includegraphics[width=6.5cm]{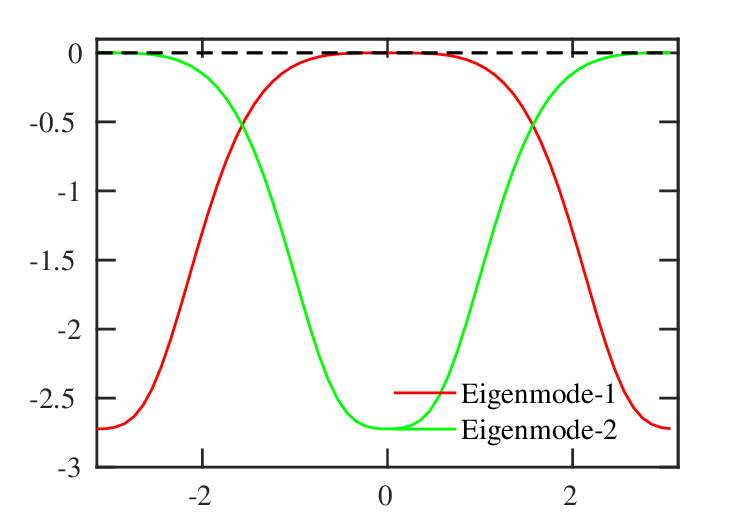}
    \begin{picture}(0,0)
    \put(-155,125){\small (d)}
    \put(-48,125){\small $\sigma = 0.1$}
    \put(-190,64){\small{\rotatebox{90}{$\tilde{\Omega}_i$}}}
    \put(-100,-8){\small{$K$}}
    \end{picture}
    \vspace{0.2cm}
    \caption{\small{Dispersion and dissipation of the fully-discrete DG(1)-RK2 scheme. Dispersion is shown in (a) for $\sigma = 0.333$ and in (c) $\sigma = 0.1$. Dissipation is shown in (b) for $\sigma = 0.333$ and in (d) for $\sigma = 0.1$. Eigenmode-1 (solid red lines) represent the primary mode, and Eigenmode-2 (solid green lines) represent the secondary modes. The dashed black lines representing the exact dispersion relation ($\tilde{\Omega}_r = K, \tilde{\Omega}_{\mathrm{i}} = 0$) are included as references.}}
	\label{fig:dg2-rk2-stability}
\end{figure}

To compute the stability limit for a particular scheme, we consider the DG(1)-RK2 scheme, for which $N_p = 1$, implemented with upwind flux ($B^+ = 1, B^- = 0; a>0$). For this configuration, $\tilde{\Omega}_\mathrm{i} \leq 0$ for all wavenumbers only for $\sigma \leq 0.333$. Figure~\ref{fig:dg2-rk2-stability} plots the real and imaginary parts of the non-dimensionalized numerical frequency for wavenumbers in the range $K \in [-\pi, \pi]$. Note that for $N_p = 1$, the size of the amplification matrix is $2\times 2$, resulting in two eigenmodes, as shown in red and green. Parts (a) and (c) of the figure illustrate the dispersion behavior for $\sigma  = 0.333$ and 0.1, respectively, where $\tilde{\Omega}_r$ is compared with $K$. The exact dispersion relation $\tilde{\Omega}_r=K$ is represented by the dashed black line for reference. As evident in the two figures, Eigenmode-1 (red color) satisfies the aforementioned relation for a considerable range of wavenumbers, signifying the physical mode, whereas the other mode (green color) is the parasite mode. It should be noted that there can be more than one parasite mode based on the size of the amplification matrix; however, the physical mode is unique. The dissipation components of the two eigenmodes for $\sigma  = 0.333$ and $0.1$ are depicted in parts (b) and (d), respectively, where the imaginary part of the numerical frequency for the physical mode remains zero for a range of wavenumbers, thereby indicating no numerical dissipation. However, the parasite mode exhibits significant numerical dissipation due to its substantial damping in the same range of wavenumbers, except at $K = 0$ for $\sigma = 0.333$. 
We observe that as $\sigma$ decreases, the range of wavenumbers for which the exact dispersion relation is valid becomes wider. It is noteworthy that the imaginary parts of the numerical frequencies for all the eigenmodes in Fig.~\ref{fig:dg2-rk2-stability}(b) remain non-positive across all wavenumbers, thereby showcasing the stability of the fully-discrete DG(1)-RK2 scheme with the upwind flux for $\sigma \leq 0.333$. A similar analysis could be performed for other combinations of schemes and fluxes.

\subsubsection{Fully-discrete asynchronous DG-RK schemes}
In the asynchronous approach, it should be noted that the updates at interior elements are performed with synchronous schemes, and at PE boundary elements, which would have delayed values at buffer nodes, asynchronous schemes are used. While the synchronous DG-RK schemes with the update equation $\boldsymbol{u}_h^{e,n+1} = \boldsymbol{\mathcal{G}}^{\text{s}} \boldsymbol{u}_h^{e,n}$ consist of only two time levels ($n$ and $n+1$), the asynchronous DG-RK schemes have multiple time levels. For example, with a delay of $\tilde{k}$ at buffer nodes, the update equation is of the form 
\begin{equation}
    \boldsymbol{u}_h^{e,n+1} = \boldsymbol{\mathcal{G}}^{\text{as},n} \boldsymbol{u}_h^{e,n} + \boldsymbol{\mathcal{G}}^{\text{as},n-\tilde{k}} \boldsymbol{u}_h^{e,n-\tilde{k}},
    \label{eq:async-dg-general}
\end{equation}
where $\boldsymbol{\mathcal{G}}^{\text{as},n}$ and $\boldsymbol{\mathcal{G}}^{\text{as},n-\tilde{k}}$ are the coefficient matrices corresponding to time levels $n$ and $n-\tilde{k}$, respectively. The presence of these two coefficient matrices poses a challenge in analyzing the stability based on the procedure used for the synchronous schemes. To overcome such an issue, Ref.~\cite{donzis2014} used a block matrix approach that provides a single amplification matrix to compute the stability for asynchronous finite difference schemes. We adopt a similar approach here.

To proceed with the analysis for the asynchronous scheme, let us consider the generic element $\Omega_e$ as the PE boundary element (see Fig.~\ref{fig:domain-decomposition-async}) that is at the left PE boundary after domain decomposition. We start with a simple case that uses a delay of $\tilde{k} = 1$ such that the fully-discrete ADG(1)-RK2 scheme uses values from time levels $n-1$, $n$, and $n+1$. For $\tilde{k} = 1$, the maximum allowable delay levels is $L=2$, and we denote this asynchronous scheme as ADG(1)-RK2-$L2$ scheme. Similar to the synchronous stability analysis, we seek a solution of the form $\boldsymbol{u}_h^{e,n} = \boldsymbol{\mu} e^{\mathrm{i}(\kappa x_e - \tilde{\omega} t_n)} $ for the element $\Omega_e$. With a delay at the left boundary, the numerical flux on the left node (at $x_e$) is computed with values from time level $n-1$. Next, substituting the numerical solution into Eq.~\eqref{eq:dg2-rk2}, we obtain the coefficient matrices $\boldsymbol{\mathcal{G}}^{\text{as},n} $ and $\boldsymbol{\mathcal{G}}^{\text{as},n-1}$ for the asynchronous scheme as
\begin{align}
    \boldsymbol{k}_1^{e} &=  2\sigma \boldsymbol{\mathcal{M}}^{-1} \left( (\boldsymbol{\mathcal{S}} + \boldsymbol{\mathcal{K}}_r) + e^{\mathrm{i} \kappa \Delta x} \boldsymbol{\mathcal{K}}^+ \right) \boldsymbol{u}_h^{e,n} + 2 \sigma \boldsymbol{\mathcal{M}}^{-1} \left( e^{-\mathrm{i} \kappa \Delta x} \boldsymbol{\mathcal{K}}^- + \boldsymbol{\mathcal{K}}_l  \right) \boldsymbol{u}_h^{e,n-1}
    \nonumber \\
    &= \hat{\boldsymbol{K}}_{10}^{e} \boldsymbol{u}_h^{e,n} + \hat{\boldsymbol{K}}_{11}^{e} \boldsymbol{u}_h^{e,n-1}
    \nonumber \\
    \boldsymbol{k}_1^{e-1} &=  2\sigma \boldsymbol{\mathcal{M}}^{-1} \left( e^{-2\mathrm{i} \kappa \Delta x} \boldsymbol{\mathcal{K}}^- + e^{-\mathrm{i} \kappa \Delta x} (\boldsymbol{\mathcal{K}}_l + \boldsymbol{\mathcal{S}} + \boldsymbol{\mathcal{K}}_r) + \boldsymbol{\mathcal{K}}^+ \right) \boldsymbol{u}_h^{e,n-1} 
    \nonumber \\
    &= \hat{\boldsymbol{K}}_{11}^{e-1} \boldsymbol{u}_h^{e,n-1}
    \nonumber \\
    \boldsymbol{k}_1^{e+1} &=  2\sigma \boldsymbol{\mathcal{M}}^{-1} \left( \boldsymbol{\mathcal{K}}^- + e^{\mathrm{i} \kappa \Delta x} (\boldsymbol{\mathcal{K}}_l + \boldsymbol{\mathcal{S}} + \boldsymbol{\mathcal{K}}_r) + e^{2\mathrm{i} \kappa \Delta x} \boldsymbol{\mathcal{K}}^+ \right) \boldsymbol{u}_h^{e,n} 
    \nonumber \\
    &= \hat{\boldsymbol{K}}_{10}^{e+1} \boldsymbol{u}_h^{e,n}
    \nonumber \\
    \boldsymbol{k}_2^{e} &= 2 \sigma \boldsymbol{\mathcal{M}}^{-1} \left( \boldsymbol{\mathcal{K}}_l \hat{\boldsymbol{K}}_{10}^{e} + (\boldsymbol{\mathcal{S}} + \boldsymbol{\mathcal{K}}_r) (\mathbb{I} + \hat{\boldsymbol{K}}_{10}^e) + \boldsymbol{\mathcal{K}}^+ (e^{\mathrm{i} \kappa \Delta x} \mathbb{I} + \hat{\boldsymbol{K}}_{10}^{e+1}) \right) \boldsymbol{u}_h^{e,n} 
    \nonumber \\
    &+ 2 \sigma \boldsymbol{\mathcal{M}}^{-1} \left( \boldsymbol{\mathcal{K}}^- (e^{-\mathrm{i} \kappa \Delta x } \mathbb{I} + \hat{\boldsymbol{K}}_{11}^{e-1}) + \boldsymbol{\mathcal{K}}_l (\mathbb{I} + \hat{\boldsymbol{K}}_{11}^{e}) + (\boldsymbol{\mathcal{S}} + \boldsymbol{\mathcal{K}}_r) \hat{\boldsymbol{K}}_{11}^e \right) \boldsymbol{u}_h^{e,n-1} 
    \nonumber \\
    &= \hat{\boldsymbol{K}}_{20}^{e} \boldsymbol{u}_h^{e,n} + \hat{\boldsymbol{K}}_{21}^{e} \boldsymbol{u}_h^{e,n-1}.
    \nonumber \\
    \boldsymbol{u}_h^{e,n+1} &= \left(\mathbb{I} + \frac{1}{2}(\hat{\boldsymbol{K}}_{10}^e + \hat{\boldsymbol{K}}_{20}^e ) \right) \boldsymbol{u}_h^{e,n} +  \frac{1}{2}(\hat{\boldsymbol{K}}_{11}^e + \hat{\boldsymbol{K}}_{21}^e )  \boldsymbol{u}_h^{e,n-1} = \boldsymbol{\mathcal{G}}^{\text{as},n} \boldsymbol{u}_h^{e,n} + \boldsymbol{\mathcal{G}}^{\text{as},n-1} \boldsymbol{u}_h^{e,n-1}.
    \label{eq:async-dg2-rk2-l2}
\end{align}
Based on the block matrix approach used in \cite{donzis2014}, to obtain an amplification matrix $\boldsymbol{\mathcal{G}}^{\text{as}}$,
we introduce a new vector $\boldsymbol{w}^{e,n}$ for the $e$th element that stores the vectors $\boldsymbol{u}_h^{e,n}$ and $\boldsymbol{u}_h^{e,n-1}$ such that a linear transformation can be defined in the following form
\begin{equation}
    \boldsymbol{w}^{e,n+1} = \boldsymbol{\mathcal{G}}^{\text{as}} \boldsymbol{w}^{e,n},
    \label{eq:one-step-async-l2}
\end{equation}
where
\begin{equation*}
    \boldsymbol{w}^{e,n+1} =
    \begin{bmatrix}
    \boldsymbol{u}_h^{e,n+1} \\
    \boldsymbol{u}_h^{e,n}
    \end{bmatrix},
    \quad \boldsymbol{w}^{e,n} =
    \begin{bmatrix}
    \boldsymbol{u}_h^{e,n} \\
    \boldsymbol{u}_h^{e,n-1}
    \end{bmatrix},
    \quad \text{and} \quad
    \boldsymbol{\mathcal{G}}^{\text{as}} = 
    \begin{bmatrix}
    \boldsymbol{\mathcal{G}}^{\text{as},n} && \boldsymbol{\mathcal{G}}^{\text{as},n-1} \\
    \mathbb{I} && \boldsymbol{0}
    \end{bmatrix}.
\end{equation*}
For basis polynomials of degree $N_p$, the dimension of the matrix $\boldsymbol{\mathcal{G}}^{\text{as}}$ is $2(N_P+1) \times 2(N_p+1) $, and $\mathbb{I}$ and $\boldsymbol{0}$ are identity and zero matrices, respectively, of size $(N_p+1) \times (N_p+1)$.
Substituting $\boldsymbol{u}_h^{e,n} = \boldsymbol{\mu} e^{\mathrm{i}(\kappa x_e - \tilde{\omega} t_n)} $ in Eq.~\eqref{eq:one-step-async-l2}, we obtain an eigenvalue problem
\begin{equation}
    e^{-\mathrm{i} \tilde{\omega} \Delta t } \hat{\boldsymbol{\mu}} = \boldsymbol{\mathcal{G}}^{\text{as}} \hat{\boldsymbol{\mu}},
    \label{eq:fully-discrete-adg-rk-eigen-eqn-w}
\end{equation}
where
$
% \begin{equation}
    \hat{\boldsymbol{\mu}}_j = \begin{bmatrix}
    \boldsymbol{\mu}_j\\
    \boldsymbol{\mu}_j e^{\mathrm{i}\tilde{\omega}_j \Delta t}
    \end{bmatrix}
% \end{equation}
$
are $2(N_p+1)$ eigenvectors of the amplification matrix $\boldsymbol{\mathcal{G}}^{\text{as}}$ along with their respective eigenvalues $\lambda_j = e^{-\mathrm{i} \tilde{\omega}_j \Delta t}, j = 0,1,\dots , 2N_p+1 $.
Using these $\hat{\boldsymbol{\mu}}_j$ as basis vectors, we can express the vector $\boldsymbol{w}^{e,n}$ in the eigenvector space as
\begin{equation}
    \boldsymbol{w}^{e,n} =
    \begin{bmatrix}
        \boldsymbol{u}_h^{e,n} \\
        % {} \\
        \boldsymbol{u}_h^{e,n-1}
    \end{bmatrix}
    = \sum_{j=0}^{2N_p+1} \vartheta_j \hat{\boldsymbol{\mu}}_j e^{\mathrm{i}(\kappa x_e - \tilde{\omega}_j n\Delta t) } =
    \begin{bmatrix}
        \sum_{j=0}^{2N_p+1} \vartheta_j \boldsymbol{\mu}_j e^{\mathrm{i}(\kappa x_e - \tilde{\omega}_j n\Delta t) } \\
        % {} \\
        \sum_{j=0}^{2N_p+1} \vartheta_j \boldsymbol{\mu}_j e^{\mathrm{i}(\kappa x_e - \tilde{\omega}_j (n-1)\Delta t) }
    \end{bmatrix}
    = 
    \begin{bmatrix}
        \sum_{j=0}^{2N_p+1} \vartheta_j \lambda_j^{n} \boldsymbol{\mu}_j e^{\mathrm{i} \kappa x_e} \\
        % {} \\
        \sum_{j=0}^{2N_p+1} \vartheta_j \lambda_j^{n-1} \boldsymbol{\mu}_j e^{\mathrm{i} \kappa x_e}
    \end{bmatrix},
    \label{eq:ADGRKsolution-eigenvectors}
\end{equation}
where the coefficients $\vartheta_j$ can be derived from the initial condition. This provides a similar eigenmode representation for $\boldsymbol{u}_h^{e,n}$ as in Eq.~\eqref{eq:DGRKsolution-eigenvectors}, but with $2(N_p+1)$ modes. As discussed earlier, each of these modes has its own dispersion and dissipation behavior, induced by the eigenvalues $\lambda_j, 0 \leq j \leq 2N_p+1$ through the numerical frequency $\tilde{\omega}_j$. To obtain the stability limit of the asynchronous scheme, the stability parameter ($\sigma$) is adjusted such that the imaginary part of the non-dimensional numerical frequency only takes non-positive values for all wavenumbers.

\begin{figure}[h!]
    \centering
    \includegraphics[width=6.5cm]{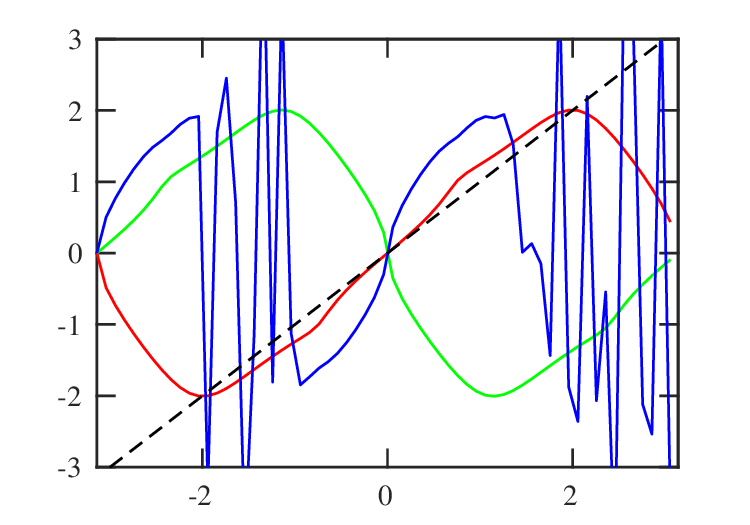}
    \begin{picture}(0,0)
    \put(-155,125){\small (a)}
    \put(-56,125){\small{$\sigma = 0.333$}}
    \put(-190,64){\small{\rotatebox{90}{$\tilde{\Omega}_{r}$}}}
    \put(-97,-8){\small{$K$}}
    \end{picture}
    \hspace{1cm}
    \includegraphics[width=6.5cm]{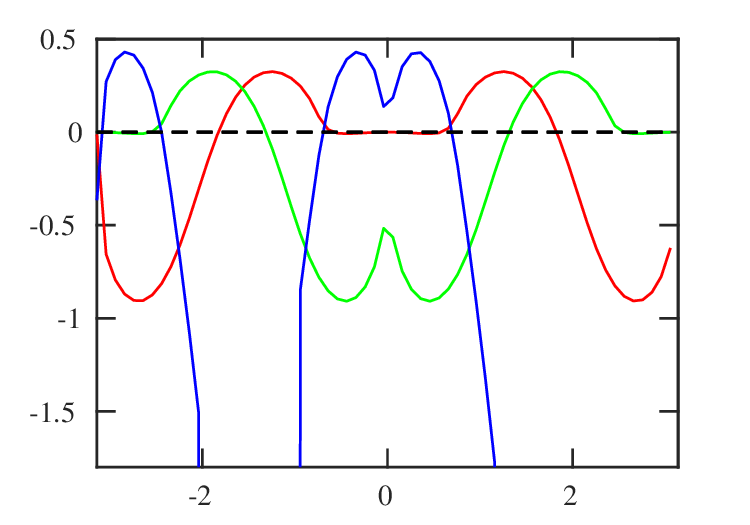}
    \begin{picture}(0,0)
    \put(-155,125){\small (b)}
    \put(-56,125){\small{$\sigma = 0.333$}}
    \put(-190,64){\small{\rotatebox{90}{$\tilde{\Omega}_{\mathrm{i}}$}}}
    \put(-97,-8){\small{$K$}}
    \end{picture}
    \vspace{0.4cm}
    
    \includegraphics[width=6.5cm]{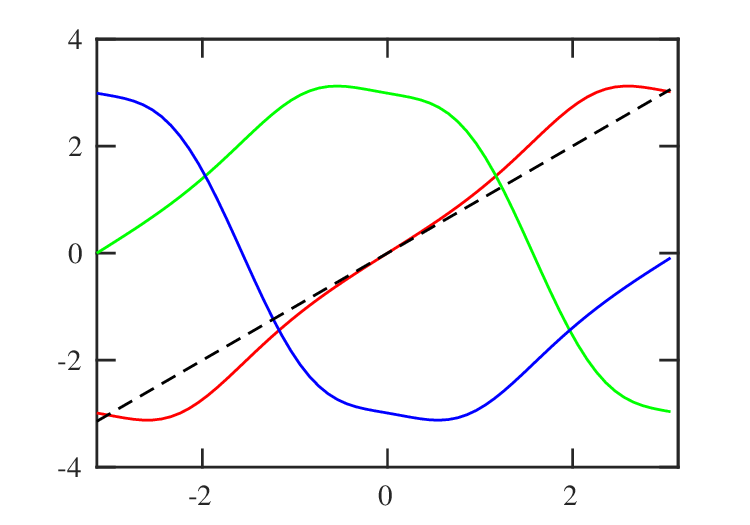}
    \begin{picture}(0,0)
    \put(-155,125){\small (c)}
    \put(-48,125){\small{$\sigma = 0.1$}}
    \put(-190,64){\small{\rotatebox{90}{$\tilde{\Omega}_{r}$}}}
    \put(-97,-8){\small{$K$}}
    \end{picture}
    \hspace{1cm}
    \includegraphics[width=6.5cm]{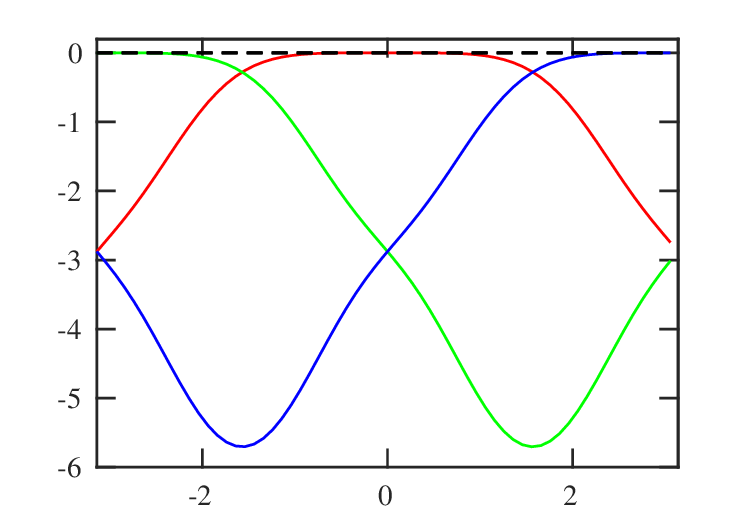}
    \begin{picture}(0,0)
    \put(-155,125){\small (d)}
    \put(-48,125){\small{$\sigma = 0.1$}}
    \put(-190,64){\small{\rotatebox{90}{$\tilde{\Omega}_{\mathrm{i}}$}}}
    \put(-97,-8){\small{$K$}}
    \end{picture}
    \vspace{0.4cm}
    
    \includegraphics[width=6.5cm]{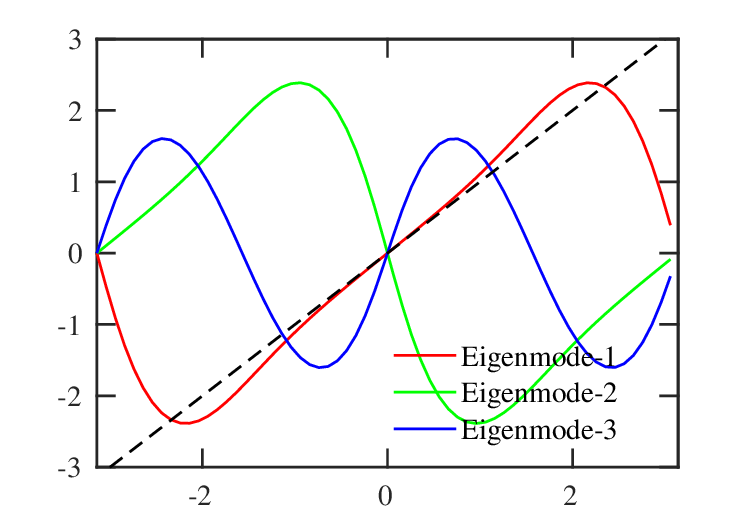}
    \begin{picture}(0,0)
    \put(-155,125){\small (e)}
    \put(-52,125){\small{$\sigma = 0.05$}}
    \put(-190,64){\small{\rotatebox{90}{$\tilde{\Omega}_{r}$}}}
    \put(-97,-8){\small{$K$}}
    \end{picture}
    \hspace{1cm}
    \includegraphics[width=6.5cm]{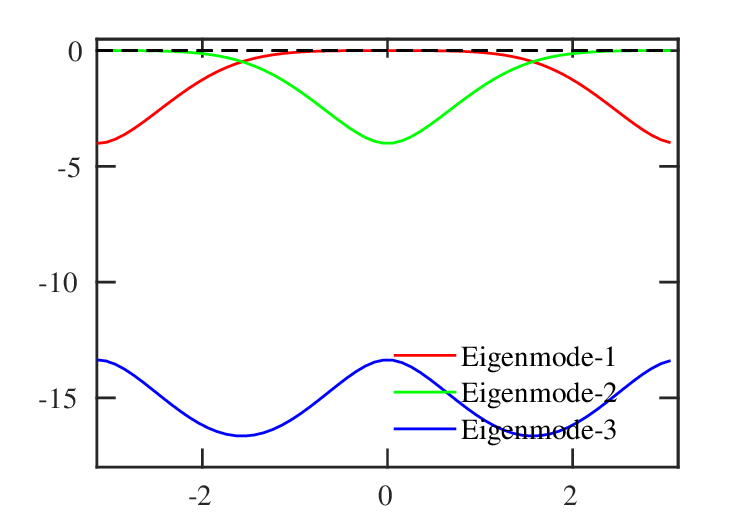}
    \begin{picture}(0,0)
    \put(-155,125){\small (f)}
    \put(-52,125){\small{$\sigma = 0.05$}}
    \put(-190,64){\small{\rotatebox{90}{$\tilde{\Omega}_{\mathrm{i}}$}}}
    \put(-97,-8){\small{$K$}}
    \end{picture}
    \vspace{0.2cm}
    \caption{\small{Dispersion and dissipation of the fully-discrete ADG(1)-RK2-$L2$ scheme. Dispersion is shown in (a) for $\sigma = 0.333$, in (c) $\sigma = 0.1$, and in (e) for $\sigma = 0.05$. Dissipation is shown in (b) for $\sigma = 0.333$, in (d) for $\sigma = 0.1$, and in (f) for $\sigma = 0.05$. Eigenmode-1 (solid red lines) represent the primary modes, and Eigenmode-2 (solid green lines) and Eigenmode-3 (solid blue lines) represent the secondary modes. The dashed black lines representing the exact dispersion relation ($\tilde{\Omega}_r = K, \tilde{\Omega}_{\mathrm{i}} = 0$) are included as references.}}
        \label{fig:async-dg2-rk2-stability}
\end{figure}

For the asynchronous ADG(1)-RK2-$L2$ scheme, the amplification matrix ($\boldsymbol{\mathcal{G}}^{\text{as}}$) is of size $4\times4$, leading to four eigenmodes. Of these four modes, one is a physical mode, and the remaining three are parasite modes. It was observed that one of the parasite modes has very low values ($<-50$) for all the wavenumbers in the dissipation plot; therefore, is excluded in the plots to show other modes on a reasonable range of y-axis. Figure~\ref{fig:async-dg2-rk2-stability} illustrates the dispersion and dissipation behavior of the ADG(1)-RK2-$L2$ scheme for different values of the stability parameter ($\sigma$). Parts (a) and (b) of the figure show the real and imaginary parts of the non-dimensional numerical frequency that demonstrate the dispersion and dissipation properties, respectively, for $\sigma=0.333$ (which is the stability limit for the DG(1)-RK2 scheme). Although the physical mode (red color) follows the exact dispersion relation for a range of wavenumbers about zero, the dissipation plot shows that all the modes for some wavenumbers have $\tilde{\Omega}_{\mathrm{i}}>0$ which indicates that the asynchronous schemes is unstable. As the stability parameter is decreased, the peak values of $\tilde{\Omega}_{\mathrm{i}}$ also decrease. For $\sigma=0.1$, part (d) of the figure shows that $\tilde{\Omega}_{\mathrm{i}}\leq 0$ for all the wavenumbers across different eigenmodes. Therefore, the asynchronous scheme is now stable. The dispersion plot (part (c)) shows that the physical mode aligns with the exact dispersion relation for a range of wavenumbers. As the Courant number is further decreased to $\sigma=0.05$, we observe that the exact dispersion relation is satisfied for a greater range of wavenumbers (see part (e)). In the dissipation plot (part (f)), it is observed that all the modes maintain the stability requirement $\tilde{\Omega}_{\mathrm{i}}\leq 0$. The above analysis establishes the stability of the asynchronous ADG(1)-RK2-$L2$ scheme when the delay $\tilde{k}=1$. Relative to the synchronous DG(1)-RK2 scheme, the asynchronous scheme has a lower stability limit.

To generalize the above analysis for an arbitrary delay $\tilde{k}$, consider the update equation $\boldsymbol{u}_h^{e,n+1} = \boldsymbol{\mathcal{G}}^{\text{as},n} \boldsymbol{u}_h^{e,n} + \boldsymbol{\mathcal{G}}^{\text{as},n-\tilde{k}} \boldsymbol{u}_h^{e,n-\tilde{k}}$. This equation can be rewritten, similar to Eq.~\eqref{eq:one-step-async-l2}, as $\boldsymbol{w}^{e,n+1} = \boldsymbol{\mathcal{G}}^{\text{as}} \boldsymbol{w}^{e,n} $, where
\begin{equation}
    \boldsymbol{w}^{e,n+1} = 
    \begin{bmatrix}
    \boldsymbol{u}_h^{e,n+1} \\
    \boldsymbol{u}_h^{e,n} \\
    \vdots \\
    \boldsymbol{u}_h^{e,n-\tilde{k}+1}
    \end{bmatrix},
    \quad
    \boldsymbol{w}^{e,n} = 
    \begin{bmatrix}
    \boldsymbol{u}_h^{e,n} \\
    \boldsymbol{u}_h^{e,n-1} \\
    \vdots \\
    \boldsymbol{u}_h^{e,n-\tilde{k}}
    \end{bmatrix},
    \quad \text{and} \quad
    \boldsymbol{\mathcal{G}}^{\text{as}} = 
    \begin{bmatrix}
    \boldsymbol{\mathcal{G}}^{\text{as},n} && \boldsymbol{0} && \dots && \boldsymbol{0} && \boldsymbol{\mathcal{G}}^{\text{as},n - \tilde{k}} \\
    \mathbb{I} && \boldsymbol{0} && \dots && \boldsymbol{0} && \boldsymbol{0} \\
    \boldsymbol{0} && \mathbb{I} && \dots && \boldsymbol{0} && \boldsymbol{0} \\
    \vdots && \ddots && \ddots && \ddots && \vdots \\
    \boldsymbol{0} && \boldsymbol{0} && \dots && \mathbb{I} && \boldsymbol{0}
    \end{bmatrix}.
\end{equation}
Now, the amplification matrix is of size $(\tilde{k}+1)(N_p+1) \times (\tilde{k}+1)(N_p+1)$, and therefore provides $(\tilde{k}+1)(N_p+1)$ eigenmodes. For such an asynchronous DG-RK scheme to be stable, the following condition should be satisfied:
\begin{itemize}
    \item The imaginary part of the non-dimensional numerical frequency ($\tilde{\Omega}_{\mathrm{i}}$) is non-positive for all the wavenumbers across different eigenmodes.
\end{itemize}
This ensures that the magnitudes of all eigenvalues of the matrix $\boldsymbol{\mathcal{G}}^{\text{as}}$ are bounded by unity for all wavenumbers.
In general, we observed that the stability limit of the asynchronous DG-RK scheme shrinks as the maximum allowable delay $L$ increases.
However, this effect is mitigated by the fact that the delay distribution typically follows a Poisson distribution, which indicates a low likelihood of encountering higher delays \cite{goswami2023lserkat-jcp}. Furthermore, the asynchronous schemes are only applied at the PE boundary elements, which are commonly a small fraction relative to the total number of elements in the computational domain. Additionally, when the errors generated due to asynchrony at the PE boundary elements propagate into the interior elements, they tend to get dissipated, thus providing stable solutions even with higher Courant number values when compared to a conservative limit imposed by the asynchronous schemes. In the subsequent sections, we show that the use of this asynchronous approach significantly affects the accuracy of the solution and derive new asynchrony-tolerant (AT) fluxes that will provide accurate solutions. These schemes use function values from multiple time levels to recover the accuracy. The stability analysis procedure described in this section can also be used for such scenarios. In summary, the asynchronous approach for DG provides stable solutions, albeit with a more conservative stability limit.

%==========
\subsection{Accuracy}
\label{sec:accuracy}
% error analysis
To quantify the accuracy of PDE solvers, it is necessary to assess the different sources of errors and identify the dominant leading-order terms. Errors in numerical solutions based on the standard DG method arise primarily because of the approximation of the differential operators in finite-dimensional spaces. For time-dependent PDEs, like Eq.~\eqref{eq:wave}, errors appear due to the spatial discretization ($E_s$) and time integration ($E_t$). Assuming a smooth solution for $u$ (in Eq.~\eqref{eq:wave}) and using the semi-discrete form in Eq.~\eqref{eq:DGscheme-modified} that uses an upwind flux, the optimal accuracy relation can be obtained as $E_s \sim \mathcal{O}(\Delta x^{N_p+1})$, where $\Delta x$ is the grid spacing and $N_p$ corresponds to the degree of the polynomial basis functions \cite{hesthaven2007,shu2009}. \rev{Note that the sharp estimate for the error is $\mathcal{O}(\Delta x^{N_p+1/2}).$} When a $q$th-order accurate Runge-Kutta scheme is used for time integration, the error scales as $E_t \sim \mathcal{O}(\Delta t^{q})$. The errors due to the two sources can propagate in space and time, depending on the nature of the PDEs. The overall accuracy of the fully discrete DG($N_p$)-RK$q$ scheme is $\mathcal{O}(\Delta x^{N_p+1}, \Delta t^{q})$. A thorough error analysis of DG-RK schemes for fully discrete equations is provided in Ref.~\cite{shu2005}. The stability parameter (Courant number $\sigma$) can be used to express the time step in terms of grid spacing as $\Delta t \sim \Delta x$. This scaling relation aids in comparing the two error components ($E_s$ and $E_t$). The overall error for the synchronous DG($N_p$)-RK$q$ scheme, now, scales as $\mathcal{O}(\Delta x^{\min(N_p+1,q)})$. In general, if $\Delta t \sim \Delta x^r$ ($r=1$ for advection and $r=2$ for diffusion), the overall error scales as $\mathcal{O}(\Delta x^{\min(N_p+1,rq)})$.

Next, we consider the asynchronous DG method. In addition to the errors introduced due to spatial schemes and time integration, errors are also incurred because of the use of delayed function values in the flux computations at PE element boundaries. As the delays in flux computations are discrete values, i.e., delayed time levels take values $\tilde{n}={n-\tilde{k}}$ where $\tilde{k}\in \{0,1,2,\dots,L-1\}$, the error analysis for the asynchronous DG method can be considered only based on fully discrete equations. Note that the values of $\tilde{k}$ can vary randomly in space and time during the simulation. Therefore, a rigorous analysis similar to that in Ref.~\cite{shu2005} is very complex and beyond the scope of this study. Here, we compute the errors from different sources separately and estimate the scaling of the overall error. Furthermore, the error scaling obtained herein is verified using the numerical experiments in Sec.~\ref{sec:numexp}. To proceed with the analysis, consider the asynchronous implementation of the DG($N_p$)-RK$q$ scheme in solving the linear advection equation (Eq.~\eqref{eq:wave}). As described in Sec.~\ref{sec:asyncDG}, the elements in the discretized domain (see Fig.~\ref{fig:domain-decomposition-async}) can be divided into two sets based on the nature of the computations: a set of interior elements $\Omega_I$ and a set of PE boundary elements $\Omega_{PE}$. At the interior elements, the solution from time level $n$ to $n+1$ evolves with values from the latest time level ($n$) using update equations, such as Eq.~\eqref{eq:EulerDG}. Note that the numerical fluxes are computed as $\hat{f}\left((u_e^-)^{{n}}, (u_e^+)^{{n}}\right) = \hat{f}^{{n}}(u_e^-,u_e^+) = a(B^+(u_e^-)^{{n}} + B^-(u_e^+)^{{n}})$ and no errors are introduced in the updated equations because of these fluxes. However, at the PE boundary elements, in the absence of communication or synchronization, the values at the PE boundary nodes can be from delayed time levels ($n-\tilde{k}$, $\tilde{k}>0$). Therefore, the numerical fluxes at these PE boundary nodes are computed as $\hat{f}\left((u_e^-)^{\tilde{n}}, (u_e^+)^{\tilde{n}}\right) = \hat{f}^{\tilde{n}}(u_e^-,u_e^+) = a(B^+(u_e^-)^{\tilde{n}} + B^-(u_e^+)^{\tilde{n}})$, which induce additional errors in the update equations that compute the DoFs, as in Eq.~\eqref{eq:asyncEulerDGconserved}.

The error incurred due to numerical fluxes computed using delayed function values can be quantified using a Taylor series expansion of the flux function about time level $n$:
\begin{equation}
    \hat{f}^{n-\tilde{k}} = \left. \hat{f}\right|^n - \tilde{k} \Delta t \left.\hat{f}^{\prime}\right|^n + \frac{(\tilde{k} \Delta t)^2}{2} \left. \hat{f}^{\prime\prime} \right|^n + \mathcal{O}\left(\tilde{k}^3\Delta t^3\right).
    \label{eq:taylor-series-f}
\end{equation}
The difference $\hat{f}^{n-\tilde{k}}-\hat{f}^n$ estimates the error due to delayed numerical flux. Substituting Eq.~\ref{eq:taylor-series-f} into the update equation, Eq.~\eqref{eq:asyncEulerDGconserved}, the error (in terms of global truncation error) incurred due to the delayed numerical flux in computing the DoFs at element $e$ can be obtained as 
\begin{equation}
    \left. \tilde{E}_{f,e}^{\tilde{k}}\right|_{\Omega_{PE}} = \dfrac{2\tilde{k}a \Delta t}{\Delta x} \left.\hat{f}' \right|_{}^n + \mathcal{O}\left(\tilde{k}^2\Delta t^2/\Delta x\right),
    \label{eq:errorU-asyncflux}
\end{equation}    
where the factor $2a/\Delta x$ comes from the update equation. This error scales as $\mathcal{O} (\tilde{k}\Delta t/\Delta x)$ based on the leading-order term. Note that, in the absence of a delay, $\tilde{k}=0$, the error incurred is zero. Furthermore, using the stability relation $\Delta t \sim \Delta x$, this scaling in terms of only $\Delta x$ represents the zeroth-order term. Again, it should be noted that this error is incurred only at the PE boundary elements, which are typically in a small fraction relative to the total number of elements per subdomain. In addition, the delay values can vary randomly across the buffer nodes, where the delay becomes zero during synchronization at a time step.

To assess the overall error due to delayed numerical fluxes in the domain, we consider a statistical description of the error due to the random nature associated with delays, which was proposed in \cite{donzis2014}. For domain decomposition, let $P$ represent the number of subdomains mapped to the same number of PEs in the distributed computing setup. As mentioned in Sec.~\ref{sec:asyncDG}, delays in time levels ($\tilde{k}$) due to communications can be random, and the probability of the occurrence of a delay $\tilde{k}=k$ is $p_k$. Note that $\tilde{k} \in {0, 1, \dots, L-1}$, where $L$ represents the maximum allowable delay, and $\sum_{k=0}^{L-1} p_k = 1$. Next, we define two averages to compute the overall error in the domain that considers the random nature of delays. The spatial average $\langle g \rangle = \sum_{e=1}^{N_E}g_i/N_E$ and an ensemble average $\overline{g}$ that is obtained over several simulations, where $g$ is a generic variable. Using these definitions, the average error over the entire domain can be expressed as
\begin{align}
    \langle \overline{{E}_f} \rangle &= \frac{1}{N_E} \sum_{e = 1}^{N_E} \overline{{E}_{f,e}} \nonumber \\ 
    &= \frac{1}{N_E} \left[ \sum_{\Omega_e \in \Omega_I} \overline{{E}_{f,e}} + \sum_{\Omega_e \in \Omega_{PE}} \overline{\tilde{E}_{f,e}}  \right].
    \label{eq:spatialavg-gen}
\end{align}
Here, we split the spatial average error due to the fluxes between the interior ($\Omega_I$) and PE boundary $(\Omega_{PE})$ elements. Obviously, the error incurred due to the numerical flux at the interior elements is zero (${E}_{f,e}=0, \forall e\in\Omega_I$) because of the absence of delays in computations. We now consider the error at the PE boundary element by using ensemble averaging. The expression for the error can be written as 
\begin{align}
     \left.\overline{\tilde{E}_{f,{e}}} \right|_{\Omega_{PE}} & \approx \sum_{k = 0}^{L-1} p_k{\tilde{E}_{f,e}^{k}}  
     \approx \sum_{k = 0}^{L-1} p_k \dfrac{2\tilde{k}a \Delta t}{\Delta x} \hat{f}^{\prime} \approx  \dfrac{2\overline{\tilde{k}} a \Delta t}{\Delta x} \hat{f}^{\prime},
\end{align}
where the leading order term in Eq.~\eqref{eq:errorU-asyncflux} was used to quantify the error due to the delay $\tilde{k}$ and the mean delay is $\overline{\tilde{k}} = \sum_{k=0}^{L-1} k p_k$. Substituting the above expression into Eq.~\eqref{eq:spatialavg-gen}, we get the spatial average as
\begin{align}
    \langle \overline{{E}_f} \rangle &\approx \frac{1}{N_E} \sum_{\Omega_e \in \Omega_{PE}} \overline{\tilde{E}_{f,e}} 
    \approx \frac{N_{PE}}{N_E}  \frac{2\overline{\tilde{k}} a \Delta t}{\Delta x} \langle \hat{f}^{\prime} \rangle \approx \frac{2P}{N_E}  2\overline{\tilde{k}} \sigma \langle \hat{f}^{\prime} \rangle,
\end{align}
where $N_{PE}$ is the number of elements that use delayed numerical flux values, which is twice the number of PEs ($P$) for a one-dimensional problem. Keeping the other parameters constant and using the relation $\Delta x=L/N_E$, the overall error incurred due to the numerical flux can be expressed as 
\begin{align}
    \langle \overline{{E}_f} \rangle & \sim \frac{P}{N_E}\overline{\tilde{k}} \sim P\overline{\tilde{k}}\Delta x.
\end{align}
The above expression demonstrates that the error introduced due to asynchrony, i.e., with the use of delayed values of $u$ for numerical flux computations, scales linearly with the number of PEs ($P$) and the mean delay ($\overline{\tilde{k}}$). The order of error depends on the manner in which the simulations are scaled on a supercomputer. In the case of strong scaling, where simulations are performed with a fixed problem size ($N_E$ is a constant) by varying the number of PEs, the overall error due to asynchrony scales as first-order ($\mathcal{O}(\Delta x)$). On the other hand, in weak scaling, where simulations are performed such that with an increase in the number of PEs, the problem size ($N_E$) is also increased to keep the problem size per PE constant ($N_E/P$ is a constant), and the error is zeroth-order in space. A similar error scaling was also observed when asynchrony was introduced in the finite difference method \cite{donzis2014}.

As mentioned earlier, in the asynchronous DG method, there are three sources of error. In this study, we estimate each separately. For the asynchronous implementation of the DG($N_p$)-RK$q$ scheme to solve the linear advection equation, the error due to spatial discretization, represented by a polynomial basis, scales as $E_s \sim \mathcal{O}(\Delta x^{N_p+1})$, the time-integration error scales as $E_t\sim \mathcal{O}(\Delta t^{q})$, and the error due to the delayed numerical fluxes is $E_f\sim \mathcal{O} (P\overline{\tilde{k}}\Delta x)$. Clearly, the error due to asynchrony dominates the overall error and results in first-order accurate solutions $E \sim \mathcal{O}(\Delta x)$ (based on strong scaling) irrespective of the order of the spatial and temporal components. In general, it should be noted that the errors from different sources propagate both in space and time. The current analysis does not include these effects but reasonably captures the leading order error. In Sec.~\ref{sec:numexp}, we verify the results of the error analysis using numerical experiments. In general, the derived error scaling also holds for any other PDE, including the time-dependent diffusion and advection-diffusion equations, when the error due to asynchrony dominates. In summary, the asynchronous DG method can provide a first-order accurate solution at the best when delayed values are used in the computations of standard numerical fluxes at PE boundaries.

%====================================
\section{Asynchrony-tolerant (AT) numerical fluxes}
\label{sec:atflux}

As the standard numerical fluxes lead to poor accuracy of the ADG method, in this section, we develop new asynchrony-tolerant (AT) numerical fluxes that provide solutions of desired accuracy. When standard numerical fluxes are used with delay function values ($u$) at the buffer nodes, the introduced error is quantified in Eq.~\eqref{eq:errorU-asyncflux}, where the leading order term is $\mathcal{O}(\tilde{k}\Delta t/\Delta x)$. If the lower order terms in this equation are eliminated, then the resulting numerical flux can provide high-order accurate solutions. Let $\hat{f}_e^{n-\tilde{k}} = \hat{f}^{n-\tilde{k}}(u_e^-, u_e^+)$ be the numerical flux computed at the PE boundary node $x_e$ using $u$ values from the time level $n-\tilde{k}$. In the absence of delay due to communication, $\tilde{k}=0$ and the numerical flux reduces to the standard synchronous flux. For $\tilde{k}>0$, i.e., in the presence of delays, the flux introduces low-order terms into the error. To eliminate these low-order terms, we consider a linear combination of the fluxes from multiple time levels, given by  
\begin{equation}
    \hat{f}_e^{\text{at},n} = \sum_{l=L_1}^{L_2} \tilde{c}^l \hat{f}_e^{n-l},
    \label{eq:at-flux-gen}
\end{equation}
where $\hat{f}_e^{\text{at},n}$ is the asynchrony-tolerant (AT) numerical flux for an update from time level $n$ to $n+1$. The coefficients $\tilde{c}^l$, for the range of $l$, are the appropriate coefficients that have to be determined. The limits $L_1$ and $L_2$ can vary across various PE boundary nodes and are functions of $\tilde{k}$, as will be shown momentarily. To determine the coefficients, consider the Taylor series expansion of $\hat{f}_e^{n-l}$ about time level $n$,
\begin{equation}
    \hat{f}_e^{n-l} = \sum_{\zeta = 0}^{\infty} \frac{(-l\Delta t)^{\zeta}}{\zeta !} \left. \hat{f}^{(\zeta)} \right|_e^n.
    \label{eq:taylor-series-f-2}
\end{equation}
This can be substituted into Eq.~\eqref{eq:at-flux-gen} to obtain the necessary constraints to recover the compromised accuracy due to asynchrony.
\begin{align}
    \sum_{l = L_1}^{L_2} \tilde{c}^l \hat{f}_e^{n-l} &= \sum_{l = L_1}^{L_2} \tilde{c}^l \sum_{\zeta = 0}^{\infty} \frac{(-l\Delta t)^{\zeta}}{\zeta !} \left. \hat{f}^{(\zeta)} \right|_e^n \nonumber \\
    &= \sum_{l=L_1}^{L_2} \tilde{c}^l \hat{f}_e^n - \sum_{l =L_1}^{L_2} l \tilde{c}^l \Delta t \left. \hat{f}' \right|_e^n + \sum_{l = L_1}^{L_2} \frac{l^2 \tilde{c}^l }{2} \Delta t^2 \left. \hat{f}'' \right|_e^n - \sum_{l=L_1}^{L_2} \frac{l^3 \tilde{c}^l}{6} \Delta t^3 \left. \hat{f}''' \right|_e^n + \cdots
    \label{eq:eqn-for-constraints}
\end{align}
Based on the above equation, the linear combination should result in the coefficient of $\hat{f}_e^n$ to be unity and the other lower order terms should be eliminated. In particular, we want to match the order of accuracy of the errors arising from the delayed numerical fluxes ($E_f$) and spatial discretization ($E_s$). For a polynomial space of degree $N_p$, assuming that the optimal spatial error scales as $E_s\sim \mathcal{O}(\Delta x^{N_p+1})$, the corresponding error in terms of the time step is $\mathcal{O}(\Delta t^{(N_p+1)/r})$, where $\Delta t \sim \Delta x^r$ is used. Accordingly, the lower order terms in Eq.~\eqref{eq:eqn-for-constraints} are terms with an exponent of $\Delta t$ between $0$ and $(N_p+1)/r$. The necessary constraints for determining $\tilde{c}^l$ can be expressed as
\begin{align}
    \sum_{l = L_1}^{L_2} \tilde{c}^l \frac{(-l \Delta t)^\zeta }{\zeta !} = \begin{cases}
        1, & \zeta = 0 \\
        0, & 0 < \zeta < \dfrac{N_p+1}{r}
    \end{cases},
    \label{eq:at-flux-constraints}
\end{align}
which provide $(N_p+1)/r$ number of equations with full rank. These equations can be solved by considering the same number of values of $l$, i.e., the lower limit $L_1 = \tilde{k}$, and the upper limit $L_2 = \tilde{k}+(N_p+1)/r-1$. By solving this set of equations, we get the coefficients of the AT flux, which approximates the standard flux with the desired order of accuracy. Note that the parameters $N_p+1$ and $r$ only determine the necessary order that is equivalent to the spatial accuracy. In general, AT fluxes of different levels of accuracy can be derived \textit{a priori}. Next, we derive an example based on the procedure outlined above.\\

\noindent\textbf{Example:} {\textit{Fourth-order accurate AT flux}}\\
In this example, we derive an AT flux for PE boundary nodes that is a fourth-order accurate $\mathcal{O}(\Delta t^4)$ approximation of the standard numerical flux $\hat{f}_e^n$ at the $n$th time level. Based on the desired accuracy, the constraints in Eq.~\eqref{eq:at-flux-constraints} should be imposed on the zeroth-, first-, second-, and third-order terms in the Taylor series (see Eq.~\eqref{eq:taylor-series-f-2}). This gives rise to four equations which can be solved by considering four time levels $\{\tilde{k}, \tilde{k}+1, \tilde{k}+2, \tilde{k}+3\}$, i.e., $L_1=\tilde{k}$ and $L_2=\tilde{k}+3$. We construct the linear system $\boldsymbol{A}\boldsymbol{\tilde{c}} = \boldsymbol{b}$, where
\begin{align}
    \boldsymbol{A} = \begin{bmatrix}
        1 & 1 & 1 & 1\\
        -\tilde{k} \Delta t & -(\tilde{k}+1)\Delta t & -(\tilde{k}+2)\Delta t & -(\tilde{k}+3)\Delta t \\
        \dfrac{\left(\tilde{k}\Delta t\right)^2}{2} & \dfrac{\left((\tilde{k}+1)\Delta t\right)^2}{2} & \dfrac{\left((\tilde{k}+2)\Delta t\right)^2}{2} & \dfrac{\left((\tilde{k}+3)\Delta t\right)^2}{2} \\
        -\dfrac{\left(\tilde{k}\Delta t\right)^3}{6} & -\dfrac{\left((\tilde{k}+1)\Delta t\right)^3}{6} & -\dfrac{\left((\tilde{k}+2)\Delta t\right)^3}{6} & -\dfrac{\left((\tilde{k}+3)\Delta t\right)^3}{6}
    \end{bmatrix},
    \boldsymbol{\tilde{c}} = \begin{bmatrix}
        \tilde{c}^{\tilde{k}} \\
        \tilde{c}^{\tilde{k}+1} \\
        \tilde{c}^{\tilde{k}+2} \\
        \tilde{c}^{\tilde{k}+3}
    \end{bmatrix},
    \text{ and }
    \boldsymbol{b} = \begin{bmatrix}
        1 \\
        0 \\
        0 \\
        0
    \end{bmatrix}.
    \label{eq:at2-flux-system-r1}
\end{align}
The solution to this linear system provides the coefficients, which are then substituted into Eq.\eqref{eq:at-flux-gen} to get the AT numerical flux,
\begin{equation}
    \hat{f}_e^{\text{at},n} = \frac{\tilde{k}^3 + 6 \tilde{k}^2 + 11 \tilde{k} + 6}{6} \hat{f}_e^{n - \tilde{k}} - \frac{\tilde{k}^3 + 5 \tilde{k}^2 + 6 \tilde{k}}{2}  \hat{f}_e^{n - \tilde{k} -1} + \frac{\tilde{k}^3 + 4 \tilde{k}^2 + 3 \tilde{k}}{2}  \hat{f}_e^{n - \tilde{k} -2} - \frac{\tilde{k}^3 + 3 \tilde{k}^2 + 2 \tilde{k}}{6}  \hat{f}_e^{n - \tilde{k} -3}.
    \label{eq:at2-flux-r1}
\end{equation} 
Note that when $\tilde{k}=0$, the expression above reduces to the standard flux. In the presence of delays, the leading order term in the error is  $({49}/{864}) k(k+1)(k+2)(k+3)\left. \hat{f}^{(iv)} \right|_e^{n} \Delta t^4$. When error due to asynchrony dominates the overall error, following the procedure in Sec.~\ref{sec:accuracy}, the overall error scales as
\begin{align}
    \langle \overline{{E}} \rangle & \sim \frac{P}{N_E}
    \left( \overline{\tilde{k}^4} + 6 \overline{\tilde{k}^3} + 11 \overline{\tilde{k}^2} +6 \overline{\tilde{k}} \right) \Delta t^4 \sim \frac{P}{N_E}
    \left( \overline{\tilde{k}^4} + 6 \overline{\tilde{k}^3} + 11 \overline{\tilde{k}^2} +6 \overline{\tilde{k}} \right) \Delta x^{4r},
    \label{eq:error-AT-example}
\end{align}
where $\Delta t \sim \Delta x^r$ is used to express the error in terms of the grid spacing. The error with AT fluxes continues to scale linearly with the number of processing elements $P$. With respect to delay statistics, scaling also depends on higher order moments of the delays, which are bounded (see Eq. (41) in \cite{konduri2017at}).

\begin{table}
    \begin{center}
	\begin{tabular}{ | c | c | c | l | l | } 
    	\hline
		Order & $L_1$ & $L_2$ & Coefficients $ \tilde{c}^l $ with $\tilde{k} = k$ & Leading order terms \\
		\hline
		\hline
		2 & $\tilde{k}$ & $\tilde{k} + 1$ & $\tilde{c}^{k} = (k + 1)$,  $\tilde{c}^{k+1} = -k $ & $\dfrac{1}{2}k(k+1)\left. \hat{f}'' \right|_e^{n}\Delta t^2$ \\
		{} & {} & {} & {} & {} \\
		3 & $\tilde{k}$ & $\tilde{k} + 2$ & $\tilde{c}^{k} = \dfrac{(k^2 + 3k + 2)}{2}$, $\tilde{c}^{k+1} = -(k^2 + 2k)$, $ \tilde{c}^{k+2} = \dfrac{(k^2 + k)}{2}$ & $\dfrac{5}{24}k(k+1)(k+2) \left. \hat{f}''' \right|_e^{n} \Delta t^3$ \\
		{} & {} & {} & {} & {} \\
		4 & $\tilde{k}$ & $\tilde{k}+3$ & $\tilde{c}^{k} = \dfrac{(k^3 + 6k^2 + 11k + 6)}{6}$, $\tilde{c}^{k+1} = - \dfrac{(k^3 + 5k^2 + 6k)}{2}$ & $\dfrac{49}{864} k(k+1)(k+2)(k+3)\left. \hat{f}^{(iv)} \right|_e^{n} \Delta t^4$ \\
		{} & {} & {} & $\tilde{c}^{k+2} = \dfrac{(k^3 + 4k^2 + 3k)}{2}$, $\tilde{c}^{k+3} = - \dfrac{(k^3 + 3k^2 + 2k)}{6} $ & {} \\
		\hline
	\end{tabular}
    \end{center}
    \caption{Coefficients of asynchrony-tolerant (AT) fluxes that provide higher order accurate solutions with the asynchronous discontinuous Galerkin (ADG) method. The leading order terms in the truncation error are also provided to show the error dependence on delay ($\tilde{k}$).}
    \label{table:at-flux}
\end{table}

Table~\ref{table:at-flux} lists the coefficients for the second-, third-, and fourth-order accurate AT fluxes, along with the leading order terms in the truncation error. In general, the error scaling due to asynchrony can be expressed as 
\begin{align}
    \langle \overline{{E_f}} \rangle & \sim \frac{P}{N_E} \Delta t^\alpha
    \sum_{m=1}^{\tau} \gamma_m \overline{\tilde{k}^m},
    \label{eq:error-AT-gen}
\end{align}
where $\tau$ depends on the number of time levels involved in the flux computation, $\alpha$ is the expected order of accuracy in time, and $\gamma_m$ is the coefficient of $m$-th moment of delay. In developing the AT fluxes, we used only fluxes from various delayed time levels to provide the desired accuracy. These fluxes are carefully crafted extrapolation schemes that preserve the conservation property of the DG method. It should be noted that the AT fluxes were developed without considering the expression/structure of the standard fluxes that are a function of $u^+$ and $u^-$ at the boundary nodes. It is also possible to derive AT fluxes by considering the delayed values of primitive variables $u^+$ and $u^-$. Furthermore, in solving PDEs with multiple spatial dimensions, we can also derive AT fluxes that use fluxes from spatial neighborhoods in addition to multiple time levels. In the next section, we verify the performance of the AT numerical fluxes derived here.

%====================================
\section{Numerical simulations}
\label{sec:numexp}

To verify the performance of the asynchronous discontinuous Galerkin (ADG) method, numerical simulations of one-dimensional linear and nonlinear partial differential equations are considered in this section. First, simulations of the linear advection equation in Eq.~\eqref{eq:wave} with a constant advection speed $a = 1$, which has a simple analytical solution to quantify the error, are used to validate the error scaling. Second, simulations of the nonlinear viscous Burgers' equation,
\begin{equation}
    \frac{\partial u}{\partial t} + u \frac{\partial u}{\partial x} = \nu \frac{\partial ^2u}{\partial x^2},
    \label{eq:burgers}
\end{equation}
\rev{
which contains both the first and second derivatives ($\nu$ is the viscosity coefficient), are used to assess the impact of asynchrony in capturing nonlinear and multi-scale features. 
Lastly, simulations of the compressible Euler equations,
\begin{align}
    \frac{\partial \rho}{\partial t} + \nabla \cdot (\rho \boldsymbol{u}) &= 0, \nonumber \\
    \frac{\partial \rho \boldsymbol{u}}{\partial t} + \nabla \cdot (\rho \boldsymbol{u} \otimes \boldsymbol{u} + \mathbb{I}p) &= \boldsymbol{0}, \nonumber \\
    \frac{\partial \rho e_0}{\partial t} + \nabla \cdot (\rho e_0 + p) \boldsymbol{u} &= 0,
    \label{eq:compressible-euler-3d}
\end{align}
where $\rho$ is the density, $\boldsymbol{u}$ is the fluid velocity in $d$-dimensions, and $e_0 = \dfrac{1}{2}\rho \boldsymbol{u}\cdot \boldsymbol{u} + \dfrac{p}{\rho (\gamma -1)}$ is the total energy, are performed to demonstrate the ability of the asynchronous DG method in capturing shocks and higher-dimensional implementations.
}
%==========
\subsection{Simulation details}
\label{sec:sd}
The linear advection and nonlinear viscous Burgers' equations are solved in a periodic domain of length $2\pi$. A multi-scale initial condition is specified using a linear combination of sinusoidal waves, with different amplitudes $A(\kappa)$ and phase angles $\phi_\kappa$ for each wave number $\kappa$, given by
\begin{equation}
    u(x,0) = \sum_{\kappa} A(\kappa)sin(\kappa x + \phi_{\kappa}). \label{eq:initialcond}
\end{equation}
Incorporating the phase angles $\phi_{\kappa}$ helps prevent scenarios in which the boundaries of the processing elements coincide with zero values of the initial condition or its gradient. This ensures that the asynchrony effect is not diluted at the PE boundaries.
The errors in the numerical solution for the linear advection equation are obtained against the analytical solution $u_a(x,t)$,
\begin{equation}
u_{a}(x,t) = \sum_{\kappa} A(\kappa) \sin(\kappa x + \phi_{\kappa} - \kappa at).
\label{eq:analytical}
\end{equation}
In the case of the nonlinear Burger's equation, due to the absence of a simple analytical solution, the error is determined using solutions obtained from finely resolved simulations performed using a fourth-order accurate numerical scheme.

\rev{
%  Compressible Euler equations
Sod's shock tube problem \cite{SOD1978JCP} is considered for the one-dimensional compressible Euler equations, where the initial conditions on the domain $x \in [0, 0.01]$ is given as,
\begin{equation}
    [\rho(x,0), u(x,0), p(x,0)] = \begin{cases}
        [1.0, 0.0, 1.0] &\text{ for } x < 0.005 \\
        [0.125, 0.0, 0.1] &\text{ for } x >= 0.005
    \end{cases}
    \label{eq:sod-initial}
\end{equation}
It has an exact solution, which can be obtained by solving Riemann problems, and is used to compute errors in the numerical solutions obtained using the synchronous and asynchronous DG schemes \cite{toro2009}.
}

% numerical schemes
The spatial derivatives in Eqs.~\eqref{eq:wave} and \eqref{eq:burgers} are discretized using the discontinuous Galerkin (DG), and local discontinuous Galerkin (LDG) methods, respectively, with Lagrange polynomials as basis functions. The numerical experiments consider basis polynomials of degrees one, two, and three that would provide an order of accuracy of two, three, and four, respectively, in the case of synchronous implementation.
For the advection term, the upwind numerical flux is used:
\begin{equation*}
    (au_h^e)^*|_{x_e} = \begin{cases}
    a u_e^-, &a \geq 0 \\
    a u_e^+, &a < 0.
    \end{cases}
\end{equation*}
Note that for Burger's equation, the advection speed $a=u$ can vary in space and time. An alternating flux \cite{cockburn1998ldg} is used for the viscous term.
\rev{
For compressible Euler equations, the discontinuous Galerkin method with the local Lax-Friedrichs flux is used, which can be expressed as
\begin{equation}
    \hat{f}(u_e^-, u_e^+) = \frac{f(u_e^-) + f(u_e^+)}{2} - \frac{\lambda}{2}(u_e^+ - u_e^-),
\end{equation}
where $\lambda = max(|f_{u}(u_e^-)|, |f_{u}(u_e^+)|)$. Additionally, the MUSCL TVBM limiter is used with the scheme to capture shocks and ensure total variation stability \cite{hesthaven2007}.
}
Time integration is performed using low-storage Runge-Kutta (RK) schemes \cite{WILLIAMSON1980, KENNEDY2000} with different orders of accuracy to match the accuracy of spatial discretization in demonstrating the error scaling.

Simulations are performed in three configurations: (1) synchronous implementation with standard fluxes (DG), (2) asynchronous implementation with standard fluxes (ADG), and (3) asynchronous implementation with asynchrony-tolerant (AT) fluxes (ADG-AT). In asynchronous implementations, delays ($\tilde{k}$) due to communication at the PE boundary nodes are simulated/emulated using a random number generator. This procedure, introduced in \cite{donzis2014}, assumes a uniform distribution in interval $[0,1]$. For a particular $L$, which is the maximum allowable delay, a probability set $\{p_0, p_1, \dots, p_{L-1} \}$ is chosen, where $p_k$ corresponds to the probability of having delay $\tilde{k}=k$ time levels. Based on the probability set, the interval $[0,1]$ is partitioned into $L$ bins such that the $k$th partition is of size $p_k, k \in \{0, 1, 2, \dots , L-1 \}$. At each time step, a random number between $0$ and $1$ is drawn at each PE boundary, which is then matched with the bin interval, determining the delay for that particular instance. For example, a maximum allowable delay $L = 3$ results in delays $\tilde{k} \in \{0,1,2\}$. If a probability set $\{0.3, 0.4, 0.3\}$ is imposed in a simulation, then the random number is mapped to one of the three bins; $[0, 0.3)$, $[0.3, 0.7)$, and $[0.7, 1]$ which correspond to delays of $0$, $1$, and $2$ time levels, respectively. The implementation based on simulated delays ensures that we have complete control over the statistics of the communication delays, allowing for a convenient comparison of the numerical results with theoretical predictions. To ensure statistical independence of the results, asynchronous simulations in each configuration are performed five times using different random seeds. All the simulations use a time step $\Delta t$ determined using a constant Courant number.

%====================================
\subsection{Results}
\label{sec:results}

\begin{figure}[h!]
    \centering
    \includegraphics[width=6.5cm]{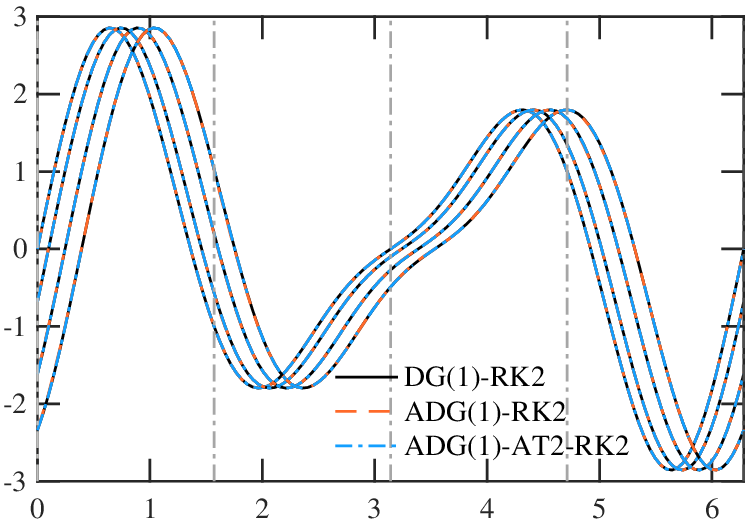}
    \begin{picture}(0,0)
    \put(-85,102){\small (a)}
    \put(-198,64){\small{\rotatebox{90}{$u_h$}}}
    \put(-100,-8){\small{$x$}}
    \put(-130,94){\large{$t$}}
    \put(-152,70){\huge{\rotatebox{30}{$\longrightarrow$}}}
    \end{picture}
    \hspace{0.8cm}
    \includegraphics[width=6.5cm]{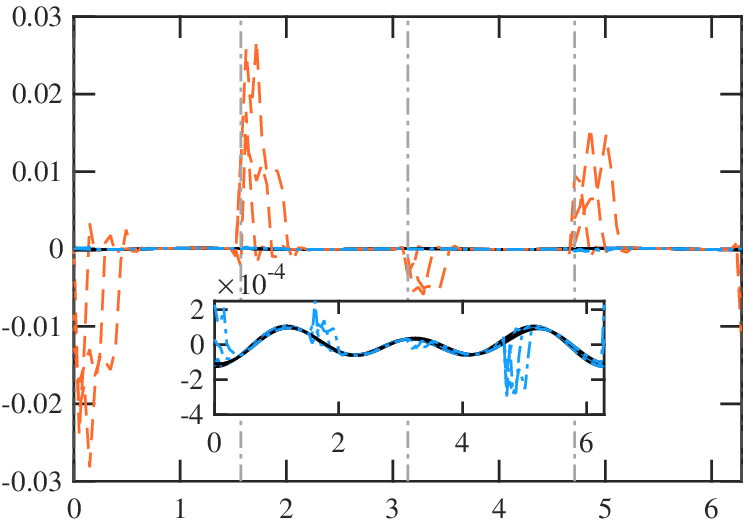}
    \begin{picture}(0,0)
    \put(-82,102){\small (b)}
    \put(-198,56){\small{\rotatebox{90}{$u_h-u_a$}}}
    \put(-100,-8){\small{$x$}}
    \put(-113,94){\large{$t$}}
    \put(-138,70){\huge{\rotatebox{30}{$\longrightarrow$}}}
    \end{picture}
    \vspace{0.2cm}
    \caption{\small{Time evolution of numerical solution of the linear advection equation using DG(1)-RK2 (solid black lines), ADG(1)-RK2 (dashed orange lines) and ADG(1)-AT2-RK2 schemes (dash-dotted blue lines). (a) Scalar field $u_h$. (b) Error $E_h^n = u_h^n - u_{a}(x_h, t^n)$. Vertical dash-dotted lines correspond to PE boundaries. Simulation parameters: $N_E = 128$, $\sigma=0.1$, $P = 4$, and $L = 3$ with $\{p_0, p_1, p_2\} = \{0.6, 0.2, 0.2\}$ for the asynchronous computations. Inset in (b): time evolution of error for DG(1)-RK2 and ADG(1)-AT2-RK2 schemes.}}
    \label{fig:SyncVsAsync}
\end{figure}

Figure~\ref{fig:SyncVsAsync} shows the temporal evolution of the numerical solution $u_h$ and the error $E_h = u_h-u_a$ for the linear advection equation using three schemes: synchronous DG(1)-RK2 (solid black lines), asynchronous ADG(1)-RK2 (dashed orange lines), and asynchronous ADG(1)-AT2-RK2 that uses a second-order AT numerical flux (dash-dotted blue lines). The initial condition consists of sinusoidal waves with wavenumbers $\kappa = \{2, 3\}$ and corresponding amplitudes $A(\kappa) = \{2, 1\}$. 
The other parameters of the simulations are $N_E = 128$, $\sigma=0.1$, $P = 4$, and the maximum allowable delay $L = 3$ time levels. A probability set $p_k=\{0.6, 0.2, 0.2\}$ is imposed for the delays. Part (a) of the figure illustrates the time evolution of solution $u_h$. As expected, the initial condition propagates to the right with a constant advection speed $a$. The three schemes exhibit negligible differences in the solution.
However, in part (b), the error plot effectively distinguishes the performance of the schemes. The asynchronous DG (ADG) scheme with delayed fluxes incurs large peaks near the PE boundaries (vertical dash-dotted lines), indicating significant errors due to asynchrony. In contrast, the errors for the synchronous DG scheme and the asynchronous DG scheme with AT fluxes are comparable and two orders of magnitude lower than those of the ADG scheme. Additionally, we observe that the errors introduced by delayed fluxes in the solutions propagate into the interior elements of the computational domain with time. This behavior is expected because the error evolution for linear problems follows the same PDE as the solution $u_h$.

\begin{figure}[h!]
    \centering
    \includegraphics[width=5.3cm]{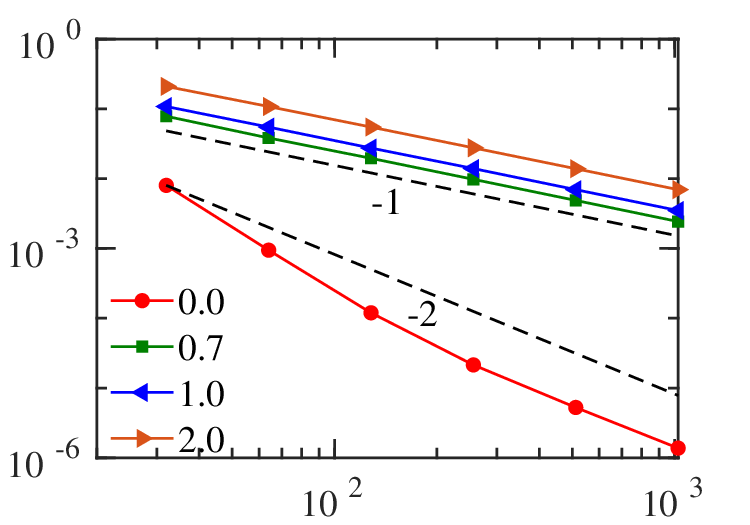}
    \begin{picture}(0,0)
    \put(-40,84){\small (a)}
    \put(-163,48){\small{\rotatebox{90}{$\langle \overline{E} \rangle $}}}
    \put(-78,-8){\small{$N_E$}}
    \put(-131,50){\scriptsize{Mean delay ($\overline{\tilde{k}}$)}}
    \end{picture}
    % \hspace{0.05cm}
    \includegraphics[width=5.3cm]{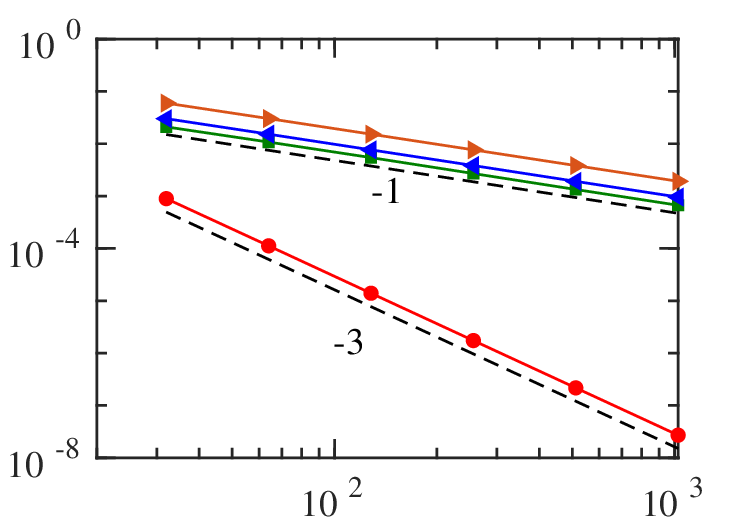}
    \begin{picture}(0,0)
    \put(-40,84){\small (b)}
    \put(-163,48){\small{\rotatebox{90}{$\langle \overline{E} \rangle $}}}
    \put(-78,-8){\small{$N_E$}}
    % \put(-160,60){\small Mean delay}
    \end{picture}
    % \hspace{0.05cm}
    \includegraphics[width=5.3cm]{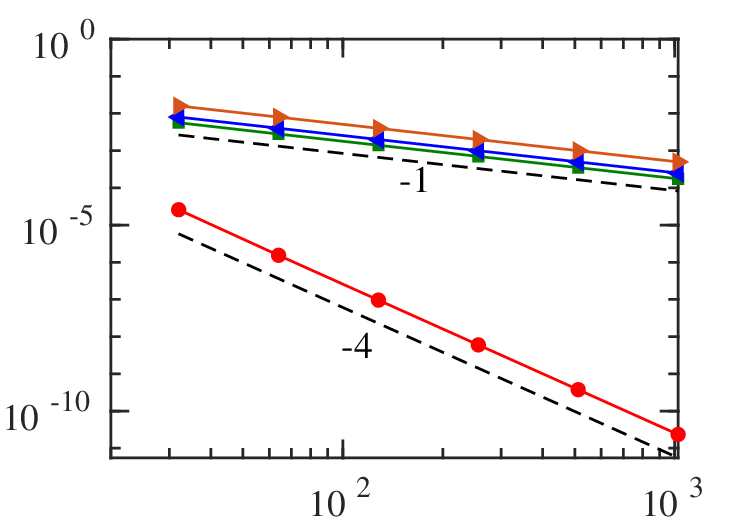}
    \begin{picture}(0,0)
    \put(-40,84){\small (c)}
    \put(-163,48){\small{\rotatebox{90}{$\langle \overline{E} \rangle $}}}
    \put(-78,-8){\small{$N_E$}}
    \end{picture}
    % \vspace{0.1cm}
    \caption{\small{Convergence plot of the average overall error $\langle \overline{E} \rangle $ with increasing grid resolution. Results are obtained from the simulations of the linear advection equation using the schemes (a) DG(1)-RK2 and ADG(1)-RK2 with $\sigma = 0.1$, (b) DG(2)-RK3 and ADG(2)-RK3 with $\sigma = 0.04$, and (c) DG(3)-RK4 and ADG(3)-RK4 with $\sigma  = 0.01$. Different lines correspond to varying degrees of asynchrony with mean delays: $\overline{\tilde{k}} = 0.0$ (red), $0.7$ (green), $1.0$ (blue), $2.0$ (orange). Dashed black lines with a slope of $-1$, $-2$, $-3$ and $-4$ are shown for reference.}}
    \label{fig:accuracyplot-asyncdg}
\end{figure}

To verify the error scaling relations obtained from the analysis presented in Sec.~\ref{sec:accuracy}, we now proceed to the statistical description of the error. The ensemble average of the error is obtained by executing each simulation configuration five times with different random seeds for sampling the delays. The spatial average is computed based on absolute error values at each point. Figure~\ref{fig:accuracyplot-asyncdg} provides a comparison of the accuracy between the synchronous and asynchronous DG schemes. The schemes in the synchronous DG cases are DG(1)-RK2, DG(2)-RK3, and DG(3)-RK4 and in the asynchronous DG cases are ADG(1)-RK2, ADG(2)-RK3, and ADG(3)-RK4, which are reported in parts (a), (b), and (c) of the figure, respectively. The same initial condition utilized in the previous figure is also employed here. In the simulations, the number of PEs is set to $P = 8$. For asynchronous cases, the maximum delay levels is restricted to $L=3$ with probability sets $\{0.5, 0.3,0.2\}$, $\{0.3, 0.4,0.3\}$, and 
$\{0.0, 0.0,1.0\}$ that provide mean delays of $\overline{\tilde{k}} =$ $0.7$, $1.0$, and $2.0$, respectively. A mean delay of $\overline{\tilde{k}} = 0.0$ corresponds to the synchronous case in all plots. From the plots, we observe that the error in the synchronous cases decreases with slopes $-2$, $-3$ and $-4$ (in parts (a), (b), and (c), respectively), as expected from the three schemes that use linear, quadratic, and cubic polynomials as basis functions. However, in the asynchronous cases, the accuracy drops to the first order regardless of the order of the basis polynomials. Furthermore, with an increase in the degree of asynchrony, i.e., as $\overline{\tilde{k}}$ increases, the magnitude of the error also increases. These observations are consistent with the scaling relation, Eq.~\eqref{eq:errorU-asyncflux}, obtained in error analysis. Evidently, the poor accuracy of the asynchronous DG method cannot be used for high-fidelity numerical simulations. The order of accuracy is expected to recover with the use of AT fluxes, which will be investigated next.

\begin{figure}[h!]
    \centering
    \includegraphics[width=5.3cm]{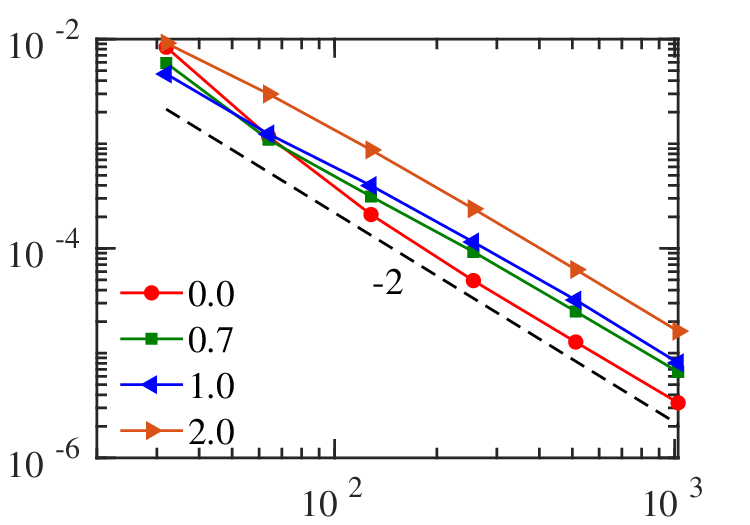}
    \begin{picture}(0,0)
    \put(-40,84){\small (a)}
    \put(-163,48){\small{\rotatebox{90}{$\langle \overline{E} \rangle $}}}
    \put(-78,-8){\small{$N_E$}}
    \put(-131,51){\scriptsize{Mean delay ($\overline{\tilde{k}}$)}}
    \end{picture}
    % \hspace{0.05cm}
    \includegraphics[width=5.3cm]{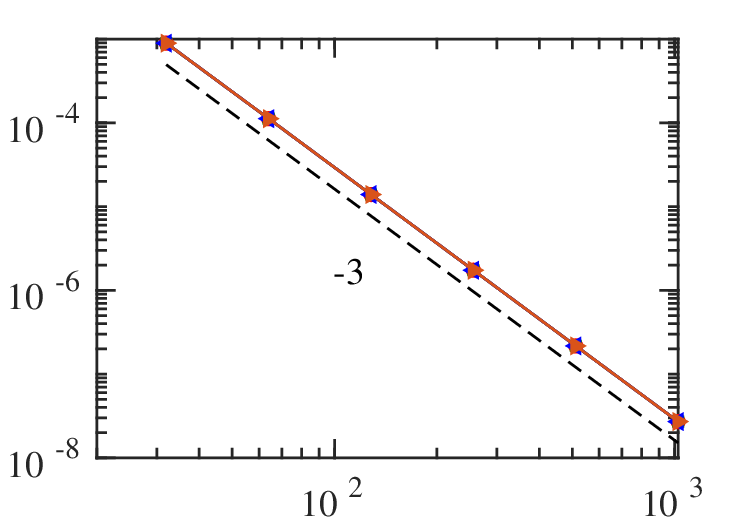}
    \begin{picture}(0,0)
    \put(-40,84){\small (b)}
    \put(-163,48){\small{\rotatebox{90}{$\langle \overline{E} \rangle $}}}
    \put(-78,-8){\small{$N_E$}}
    % \put(-160,60){\small Mean delay}
    \end{picture}
    % \hspace{0.05cm}
    \includegraphics[width=5.3cm]{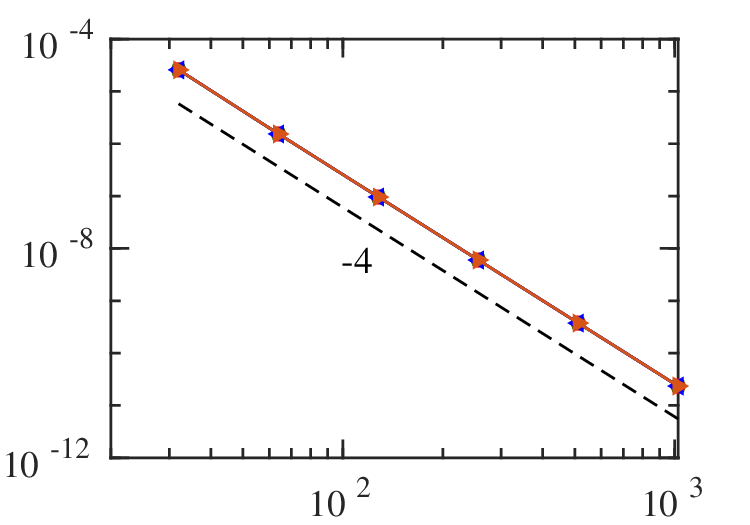}
    \begin{picture}(0,0)
    \put(-40,84){\small (c)}
    \put(-163,48){\small{\rotatebox{90}{$\langle \overline{E} \rangle $}}}
    \put(-78,-8){\small{$N_E$}}
    \end{picture}
    % \vspace{0.2cm}
    \caption{\small{Convergence plot of the average overall error $\langle \overline{E} \rangle$ with increasing grid resolution. Results are obtained from the simulations of the linear advection equation using the schemes (a) DG(1)-RK2 and ADG(1)-AT2-RK2 with $\sigma  = 0.1$, (b) DG(2)-RK3 and ADG(2)-AT3-RK3 with $\sigma = 0.04$, and (c) DG(3)-RK4 and ADG(3)-AT4-RK4 with $\sigma = 0.01$. Different lines correspond to varying degrees of asynchrony with mean delays: $\overline{\tilde{k}} = 0.0$ (red), $0.7$ (green), $1.0$ (blue), $2.0$ (orange). Dashed black lines with a slope of $-2$, $-3$ and $-4$ are shown for reference.}}
    \label{fig:accuracyplot-atdg}
\end{figure}

Figure~\ref{fig:accuracyplot-atdg} shows the results obtained from the synchronous schemes and the asynchronous schemes with AT fluxes to improve the accuracy of the asynchronous DG method. The simulation parameters are the same as those used in the previous set of experiments (from Fig.~\ref{fig:accuracyplot-asyncdg}). The asynchronous DG method uses the AT numerical fluxes from Tab.~\ref{table:at-flux}. The references to the asynchronous cases are the ADG(1)-AT2-RK2, ADG(2)-AT3-RK3, and ADG(3)-AT4-RK4 schemes, which are expected to provide second-, third-, and fourth-order accurate solutions based on the error scaling relation in Eq.~\eqref{eq:errorU-asyncflux}. In part (a) of the figure, we observe that the error in all cases asymptotically converges with a slope of $-2$, indicating that the schemes provide a second-order accurate solution in space. However, with an increase in the amount of asynchrony, the error increases reasonably. Parts (b) and (c) verify the error convergence for the third- ($\sim \mathcal{O}(\Delta x^3)$) and fourth- ($\sim \mathcal{O}(\Delta x^4)$) order accurate schemes, respectively. In these plots, the effect of asynchrony is hardly noticeable.

\begin{figure}[h!]
    \centering
    \includegraphics[width=6.5cm]{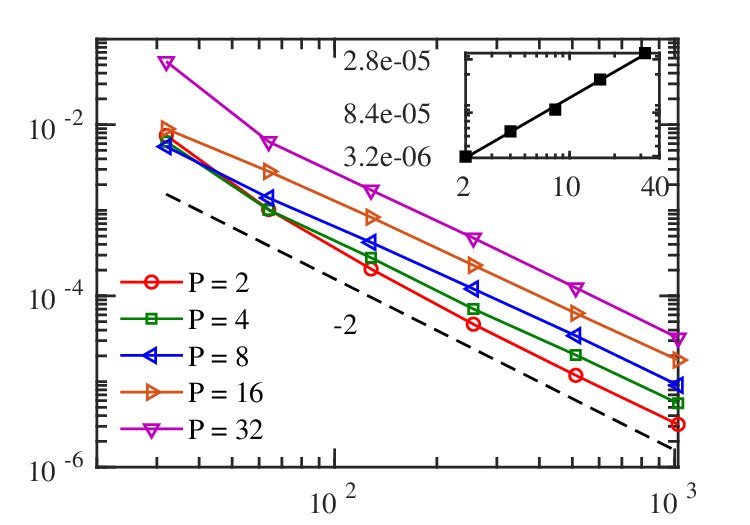}
    \begin{picture}(0,0)
    \put(-130,110){\small (a)}
    \put(-198,64){\small{\rotatebox{90}{$\langle \overline{E} \rangle$}}}
    \put(-100,-8){\small{$N_E$}}
    % \put(-160,60){\small Mean delay}
    \end{picture}
    \hspace{1cm}
    \includegraphics[width=6.5cm]{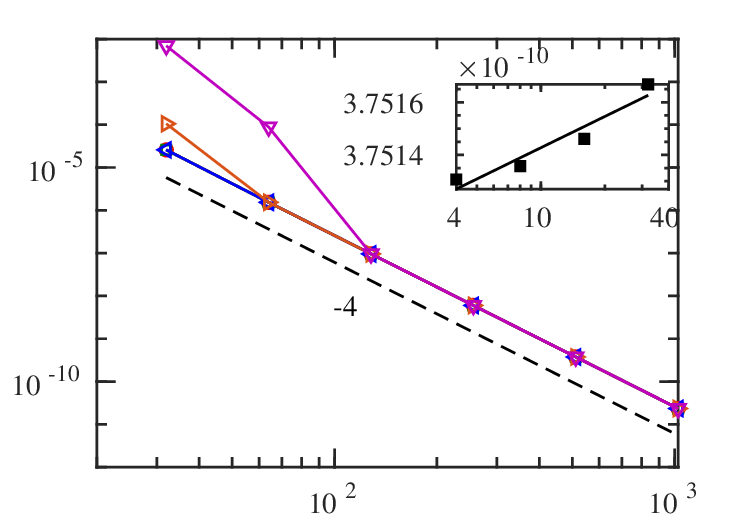}
    \begin{picture}(0,0)
    \put(-130,110){\small (b)}
    \put(-198,64){\small{\rotatebox{90}{$\langle \overline{E} \rangle$}}}
    \put(-100,-8){\small{$N_E$}}
    \end{picture}
    \vspace{0.2cm}
    \caption{\small{Convergence plot of the average overall error $\langle \overline{E} \rangle $ with increasing grid resolution. Results are obtained from the simulations of the linear advection equation. (a) ADG(1)-AT2-RK2 scheme, (b) ADG(3)-AT4-RK4 scheme. Different lines correspond to different numbers of processing elements: $P = 2$ (red), $4$ (green), $8$ (blue), $16$ (orange), $32$ (magenta). Simulation parameters: $L = 3$ with $\{p_0, p_1, p_2\} = \{0.3, 0.4, 0.3 \}$, $\sigma=$ $0.1$ (a), $0.01$ (b). Inset: plots of the average error $\langle \overline{E} \rangle $ with $P$ at (a) $N_E = 1024$ and (b) $N_E = 512$. Dashed lines with slopes of $-2$ and $-4$ are shown for reference.}}
    \label{fig:ErrorvsP-atdg}
\end{figure}

Let us recall the error scaling relation obtained in Eq.~\eqref{eq:error-AT-gen}, which illustrates the error dependence on the simulation parameters, such as the number of PE ($P$) used and the delay statistics:
\begin{align}
    \langle \overline{{E_f}} \rangle & \sim \frac{P}{N_E} \Delta t^a
    \sum_{m=1}^{\tau} \gamma_m \overline{\tilde{k}^m}.
    \label{eq:error-AT-gen2}
\end{align}
When the error due to asynchrony dominates, the above expression indicates that the error varies linearly with $P$, which is confirmed in Fig.~\ref{fig:ErrorvsP-atdg}. Parts (a) and (b) show the results for the ADG(1)-AT2-RK2 and ADG(3)-AT4-RK4 schemes, respectively. The simulations use a probability set $\{p_0, p_1, p_2\} = \{0.3, 0.4, 0.3 \}$ for delays. The different lines in the two plots indicate the error convergence for different values of $P$. It is observed that the ADG(1)-AT2-RK2 and ADG(3)-AT4-RK4 schemes provide second- and fourth-order accurate solutions, respectively, which is consistent with the theoretical predictions. Furthermore, the insets in both plots show a linear dependence of the error on the number of PEs.

%%%%%%%%%%%%%%%%%%%%% Error vs L %%%%%%%%%%%%%%%%%%%%%%%%%%%%%%

\begin{figure}[h!]
    \centering
    \includegraphics[width=6.5cm]{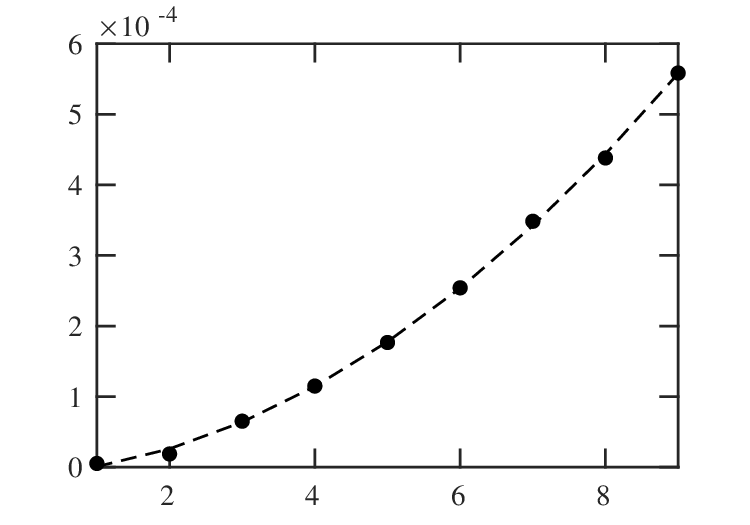}
    \begin{picture}(0,0)
    \put(-65,105){\small (a)}
    \put(-198,64){\small{\rotatebox{90}{$\langle \overline{E} \rangle $}}}
    \put(-100,-8){\small{$L$}}
    \end{picture}
    \hspace{1cm}
    \includegraphics[width=6.5cm]{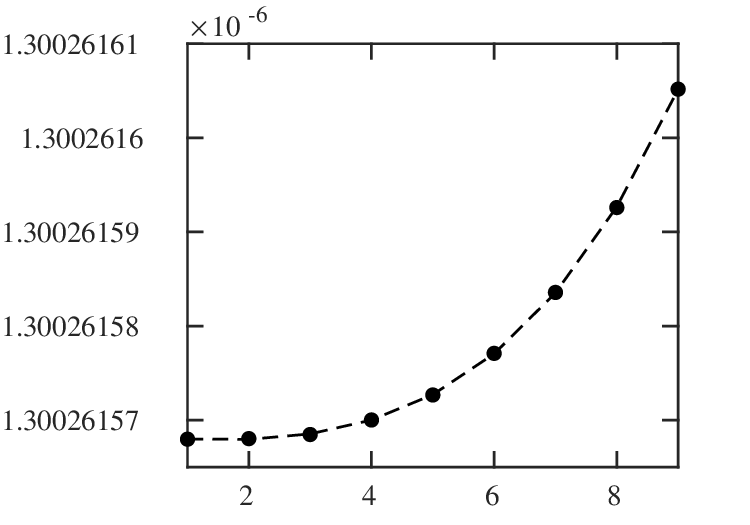}
    \begin{picture}(0,0)
    \put(-65,105){\small (b)}
    \put(-202,64){\small{\rotatebox{90}{$\langle \overline{E} \rangle $}}}
    \put(-100,-8){\small{$L$}}
    \end{picture}
    \vspace{0.2cm}
    \caption{\small{Scaling of the average overall error $\langle \overline{E} \rangle $ with moments of delay $(\tilde{k})$. In parts (a) and (b), circles are obtained from simulations of linear advection equation using ADG(1)-AT2-RK2 and ADG(3)-AT4-RK4, respectively, with parameters (a) $N_E = 512$, $\sigma = 0.1$, (b) $N_E = 128$, $\sigma = 0.01$, and $P = 16$. The dashed curves in (a) and (b) are second and fourth-order polynomial fits, respectively.}}
    \label{fig:ErrorvsL-atdg}
\end{figure}

It is evident from Eq.~\eqref{eq:error-AT-gen2} that for an asynchronous DG scheme with AT fluxes, the error due to asynchrony depends on higher order moments of the delay $\tilde{k}$. To validate this, we consider simulations with the ADG(1)-AT2-RK2 and ADG(3)-AT4-RK4 schemes. These two schemes use two and four time levels in numerical flux computations. Accordingly, the average error, when asynchrony dominates, scales as 
\begin{align}
    \langle \overline{E} \rangle &\sim \left( \overline{\tilde{k}^2} + \overline{\tilde{k}} \right) \hspace{2.4cm} \text{for ADG(1)-AT2-RK2 scheme} \nonumber \\
    &\sim \left( \overline{\tilde{k}^4} + 6\overline{\tilde{k}^3} + 11\overline{\tilde{k}^2} + 6\overline{\tilde{k}} \right) \quad \text{for ADG(3)-AT4-RK4 scheme.} 
\end{align}
The above relations can be further simplified by assuming that the probability of occurrence of a level $k$ for a given $L$ is $p_k = 1/L$. Using this probability distribution, the error scaling can be re-written as
\begin{align}
    \langle \overline{E} \rangle &\sim \left( L^2 -1 \right) \hspace{3.25cm} \text{for ADG(1)-AT2-RK2 scheme} \nonumber \\
    &\sim \left( L^4 + 8L^3 + 14L^2 -8L -15 \right) \quad \text{for ADG(3)-AT4-RK4 scheme.}
\end{align}
Figure~\ref{fig:ErrorvsL-atdg} presents the variation in the average error with $L$ for the ADG(1)-AT2-RK2 (part(a)) and ADG(3)-AT4-RK4 (part(b)) schemes. The solid black circles in the plots are obtained from numerical simulations, and the dashed black lines are polynomial fits of orders two and four in parts (a) and (b), respectively. Good agreement is observed between the numerical experiments and theoretical predictions.

\begin{figure}[h!]
    \centering
    \includegraphics[width=6.8cm]{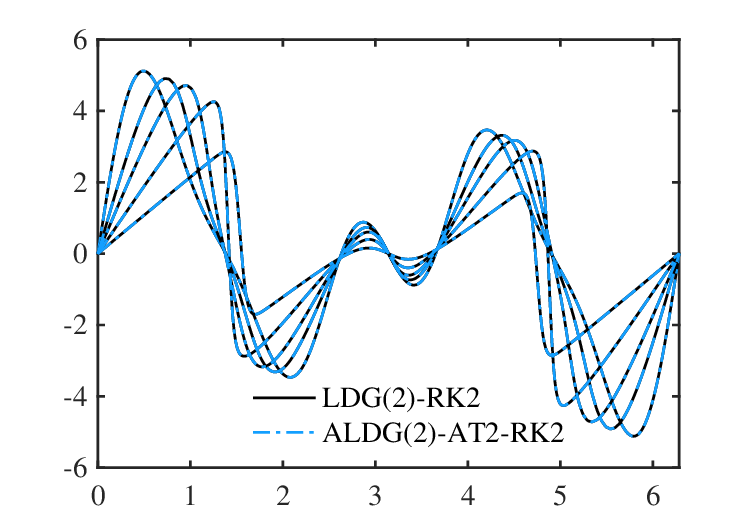}
    \begin{picture}(0,0)
    \put(-45,105){\small (a)}
    \put(-198,68){\small{\rotatebox{90}{$u$}}}
    \put(-100,-8){\small{$x$}}
    \put(-128,94){\large{$t$}}
    \put(-153,70){\huge{\rotatebox{30}{$\longrightarrow$}}}
    \end{picture}
    \hspace{1cm}
    \includegraphics[width=6.5cm]{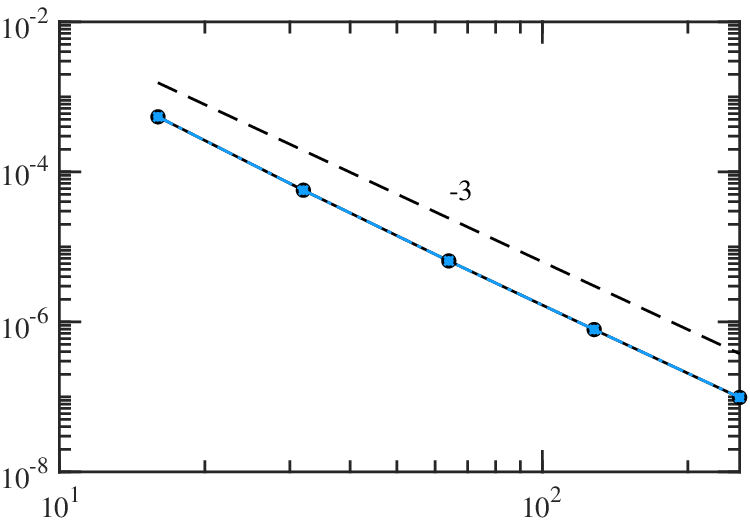}
    \begin{picture}(0,0)
    \put(-45,105){\small (b)}
    \put(-198,66){\small{\rotatebox{90}{$\langle E \rangle$}}}
    \put(-100,-8){\small{$N_E$}}
    \end{picture}
    \vspace{0.2cm}
    \caption{\small{(a) Time evolution of the numerical solution using LDG(2)-RK2 (solid black lines) and ALDG(2)-AT2-RK2 (dash-dotted blue lines) schemes for 128 elements. (b) Convergence plot of the average error with increasing mesh resolution for LDG(2)-RK2 (solid black) and ALDG(2)-AT2-RK2 (dash-dotted blue) schemes. Results are obtained from the simulations of the nonlinear viscous Burgers' equation. Simulation parameters: $\kappa = \{2, 3, 5 \}$, $A(\kappa) = \{3, 2, 1\}$, $\nu = 0.1$, $P = 4$, $L = 3$ with $\{p_0, p_1, p_2\} = \{0.3, 0.4, 0.3\}$ for asynchronous computations. A dashed black line in (b) with a slope of $-3$ is shown for reference.}}
	\label{fig:burgers-solution}
\end{figure}

% Through simulations of the viscous Burgers' equation, we aim to gain an understanding on the performance of ADG-AT schemes under highly nonlinear conditions. 
% By evaluating their ability to capture the complex dynamics of fluid turbulence accurately, we can extend our confidence in the practical applicability of these schemes to a broader range of nonlinear systems beyond linear equations.

The above experiments validate the accuracy of the ADG-AT schemes for linear equations. However, it is essential to extend our assessment to systems governed by nonlinear processes, as they are prevalent in modeling various natural processes and engineered systems. One prominent example of such nonlinear behavior is observed in fluid turbulence phenomena, which play a crucial role in understanding complex fluid flows. To investigate the capabilities of the ADG-AT schemes in capturing the nonlinear effects encountered in turbulent fluid flows, we employ the viscous Burgers' equation (Eq.~\eqref{eq:burgers}) as a representative model.
% This equation serves as a useful proxy for exploring the intricate dynamics of fluid turbulence, while disregarding significant pressure effects. 
We use the local discontinuous Galerkin (LDG) method \cite{ldg-burgers-zhang2011} to approximate the solution in space with a quadratic polynomial basis ($N_p=2$). For time integration, we employ a second-order accurate Runge-Kutta (RK2) scheme. The advection and diffusion terms within the interior elements are handled using Lax-Friedrichs \cite{dg-book-li2006} and alternating fluxes, respectively, whereas second-order accurate asynchrony-tolerant (AT) fluxes are incorporated at the PE boundary elements. The synchronous and asynchronous schemes are referred to as LDG(2)-RK2 and ALDG(2)-AT2-RK2, respectively. In the numerical simulations, the parameters are $\kappa = \{2, 3, 5 \}$, $A(\kappa) = \{3, 2, 1\}$, $\nu = 0.1$, $\sigma = 0.0005$, $P = 4$, and $L = 3$ with $\{p_0, p_1, p_2\} = \{0.3, 0.4, 0.3\}$. Figure~\ref{fig:burgers-solution}(a) illustrates the time evolution of the numerical solutions, demonstrating a good agreement in both space and time between the synchronous (solid black line) and asynchronous (dash-dotted blue line) schemes. Moreover, to ascertain the order of accuracy of the schemes, we plot the average error against the increasing grid resolution in Fig.~\ref{fig:burgers-solution}(b). In both the synchronous and asynchronous cases, the error decreases with a slope of $-3$, as expected from schemes that use quadratic basis functions. These results affirm the effectiveness of the asynchronous approach in capturing the solutions of the nonlinear viscous Burgers' equation.

% \begin{figure}[ht!]
%     \centering
%     \includegraphics[width=6.5cm]{figures/BurgersDG3_solution.eps}
%     \begin{picture}(0,0)
%     \put(-110,125){\small (a)}
%     \put(-188,70){\small{\rotatebox{90}{$u$}}}
%     \put(-100,-8){\small{$x$}}
%     \end{picture}
%     \hspace{1cm}
%     \includegraphics[width=6.5cm]{figures/BurgersDG3_accuracy.eps}
%     \begin{picture}(0,0)
%     \put(-110,125){\small (b)}
%     \put(-198,70){\small{\rotatebox{90}{$\|e\|_{L_2}$}}}
%     \put(-100,-8){\small{$N_E$}}
%     \end{picture}
%     \vspace{0.2cm}
%     \caption{\small{(a) Time evolution of the numerical solution using synchronous LDG(2)-AB2 (solid blue lines) and ALDG(2)-AT2-AB2 (dashed magenta lines) schemes for 128 elements, and (b) convergence plot of the $L2$-error with increasing mesh resolution of synchronous ALDG(2)-AT2-AB2 (black) and ALDG(2)-AT2-AB2 (red) schemes. Results are obtained from the simulations of the nonlinear viscous Burgers' equation. Simulation parameters: $P = 4, L = 3$ with $\{p_0, p_1, p_2\} = \{0.3, 0.4, 0.3\}$ for asynchronous computations. A dashed line in (b) with the slope of $-3$ is shown for reference.}}
% 	\label{fig:Burgers}
% \end{figure}

\begin{figure}[h!]
    \centering
    \includegraphics[width=6.5cm]{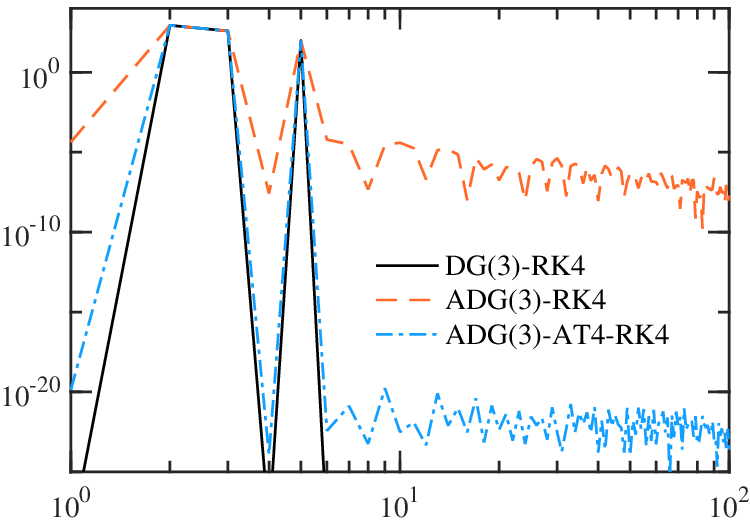}
    \begin{picture}(0,0)
    \put(-45,105){\small (a)}
    \put(-198,66){\small{\rotatebox{90}{$\mathcal{E}(\kappa)$}}}
    \put(-100,-8){\small{$\kappa$}}
    \end{picture}
    \hspace{1cm}
    \includegraphics[width=6.5cm]{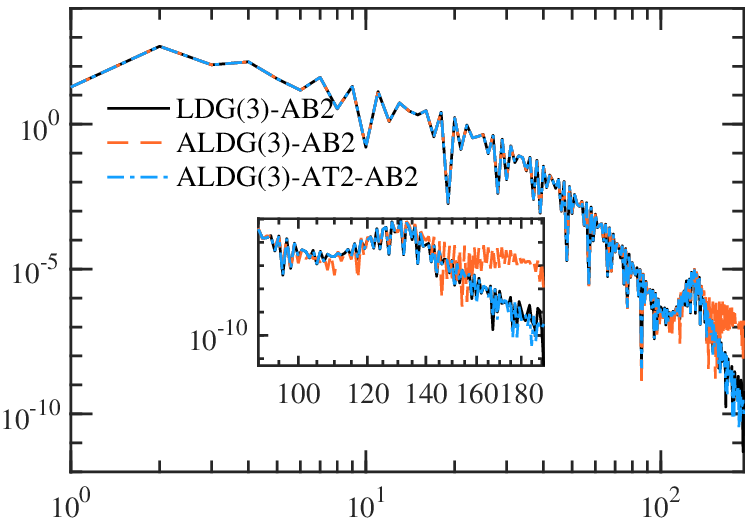}
    \begin{picture}(0,0)
    \put(-45,105){\small (b)}
    \put(-198,66){\small{\rotatebox{90}{$\mathcal{E}(\kappa)$}}}
    \put(-100,-8){\small{$\kappa$}}
    \end{picture}
    \vspace{0.2cm}
\caption{\small{Energy spectra $\mathcal{E}(\kappa)$ of $u$ for the (a) linear advection and (b) nonlinear viscous Burgers' equations. The linear equation is solved using the DG(3)-RK4 (solid black line), ADG(3)-RK4 (dashed red line) and ADG(3)-AT4-RK4 (dash-dotted blue line) schemes. The nonlinear equation is solved using LDG(3)-AB2 (solid black line), ALDG(3)-AB2 (dashed red line) and ALDG(3)-AT2-AB2 (dash-dotted blue line) schemes. Simulation parameters: $N_E=128$, $t_{End} = 0.2$, $\kappa = \{2, 3, 5 \}$, $A(\kappa) = \{3, 2, 1\}$, $\nu = 0.1$, $P = 4$, $L = 3$ with $\{p_0, p_1, p_2\} = \{0.3, 0.4, 0.3\}$ for asynchronous computations. Inset in (b): zoomed-in energy spectra at high wavenumbers.}}
\label{fig:spectra}
\end{figure}

In the preceding figures, we analyzed the accuracy of the asynchronous schemes using the solutions in the physical space. The spatial distribution of the errors, shown in Fig.~\ref{fig:SyncVsAsync}, indicates that asynchrony introduces localized errors near PE boundaries, which could affect the high wavenumber content of the solutions. To assess this effect due to asynchrony, we compute the spectra of the solutions for the linear advection and nonlinear viscous Burgers equations. First, consider the linear case that uses the DG(3)-RK4, ADG(3)-RK4, and ADG(3)-AT4-RK4 schemes. The simulation parameters are $N_E=128$, $t_{End} = 0.2$, $\kappa = \{2, 3, 5 \}$, $A(\kappa) = \{3, 2, 1\}$, $P = 4$, and $L = 3$ with $\{p_0, p_1, p_2\} = \{0.3, 0.4, 0.3\}$. \rev{Part (a) of Fig.~\ref{fig:spectra} demonstrates the spectra for the linear case.} We observe that the spectrum for the synchronous DG(3)-RK4 scheme (black line) has energy at wavenumbers $2$, $3$, and $5$, which is consistent with the imposed initial condition. Note that for the linear advection equation, the spectrum remains unchanged over time. For the asynchronous ADG(3)-RK4 scheme (dashed orange line), we observe that a significant amount of energy is present at the same wavenumbers ($2$, $3$, and $5$). However, the spectrum also exhibits a spurious energy of lower magnitude for the remainder of the wavenumbers (broadband). This is attributed to the localized errors introduced due to asynchrony at the PE boundaries. This effect is significantly mitigated with the use of AT fluxes, as observed for the ADG(3)-AT4-RK4 scheme (dash-dotted blue line).

For the simulations of the nonlinear Burgers' equation, we use the LDG scheme with a quadratic basis polynomial and a second-order Adams-Bashforth scheme for time integration. The synchronous and asynchronous schemes are LDG(3)-AB2, ALDG(3)-AB2, and ALDG(3)-AT2-AB2. The simulation parameters are the same as those in the linear case. Additionally, viscosity of $\nu = 0.1$ is imposed. Figure~\ref{fig:spectra}(b) shows the spectra of the three schemes. Unlike the linear case, the nonlinear equation gives rise to higher harmonics and distributes the energy across a wide range of wavenumbers. Overall, the synchronous scheme (black line) exhibits a (nearly) decaying spectrum. In the case of the asynchronous ALDG(3)-AB2 scheme (dashed orange line), we observe a deviation from the synchronous spectrum at high wavenumbers. The higher energy at these high wavenumbers is due to the errors introduced by asynchrony. Again, the use of AT fluxes (dash-dotted blue line) mitigates this effect and shows a good agreement with the synchronous spectrum.

\rev{
Having established the accuracy of the ADG-AT schemes for both linear and nonlinear equations, it is essential to evaluate their ability to capture sharp discontinuities, such as shock waves, which are a critical element of hyperbolic problems. To test the robustness and shock-capturing ability of the ADG-AT schemes, we utilize the compressible Euler equations (Eq.~\eqref{eq:compressible-euler-3d}) in a one-dimensional domain which impose the conservation of mass, momentum, and energy in compressible flows. Specifically, we implement the Sod's shock tube problem (Eq.~\eqref{eq:sod-initial}, a standard benchmark that features contact discontinuity as initial conditions. We implement the discontinuous Galerkin method with linear basis polynomials and Lax-Friderichs flux for spatial discretization and utilize a second-order strong-stability preserving Runge-Kutta scheme for time integration. For the asynchronous case, a second-order asynchrony-tolerant flux is employed.  In the presence of shocks, limiters are necessary to ensure non-oscillatory solutions. In this particular case, we use the MUSCL TVBM limiter, which guarantees total variation stability \cite{hesthaven2007}. Figure~\ref{fig:sod-euler} presents the solution profiles for density, pressure, and velocity at $t_{End} = 0.002$ for 512 elements, along with the initial conditions. Exact solutions (solid green lines) obtained using the Riemann solver are added in the three plots for reference. Simulation parameters for the asynchronous implementation are $P = 4$ and $L = 3$ with $\{p_0, p_1, p_2\} = \{0.3, 0.4, 0.3\}$. The results indicate that both the synchronous DG(1)-RK2 (black squares) and ADG(1)-AT2-RK2 (blue crosses) schemes are successful in capturing shocks. A very good agreement is evident between the solutions generated by the two schemes. These results highlight the ability of the ADG method to resolve shocks and complex wave interactions accurately, demonstrating its potential for effectively handling hyperbolic systems with discontinuities.
\begin{figure}[h!]
    \centering
    \includegraphics[width=5.3cm]{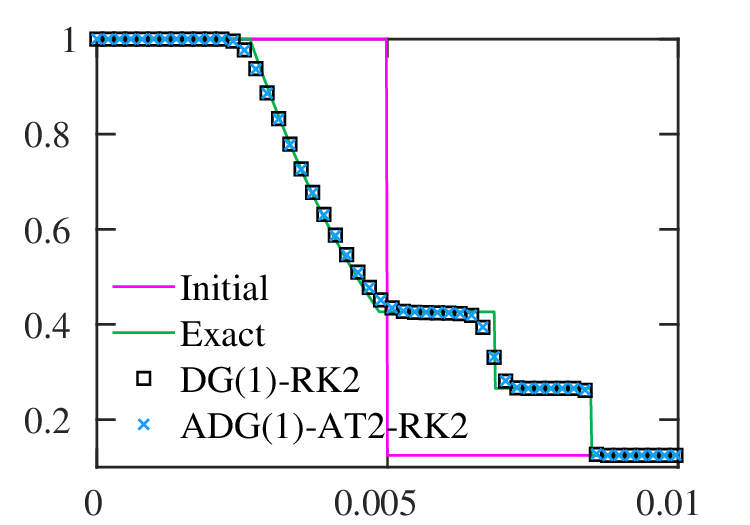}
    \begin{picture}(0,0)
    \put(-40,80){\small (a)}
    \put(-160,48){\small{\rotatebox{90}{$\rho $}}}
    \put(-78,-8){\small{$x$}}
    \end{picture}
    % \hspace{0.05cm}
    \includegraphics[width=5.3cm]{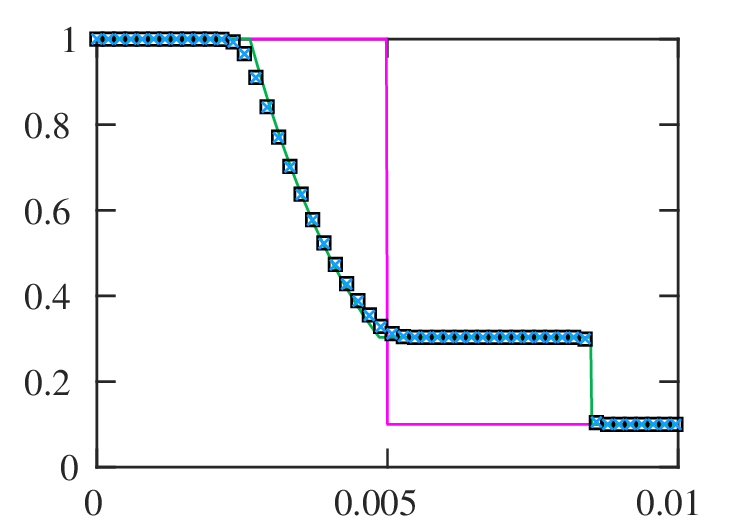}
    \begin{picture}(0,0)
    \put(-50,80){\small (b)}
    \put(-160,48){\small{\rotatebox{90}{$p $}}}
    \put(-78,-8){\small{$x$}}
    % \put(-160,60){\small Mean delay}
    \end{picture}
    % \hspace{0.05cm}
    \includegraphics[width=5.3cm]{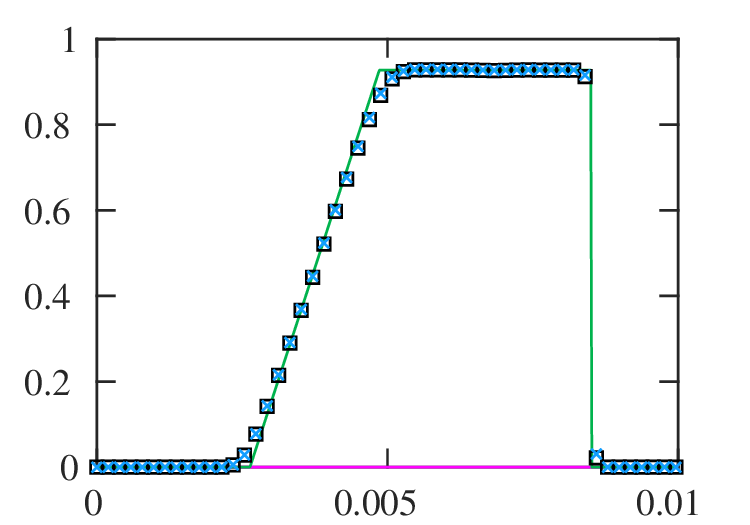}
    \begin{picture}(0,0)
    \put(-50,80){\small (c)}
    \put(-160,48){\small{\rotatebox{90}{$u$}}}
    \put(-78,-8){\small{$x$}}
    \end{picture}
    % \vspace{0.2cm}
    \caption{\small{
    Numerical solutions of one-dimensional compressible Euler equations for the Sod’s shock tube problem using DG(1)-RK2 (black squares) and ADG(1)-AT2-RK2 schemes (blue crosses) for (a) density, (b) pressure, and (c) velocity at $t = 0.002$. Initial (solid magenta lines) and exact solutions (solid green lines) are added for reference. Simulation parameters: MUSCL TVBM limiter with $M = 10$, $N_E = 512$, $\sigma  = 0.1$, and $P = 4$. For asynchronous computations $L = 3$ with $\{p_0, p_1, p_2\} = \{0.3, 0.4, 0.3\}$.}}
    \label{fig:sod-euler}
\end{figure}
}

\subsection{Extension to higher dimensions}
\begin{figure}[h!]
    \centering
    \includegraphics[width=6.5cm]{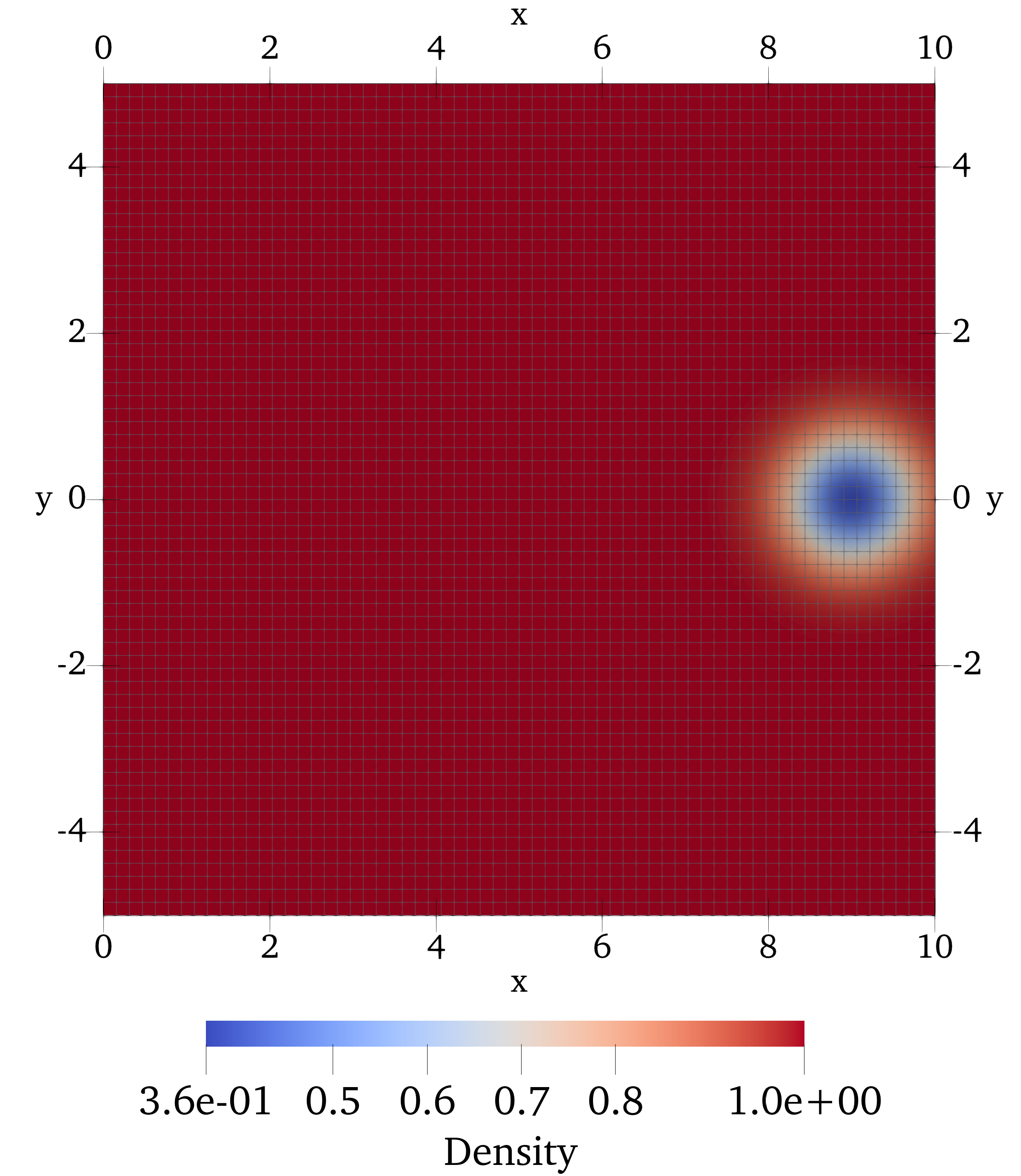}
    \begin{picture}(0,0)
    \put(-130,110){\small (a)}
    \end{picture}
    \hspace{1cm}
    \includegraphics[width=6.5cm]{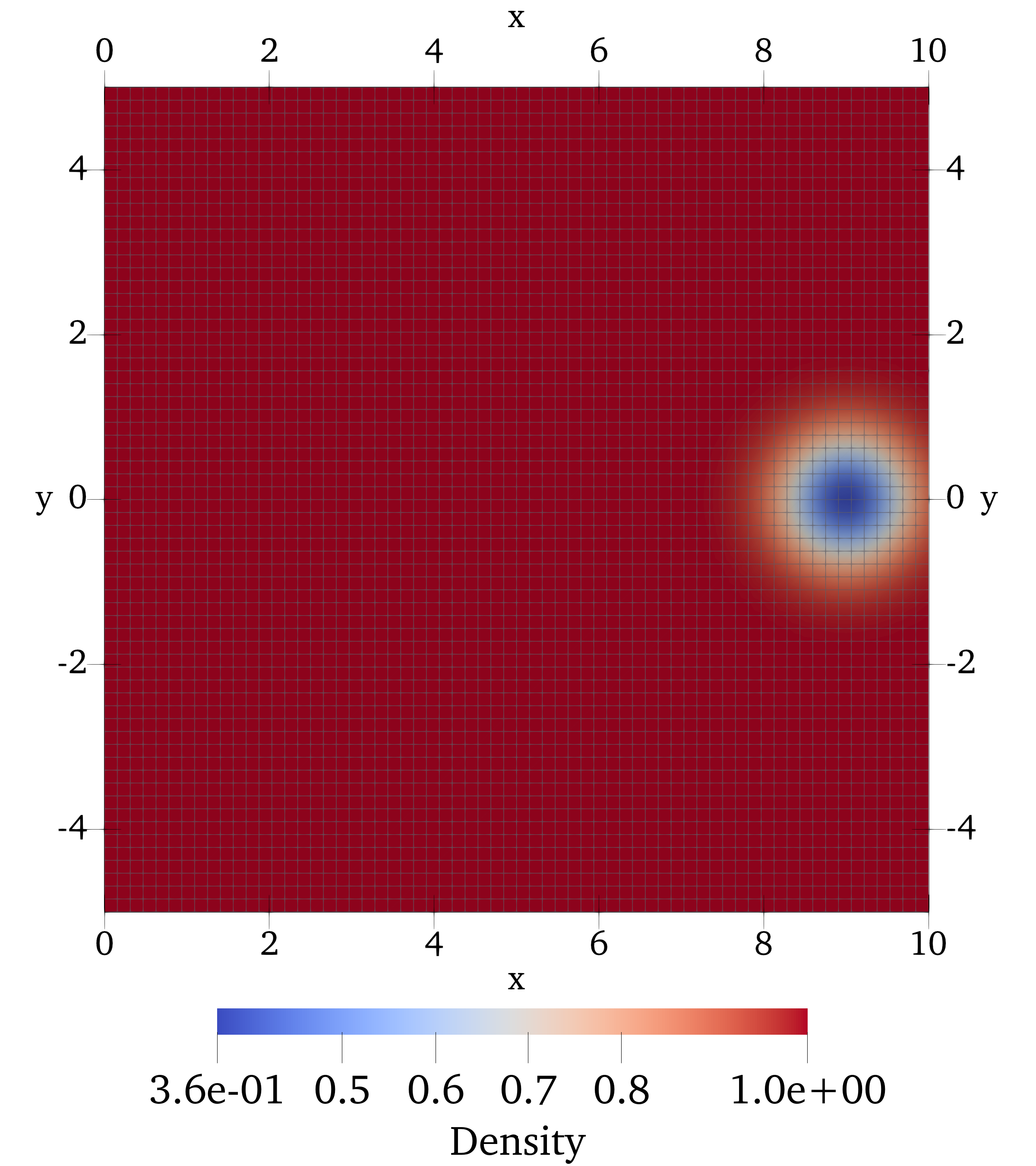}
    \begin{picture}(0,0)
    \put(-130,110){\small (b)}
    \end{picture}
    \vspace{0.2cm}
    \includegraphics[width=6.5cm]{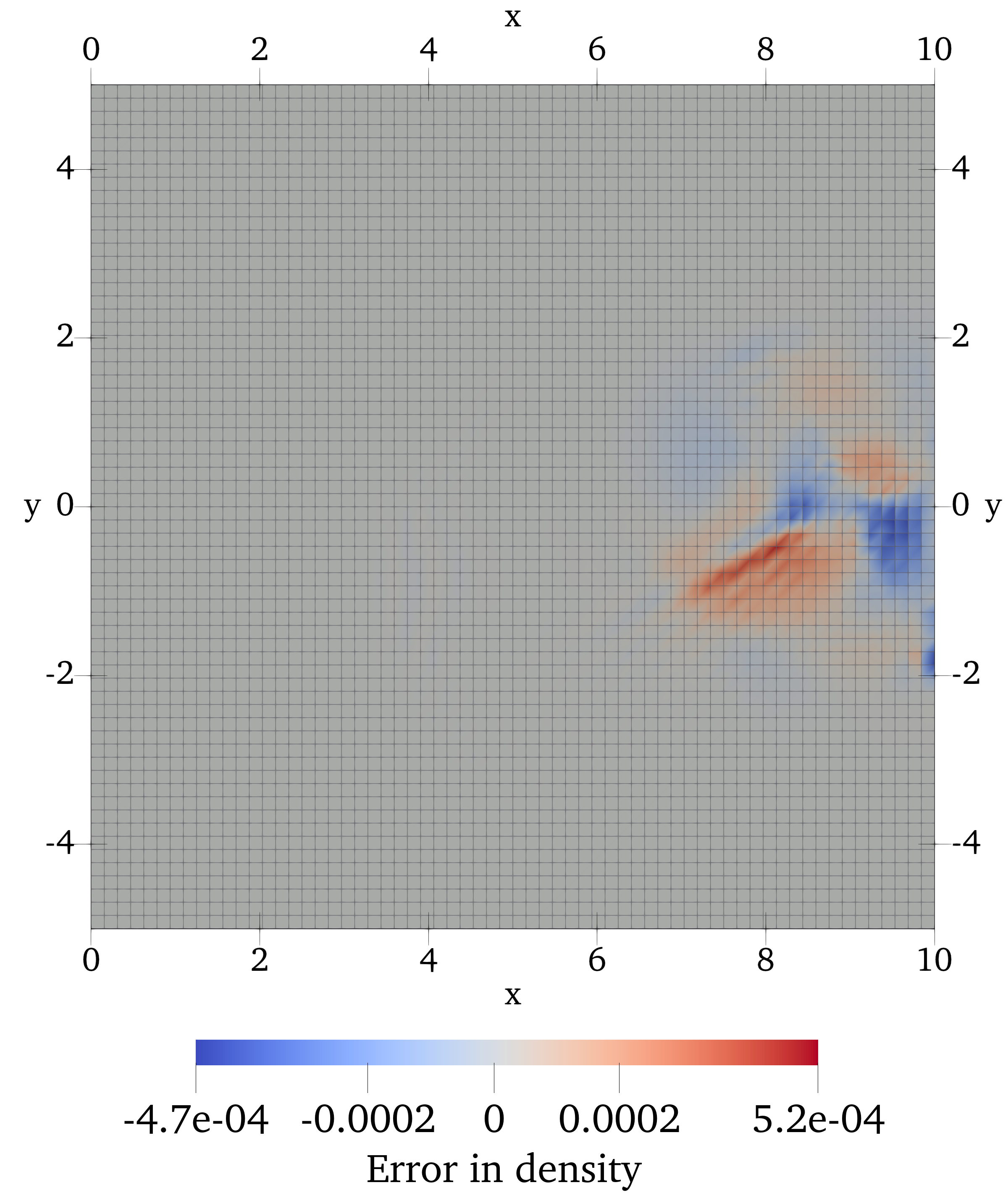}
    \begin{picture}(0,0)
    \put(-130,110){\small (c)}
    \end{picture}
    \hspace{1cm}
    \includegraphics[width=6.5cm]{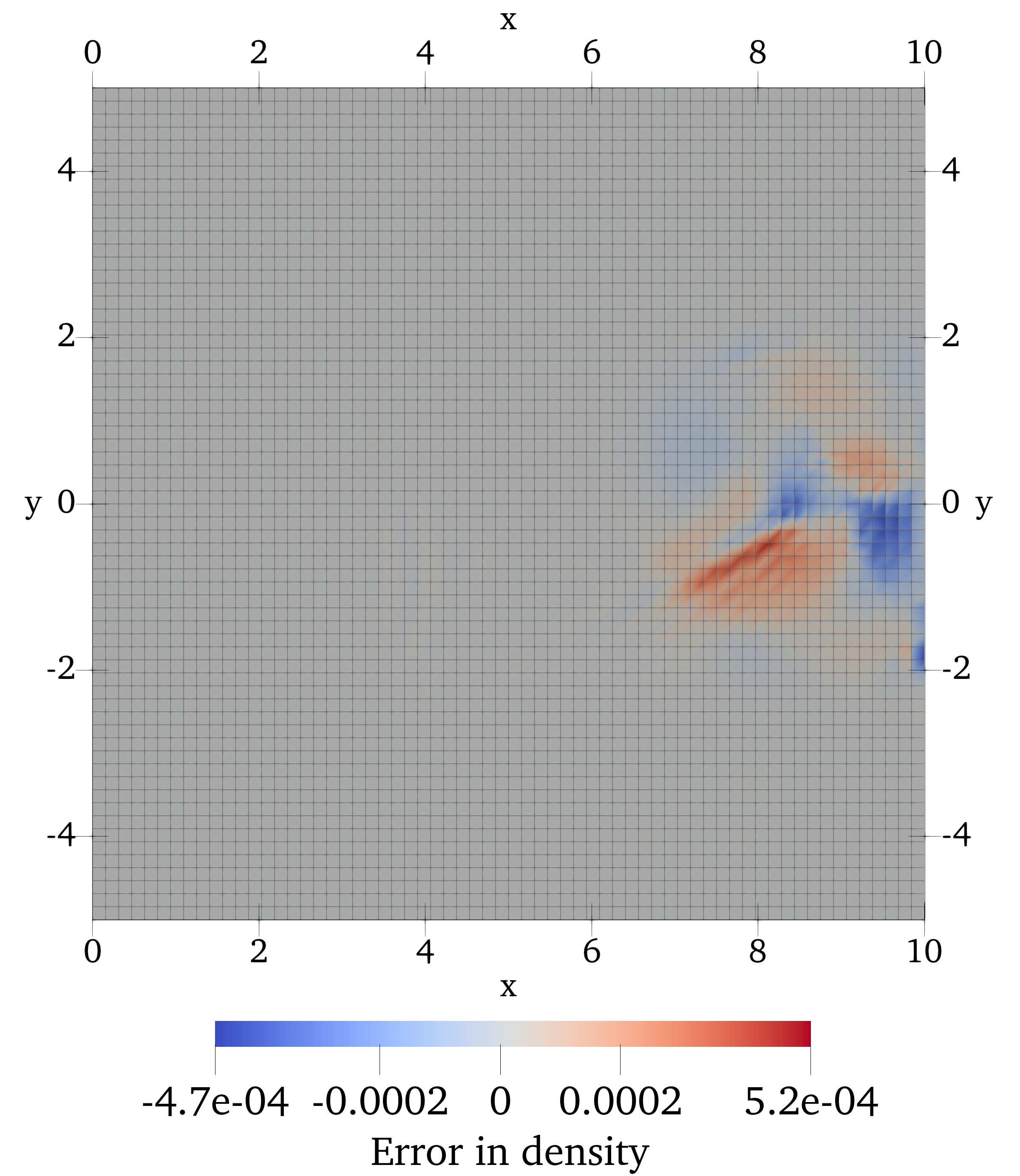}
    \begin{picture}(0,0)
    \put(-130,110){\small (d)}
    \end{picture}
    \caption{\small{Plots for a vortex transported through a background flow field based on compressible Euler equations on a two-dimensional domain. (a) and (b) are the solution plots for the density variable at $t=4$ obtained using DG(1)-RK2 and ADG(1)-AT2-RK2 schemes, respectively. (c) and (d) are the respective errors for the two numerical solutions. Simulation parameters: $N_E = 4096$, $\sigma = 0.01$, $P = 256$, and $L = 3$. The asynchronous implementation is based on the communication-avoiding algorithm.}}
    \label{fig:solution-2d}
\end{figure}

\begin{figure}[h!]
    \centering
    \includegraphics[width=5.3cm]{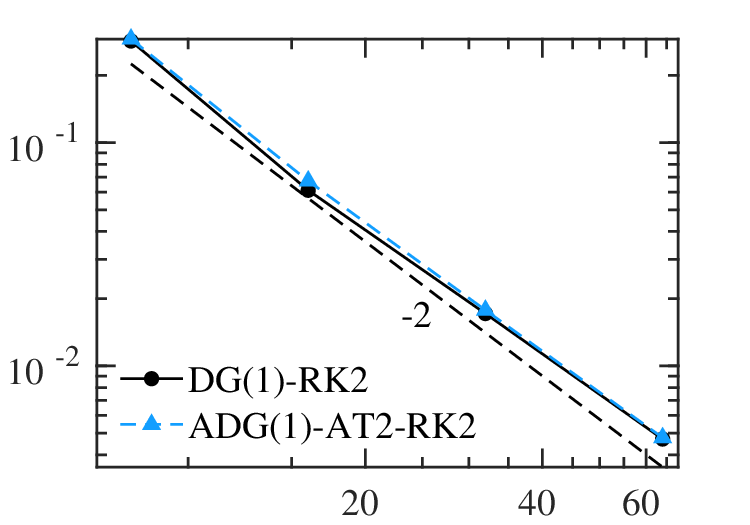}
    \begin{picture}(0,0)
    \put(-40,84){\small (a)}
    \put(-163,48){\small{\rotatebox{90}{$\langle \overline{E} \rangle $}}}
    \put(-137,-8){\small{Number of elements in $x$-direction}}
    \end{picture}
    % \hspace{0.05cm}
    \includegraphics[width=5.3cm]{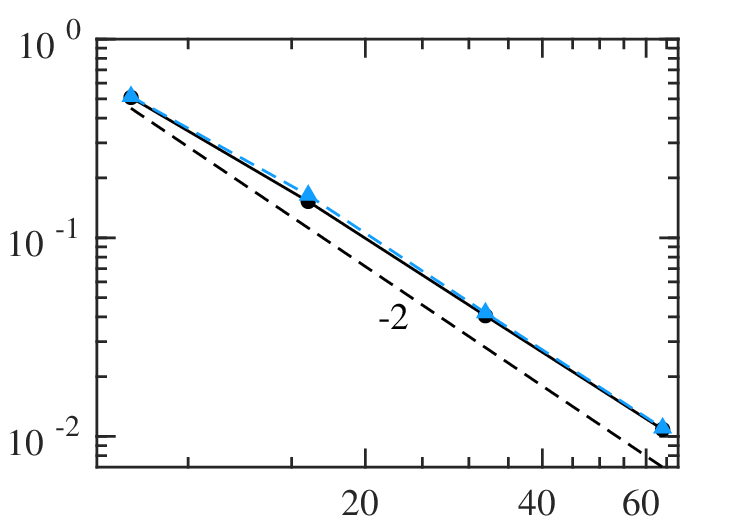}
    \begin{picture}(0,0)
    \put(-40,84){\small (b)}
    \put(-163,48){\small{\rotatebox{90}{$\langle \overline{E} \rangle $}}}
    \put(-137,-8){\small{Number of elements in $x$-direction}}
    % \put(-160,60){\small Mean delay}
    \end{picture}
    % \hspace{0.05cm}
    \includegraphics[width=5.3cm]{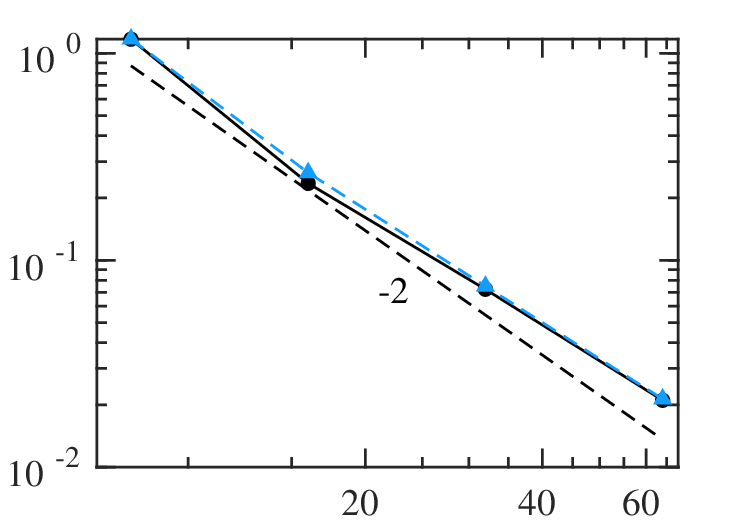}
    \begin{picture}(0,0)
    \put(-40,84){\small (c)}
    \put(-163,48){\small{\rotatebox{90}{$\langle \overline{E} \rangle $}}}
    \put(-137,-8){\small{Number of elements in $x$-direction}}
    \end{picture}
    % \vspace{0.2cm}
    \caption{\small{Convergence plot of the $L2$-norm errors with increasing grid resolutions. Results are obtained from the simulations of compressible Euler equations for (a) density, (b) momentum, and (c) energy variables. Solid black lines with circles represent the DG(1)-RK2 scheme, and dashed blue lines with triangles represent the ADG(1)-AT2-RK2 scheme. Black dashed lines with the slope of -2 are added for reference. Simulation parameters: $\sigma = 0.01$, $P = 8$, and $L = 3$. The asynchronous implementation is based on the communication-avoiding algorithm.}}
    \label{fig:accuracy-2d}
\end{figure}

\rev{
In the above numerical experiments, the performance of the asynchronous DG method has been investigated for one-dimensional problems. Additionally, it is important to assess its efficacy for higher-dimensional cases.
Here, we consider the inviscid compressible Euler equations in a two-dimensional domain. The governing equations are
\begin{align}
    \frac{\partial \rho}{\partial t} + \frac{\partial \rho u}{\partial x} + \frac{\partial \rho v}{\partial y} &= 0, \nonumber \\
    \frac{\partial (\rho u)}{\partial t} + \frac{\partial (\rho u^2 + p)}{\partial x} + \frac{\partial \rho u v}{\partial y} &= 0, \nonumber \\
    \frac{\partial (\rho v)}{\partial t} + \frac{\partial (\rho u v)}{\partial x} + \frac{\partial (\rho v^2 + p)}{\partial y} &= 0, \nonumber \\
    \frac{\partial (\rho e_0)}{\partial t} + \frac{ \partial (u(\rho e_0 + p))}{\partial x} + \frac{\partial (v(\rho e_0 + p)}{\partial y} &= 0.
\end{align}
Here, $\rho$ is the density, $u$ and $v$ are the velocity components in the $x$- and $y$- directions, $p$ is the pressure, and $e_0$ is the total energy.
The setup features an isentropic vortex transported through a background flow field in a square domain of size $[0,10] \times [-5,5]$ with an exact solution given by
\begin{align}
    u &= 1 - \beta \exp{(1-r^2)} \frac{y - y_0}{2\pi}, \nonumber \\
    v &= \beta \exp{(1-r^2)} \frac{x - x_0}{2\pi}, \nonumber \\
    \rho &= \left(1 - \left(\frac{\gamma - 1}{16 \gamma \pi^2} \right) \beta^2 \exp{\left(2(1 - r^2)\right)} \right)^{\frac{1}{\gamma -1}}, \nonumber \\
    p &= \rho^\gamma,
\end{align}
where $r = \sqrt{(x-t-x_0)^2 + (y-y_0)^2}$, $x_0 = 5$, $y_0 = 0$, $\beta  = 5$, and $\gamma = 1.4$ \cite{hesthaven2007}. This exact solution also provides the initial and boundary conditions.}

\rev{To simulate this problem, we implemented the asynchronous discontinuous Galerkin method with AT fluxes in one of the DG solvers (\textit{step-76}) in the open-source finite-element library \texttt{deal.II} \cite{dealII94}. This particular implementation uses basis polynomials of degree one and the Lax-Friedrichs flux for spatial discretization along with a second-order low-storage explicit Runge-Kutta scheme for time integration \cite{WILLIAMSON1980, goswami2023lserkat-jcp}. The asynchronous implementation is based on the communication-avoiding algorithm (CAA) with $L = 3$, where every five time steps, the first two steps comprise communications with synchronization, and the subsequent three steps are performed without any communication. In this setup, communication is reduced by $60\%$ compared to the standard synchronous approach. This results in delays $\tilde{k}=0, 1, 2$ for the three time levels, during which communication is absent. In the presence of these delays, we employed the asynchronous discontinuous Galerkin method with the second-order AT flux (ADG(1)-AT2).
Figure~\ref{fig:solution-2d} compares the density profiles obtained using the two methods and their associated errors at $t = 4$. The mesh consists of 4096 quadrilateral elements divided into 256 PEs. The solution and error contours exhibit an excellent match between the two methods, with both synchronous and asynchronous schemes providing similar error structure. Furthermore, to validate the accuracy of the asynchronous implementation, we computed $L2$-norm errors for three variables -- density, momentum, and energy -- on eight PEs with increasing grid resolution. In Fig.~\ref{fig:accuracy-2d}, we observe that the errors for both DG(1)-RK2 (solid black lines) and ADG(1)-AT2-RK2 (dashed blue lines) decrease with the same slope and similar magnitude, indicating that the ADG(1)-AT2-RK2 scheme provides second-order accurate solution. These results suggest that the asynchronous DG method with AT fluxes is effective for high-dimensional problems. Further analysis with more complex cases is a part of our ongoing effort.
}

%====================================
\section{Conclusions}
\label{sec:conclusions}

% NEED TO REVIEW
To simulate complex nonlinear PDEs of practical relevance, numerical schemes capable of providing accurate solutions in complex geometries are crucial. However, the computational demands of such simulations are substantial, necessitating massive parallelism and excellent scalability. Unfortunately, the conventional approach of global communications and bulk synchronizations among processing elements (PEs) at each time step for solving time-dependent partial differential equations (PDEs) can become a major bottleneck at extreme scales. Motivated by the need to overcome these limitations, we have developed an asynchronous computing approach based on the discontinuous Galerkin (DG) method, the asynchronous discontinuous Galerkin (ADG) method. The method relaxes communication and synchronization requirements at a mathematical level, enabling more efficient and scalable simulations. The method incorporates delayed data at the buffer/ghost elements. However, we find that such an asynchronous implementation compromises local conservation at the boundary elements of PEs when previous/older time level values are used to compute fluxes. To address this issue, we enforce PEs to use values from a common and delayed time level at all the nodes and present the asynchronous discontinuous Galerkin (ADG) method that preserves the conservation property.

To ascertain the stability and performance of the asynchronous DG schemes, we conduct a comprehensive Fourier mode analysis, which provides detailed insights into the nature of numerical errors, including dissipative and dispersive errors. In our analysis, we employ a block-matrix method to assess stability by bounding the numerical frequency for each eigenmode for all wavenumbers obtained from the amplification matrix. Here the amplification matrix of a system describes the evolution of the solution at each time step. This approach is necessary due to the involvement of multiple time levels in the scheme, rendering the classical von Neumann analysis inadequate.
The analysis has confirmed the stability of the asynchronous DG schemes. However, it is observed that these schemes impose more stringent Courant-Friedrichs-Lewy (CFL) constraints compared to their synchronous counterparts. Specifically, the stability region of the asynchronous DG scheme is found to shrink as the maximum allowable delay is increased. Nevertheless, it is important to note that the impact of these constraints is mitigated by the characteristic behavior of the delay distribution, which typically follows a Poisson distribution. This distribution indicates a low likelihood of encountering higher delays, thereby reducing the practical implications of the more restrictive CFL constraints.
Additionally, it is not a concern for problems involving reactions where the chemical time scales are significantly smaller than the time scales of the physical processes. In such cases, both synchronous and asynchronous DG schemes can utilize the same step size to accurately capture reaction dynamics, rendering the stability limit imposed by the asynchronous DG method irrelevant. As a result, the increased parallel efficiency offered by the asynchronous DG method can be leveraged for reacting flow problems without sacrificing stability or computational efficiency. Nonetheless, to address the restricted stability of asynchronous computing, data-driven discretization methods can be employed, which have the potential to significantly improve the stability limit by using dynamic weights or coefficients that adapt to local gradients in the function \rev{\cite{data-driven-pnas2019}}.

Furthermore, we have conducted an analysis of the errors introduced by the numerical flux due to asynchrony. These errors are incurred solely at PE boundaries and depend on the extent of delays in communication. The analysis is performed within a statistical framework that considers the stochastic nature of delays and the non-uniformity of delays in space. The results of this analysis indicate that the asynchronous DG method achieves, at most, first-order accuracy regardless of the degree of the polynomial basis functions employed.
To overcome the limitation of reduced accuracy in the ADG schemes, we developed novel asynchrony-tolerant (AT) fluxes. These AT fluxes incorporate additional values from previous time steps, already available in the memory of the processing elements. The theoretical predictions regarding the accuracy of the schemes are substantiated through extensive numerical experiments conducted for both linear and nonlinear equations \rev{including the Sod's shock tube problem and a two-dimensional test case.} The excellent agreement observed between the theoretical predictions and the numerical results across different parameter spaces further strengthens the foundation of mathematically asynchronous computing methods for solving PDEs at extreme scales.

It should be noted that the results presented here can be directly extended to the finite volume method, due to its similarities with the discontinuous Galerkin (DG) method. \rev{Additionally, the current developments were based on the one-dimensional linear advection equation, the nonlinear viscous Burgers' equation, and compressible Euler equations in one and two dimensions.} It is of natural interest to extend these developments to more complex partial differential equations, such as compressible Navier-Stokes equations, in both two- and three-dimensional domains, including complex geometries and unstructured meshes. An important next step would be to apply our method to develop a fully compressible reacting flow solver, where we can demonstrate the scalability gains of the asynchronous discontinuous Galerkin (ADG) method compared to the standard DG method. 

% Furthermore, as discussed earlier, we must incorporate data-discretization methods to improve the stability limits of the ADG schemes. At the compiler level, the ADG method can be combined with dynamic runtime programming models that support dynamic task parallelism, allowing for greater flexibility and adaptability during the execution of a solver. These models can manage load balancing and fault tolerance, and can also adjust the maximum allowable delay in communication by analyzing the solution behavior at runtime.

In conclusion, this study represents a significant step forward in the development of asynchronous computing approaches for PDEs at extreme scales. Through the introduction of the conservative asynchronous DG method with the utilization of asynchrony-tolerant fluxes, we addressed the challenges associated with communication and synchronization bottlenecks, enabling more efficient and accurate simulations. 

% These findings provide a robust foundation for advancing and applying asynchronous computing methods in tackling complex natural and engineering problems on exascale systems, ultimately enhancing our understanding and ability to solve real-world multiscale phenomena.

\section*{Acknowledgments}
The authors gratefully acknowledge the financial support from the SERB Start-up Research Grant, the MoE-STARS grant, and the National Supercomputing Mission, India. Special acknowledgment is also due to the Council of Scientific and Industrial Research (CSIR), India, for awarding the doctoral fellowship to SKG. KA is also supported by the Arcot Ramachandran Young Investigator award.
% The authors acknowledge the valuable contributions of their discussions with Phani Motamarri and Praveen Chandrashekar.
The authors beneﬁted from discussions with Phani Motamarri and Praveen Chandrashekar.

% \appendix
% \section{}
% \label{}

% References
%
% Following citation commands can be used in the body text:
% Usage of \cite is as follows:
%   \cite{key}          ==>>  [#]
%   \cite[chap. 2]{key} ==>>  [#, chap. 2]
%   \citet{key}         ==>>  Author [#]

% References with bibTeX database:

\bibliographystyle{model1-num-names}
\bibliography{main.bib}

\begin{thebibliography}{54}
\expandafter\ifx\csname natexlab\endcsname\relax\def\natexlab#1{#1}\fi
\providecommand{\bibinfo}[2]{#2}
\ifx\xfnm\relax \def\xfnm[#1]{\unskip,\space#1}\fi
%Type = Book
\bibitem[{Hesthaven and Warburton(2007)}]{hesthaven2007}
\bibinfo{author}{J.~S. Hesthaven}, \bibinfo{author}{T.~Warburton},
  \bibinfo{title}{{Nodal Discontinuous Galerkin Methods: Algorithms, Analysis,
  and Applications}}, \bibinfo{publisher}{Springer Publishing Company,
  Incorporated}, \bibinfo{edition}{1st} edition, \bibinfo{year}{2007}.
%Type = Article
\bibitem[{Cockburn et~al.(2009)Cockburn, Gopalakrishnan, and
  Lazarov}]{cockburn2009unified}
\bibinfo{author}{B.~Cockburn}, \bibinfo{author}{J.~Gopalakrishnan},
  \bibinfo{author}{R.~Lazarov},
\newblock \bibinfo{title}{{Unified Hybridization of Discontinuous Galerkin,
  Mixed, and Continuous Galerkin Methods for Second Order Elliptic Problems}},
\newblock \bibinfo{journal}{SIAM Journal on Numerical Analysis}
  \bibinfo{volume}{47} (\bibinfo{year}{2009}) \bibinfo{pages}{1319--1365}.
%Type = Inproceedings
\bibitem[{Roca et~al.(2013)Roca, Nguyen, and Peraire}]{roca2013scalable}
\bibinfo{author}{X.~Roca}, \bibinfo{author}{C.~Nguyen},
  \bibinfo{author}{J.~Peraire},
\newblock \bibinfo{title}{{Scalable parallelization of the hybridized
  discontinuous Galerkin method for compressible flow}},
\newblock in: \bibinfo{booktitle}{21st AIAA Computational fluid dynamics
  conference}, p. \bibinfo{pages}{2939}.
%Type = Article
\bibitem[{Lions et~al.(2001)Lions, Maday, and Turinici}]{LIONS2001661}
\bibinfo{author}{J.-L. Lions}, \bibinfo{author}{Y.~Maday},
  \bibinfo{author}{G.~Turinici},
\newblock \bibinfo{title}{{Résolution d'EDP par un schéma en temps
  pararéel}},
\newblock \bibinfo{journal}{Comptes Rendus de l'Académie des Sciences - Series
  I - Mathematics} \bibinfo{volume}{332} (\bibinfo{year}{2001})
  \bibinfo{pages}{661--668}.
%Type = Book
\bibitem[{Burrage(1997)}]{burrage1997parallel}
\bibinfo{author}{K.~Burrage}, \bibinfo{title}{{Parallel Methods for ODEs}},
  Advances in computational mathematics, \bibinfo{publisher}{Baltzer Science
  Publishers}, \bibinfo{year}{1997}.
%Type = Article
\bibitem[{Gander and Vandewalle(1 01)}]{lirias1119269}
\bibinfo{author}{M.~J. Gander}, \bibinfo{author}{S.~Vandewalle},
\newblock \bibinfo{title}{{Analysis of the parareal time-parallel
  time-integration method}},
\newblock \bibinfo{journal}{SIAM Journal on Scientific Computing}
  \bibinfo{volume}{29} (\bibinfo{year}{2007-01-01}).
%Type = Article
\bibitem[{Xia et~al.(2015)Xia, Lou, Luo, Edwards, and Mueller}]{xia2015openacc}
\bibinfo{author}{Y.~Xia}, \bibinfo{author}{J.~Lou}, \bibinfo{author}{H.~Luo},
  \bibinfo{author}{J.~Edwards}, \bibinfo{author}{F.~Mueller},
\newblock \bibinfo{title}{{OpenACC acceleration of an unstructured CFD solver
  based on a reconstructed discontinuous Galerkin method for compressible
  flows}},
\newblock \bibinfo{journal}{International Journal for Numerical Methods in
  Fluids} \bibinfo{volume}{78} (\bibinfo{year}{2015})
  \bibinfo{pages}{123--139}.
%Type = Inproceedings
\bibitem[{Kirby and Mavriplis(2020)}]{kirby2020gpu}
\bibinfo{author}{A.~C. Kirby}, \bibinfo{author}{D.~J. Mavriplis},
\newblock \bibinfo{title}{{GPU-Accelerated Discontinuous Galerkin Methods: 30x
  Speedup on 345 Billion Unknowns}},
\newblock in: \bibinfo{booktitle}{2020 IEEE High Performance Extreme Computing
  Conference (HPEC)}, pp. \bibinfo{pages}{1--7}.
%Type = Inproceedings
\bibitem[{Nguyen et~al.(????)Nguyen, Terrana, and Peraire}]{nguyen2023implicit}
\bibinfo{author}{C.~Nguyen}, \bibinfo{author}{S.~Terrana},
  \bibinfo{author}{J.~Peraire},
\newblock \bibinfo{title}{{Implicit Large eddy simulation of hypersonic
  boundary-layer transition for a flared cone}},
\newblock in: \bibinfo{booktitle}{AIAA SCITECH 2023 Forum}.
%Type = Article
\bibitem[{Kronbichler and Kormann(2019)}]{kronbichler2019fast}
\bibinfo{author}{M.~Kronbichler}, \bibinfo{author}{K.~Kormann},
\newblock \bibinfo{title}{{Fast Matrix-Free Evaluation of Discontinuous
  Galerkin Finite Element Operators}},
\newblock \bibinfo{journal}{ACM Trans. Math. Softw.} \bibinfo{volume}{45}
  (\bibinfo{year}{2019}).
%Type = Article
\bibitem[{Kronbichler and Wall(2018)}]{kronbichler2018performance}
\bibinfo{author}{M.~Kronbichler}, \bibinfo{author}{W.~A. Wall},
\newblock \bibinfo{title}{{A Performance Comparison of Continuous and
  Discontinuous Galerkin Methods with Fast Multigrid Solvers}},
\newblock \bibinfo{journal}{SIAM Journal on Scientific Computing}
  \bibinfo{volume}{40} (\bibinfo{year}{2018}) \bibinfo{pages}{A3423--A3448}.
%Type = Inproceedings
\bibitem[{Arndt et~al.(2020)Arndt, Fehn, Kanschat, Kormann, Kronbichler, Munch,
  Wall, and Witte}]{exadg2020}
\bibinfo{author}{D.~Arndt}, \bibinfo{author}{N.~Fehn},
  \bibinfo{author}{G.~Kanschat}, \bibinfo{author}{K.~Kormann},
  \bibinfo{author}{M.~Kronbichler}, \bibinfo{author}{P.~Munch},
  \bibinfo{author}{W.~A. Wall}, \bibinfo{author}{J.~Witte},
\newblock \bibinfo{title}{{ExaDG: High-Order Discontinuous Galerkin for the
  Exa-Scale}},
\newblock in: \bibinfo{editor}{H.-J. Bungartz}, \bibinfo{editor}{S.~Reiz},
  \bibinfo{editor}{B.~Uekermann}, \bibinfo{editor}{P.~Neumann},
  \bibinfo{editor}{W.~E. Nagel} (Eds.), \bibinfo{booktitle}{Software for
  Exascale Computing - SPPEXA 2016-2019}, \bibinfo{publisher}{Springer
  International Publishing}, \bibinfo{address}{Cham}, \bibinfo{year}{2020}, pp.
  \bibinfo{pages}{189--224}.
%Type = Article
\bibitem[{Arndt et~al.(2022)Arndt, Bangerth, Feder, Fehling, Gassm{\"o}ller,
  Heister, Heltai, Kronbichler, Maier, Munch, Pelteret, Sticko, Turcksin, and
  Wells}]{dealII94}
\bibinfo{author}{D.~Arndt}, \bibinfo{author}{W.~Bangerth},
  \bibinfo{author}{M.~Feder}, \bibinfo{author}{M.~Fehling},
  \bibinfo{author}{R.~Gassm{\"o}ller}, \bibinfo{author}{T.~Heister},
  \bibinfo{author}{L.~Heltai}, \bibinfo{author}{M.~Kronbichler},
  \bibinfo{author}{M.~Maier}, \bibinfo{author}{P.~Munch},
  \bibinfo{author}{J.-P. Pelteret}, \bibinfo{author}{S.~Sticko},
  \bibinfo{author}{B.~Turcksin}, \bibinfo{author}{D.~Wells},
\newblock \bibinfo{title}{{The \texttt{deal.II} Library, Version 9.4}},
\newblock \bibinfo{journal}{Journal of Numerical Mathematics}
  \bibinfo{volume}{30} (\bibinfo{year}{2022}) \bibinfo{pages}{231--246}.
%Type = Article
\bibitem[{Bastian et~al.(2008{\natexlab{a}})Bastian, Blatt, Dedner, Engwer,
  Kl{\"o}fkorn, Ohlberger, and Sander}]{bastian2008-dune1}
\bibinfo{author}{P.~Bastian}, \bibinfo{author}{M.~Blatt},
  \bibinfo{author}{A.~Dedner}, \bibinfo{author}{C.~Engwer},
  \bibinfo{author}{R.~Kl{\"o}fkorn}, \bibinfo{author}{M.~Ohlberger},
  \bibinfo{author}{O.~Sander},
\newblock \bibinfo{title}{{A generic grid interface for parallel and adaptive
  scientific computing. Part I: abstract framework}},
\newblock \bibinfo{journal}{Computing} \bibinfo{volume}{82}
  (\bibinfo{year}{2008}{\natexlab{a}}) \bibinfo{pages}{103--119}.
%Type = Article
\bibitem[{Bastian et~al.(2008{\natexlab{b}})Bastian, Blatt, Dedner, Engwer,
  Kl{\"o}fkorn, Kornhuber, Ohlberger, and Sander}]{bastian2008-dune2}
\bibinfo{author}{P.~Bastian}, \bibinfo{author}{M.~Blatt},
  \bibinfo{author}{A.~Dedner}, \bibinfo{author}{C.~Engwer},
  \bibinfo{author}{R.~Kl{\"o}fkorn}, \bibinfo{author}{R.~Kornhuber},
  \bibinfo{author}{M.~Ohlberger}, \bibinfo{author}{O.~Sander},
\newblock \bibinfo{title}{{A generic grid interface for parallel and adaptive
  scientific computing. Part II: implementation and tests in DUNE}},
\newblock \bibinfo{journal}{Computing} \bibinfo{volume}{82}
  (\bibinfo{year}{2008}{\natexlab{b}}) \bibinfo{pages}{121--138}.
%Type = Article
\bibitem[{Bastian et~al.(2021)Bastian, Blatt, Dedner, Dreier, Engwer, Fritze,
  Gräser, Grüninger, Kempf, Klöfkorn, Ohlberger, and Sander}]{BASTIAN2021}
\bibinfo{author}{P.~Bastian}, \bibinfo{author}{M.~Blatt},
  \bibinfo{author}{A.~Dedner}, \bibinfo{author}{N.-A. Dreier},
  \bibinfo{author}{C.~Engwer}, \bibinfo{author}{R.~Fritze},
  \bibinfo{author}{C.~Gräser}, \bibinfo{author}{C.~Grüninger},
  \bibinfo{author}{D.~Kempf}, \bibinfo{author}{R.~Klöfkorn},
  \bibinfo{author}{M.~Ohlberger}, \bibinfo{author}{O.~Sander},
\newblock \bibinfo{title}{{The Dune framework: Basic concepts and recent
  developments}},
\newblock \bibinfo{journal}{Computers \& Mathematics with Applications}
  \bibinfo{volume}{81} (\bibinfo{year}{2021}) \bibinfo{pages}{75--112}.
  \bibinfo{note}{Development and Application of Open-source Software for
  Problems with Numerical PDEs}.
%Type = Inproceedings
\bibitem[{Kl{\"o}fkorn(2012)}]{klofkorn2012-dune-dg}
\bibinfo{author}{R.~Kl{\"o}fkorn},
\newblock \bibinfo{title}{Efficient matrix-free implementation of discontinuous
  galerkin methods for compressible flow problems},
\newblock in: \bibinfo{editor}{A.~al.} (Ed.), \bibinfo{booktitle}{Proceedings
  of the ALGORITMY 2012}, pp. \bibinfo{pages}{11--21}.
%Type = Inproceedings
\bibitem[{Bastian et~al.(2014)Bastian, Engwer, G{\"o}ddeke, Iliev, Ippisch,
  Ohlberger, Turek, Fahlke, Kaulmann, M{\"u}thing, and
  Ribbrock}]{bastian2014exadune}
\bibinfo{author}{P.~Bastian}, \bibinfo{author}{C.~Engwer},
  \bibinfo{author}{D.~G{\"o}ddeke}, \bibinfo{author}{O.~Iliev},
  \bibinfo{author}{O.~Ippisch}, \bibinfo{author}{M.~Ohlberger},
  \bibinfo{author}{S.~Turek}, \bibinfo{author}{J.~Fahlke},
  \bibinfo{author}{S.~Kaulmann}, \bibinfo{author}{S.~M{\"u}thing},
  \bibinfo{author}{D.~Ribbrock},
\newblock \bibinfo{title}{Exa-dune: Flexible pde solvers, numerical methods and
  applications},
\newblock in: \bibinfo{editor}{L.~Lopes}, \bibinfo{editor}{J.~{\v{Z}}ilinskas},
  \bibinfo{editor}{A.~Costan}, \bibinfo{editor}{R.~G. Cascella},
  \bibinfo{editor}{G.~Kecskemeti}, \bibinfo{editor}{E.~Jeannot},
  \bibinfo{editor}{M.~Cannataro}, \bibinfo{editor}{L.~Ricci},
  \bibinfo{editor}{S.~Benkner}, \bibinfo{editor}{S.~Petit},
  \bibinfo{editor}{V.~Scarano}, \bibinfo{editor}{J.~Gracia},
  \bibinfo{editor}{S.~Hunold}, \bibinfo{editor}{S.~L. Scott},
  \bibinfo{editor}{S.~Lankes}, \bibinfo{editor}{C.~Lengauer},
  \bibinfo{editor}{J.~Carretero}, \bibinfo{editor}{J.~Breitbart},
  \bibinfo{editor}{M.~Alexander} (Eds.), \bibinfo{booktitle}{Euro-Par 2014:
  Parallel Processing Workshops}, \bibinfo{publisher}{Springer International
  Publishing}, \bibinfo{address}{Cham}, \bibinfo{year}{2014}, pp.
  \bibinfo{pages}{530--541}.
%Type = Inproceedings
\bibitem[{Bastian et~al.(2016{\natexlab{a}})Bastian, Engwer, Fahlke, Geveler,
  G{\"o}ddeke, Iliev, Ippisch, Milk, Mohring, M{\"u}thing, Ohlberger, Ribbrock,
  and Turek}]{exadune1-bastian2016}
\bibinfo{author}{P.~Bastian}, \bibinfo{author}{C.~Engwer},
  \bibinfo{author}{J.~Fahlke}, \bibinfo{author}{M.~Geveler},
  \bibinfo{author}{D.~G{\"o}ddeke}, \bibinfo{author}{O.~Iliev},
  \bibinfo{author}{O.~Ippisch}, \bibinfo{author}{R.~Milk},
  \bibinfo{author}{J.~Mohring}, \bibinfo{author}{S.~M{\"u}thing},
  \bibinfo{author}{M.~Ohlberger}, \bibinfo{author}{D.~Ribbrock},
  \bibinfo{author}{S.~Turek},
\newblock \bibinfo{title}{Advances concerning multiscale methods and
  uncertainty quantification in exa-dune},
\newblock in: \bibinfo{editor}{H.-J. Bungartz}, \bibinfo{editor}{P.~Neumann},
  \bibinfo{editor}{W.~E. Nagel} (Eds.), \bibinfo{booktitle}{Software for
  Exascale Computing - SPPEXA 2013-2015}, \bibinfo{publisher}{Springer
  International Publishing}, \bibinfo{address}{Cham},
  \bibinfo{year}{2016}{\natexlab{a}}, pp. \bibinfo{pages}{25--43}.
%Type = Inproceedings
\bibitem[{Bastian et~al.(2016{\natexlab{b}})Bastian, Engwer, Fahlke, Geveler,
  G{\"o}ddeke, Iliev, Ippisch, Milk, Mohring, M{\"u}thing, Ohlberger, Ribbrock,
  and Turek}]{exadune2-bastian2016}
\bibinfo{author}{P.~Bastian}, \bibinfo{author}{C.~Engwer},
  \bibinfo{author}{J.~Fahlke}, \bibinfo{author}{M.~Geveler},
  \bibinfo{author}{D.~G{\"o}ddeke}, \bibinfo{author}{O.~Iliev},
  \bibinfo{author}{O.~Ippisch}, \bibinfo{author}{R.~Milk},
  \bibinfo{author}{J.~Mohring}, \bibinfo{author}{S.~M{\"u}thing},
  \bibinfo{author}{M.~Ohlberger}, \bibinfo{author}{D.~Ribbrock},
  \bibinfo{author}{S.~Turek},
\newblock \bibinfo{title}{Hardware-based efficiency advances in the exa-dune
  project},
\newblock in: \bibinfo{editor}{H.-J. Bungartz}, \bibinfo{editor}{P.~Neumann},
  \bibinfo{editor}{W.~E. Nagel} (Eds.), \bibinfo{booktitle}{Software for
  Exascale Computing - SPPEXA 2013-2015}, \bibinfo{publisher}{Springer
  International Publishing}, \bibinfo{address}{Cham},
  \bibinfo{year}{2016}{\natexlab{b}}, pp. \bibinfo{pages}{3--23}.
%Type = Article
\bibitem[{Krais et~al.(2021)Krais, Beck, Bolemann, Frank, Flad, Gassner,
  Hindenlang, Hoffmann, Kuhn, Sonntag, and Munz}]{KRAIS2021186-flexi}
\bibinfo{author}{N.~Krais}, \bibinfo{author}{A.~Beck},
  \bibinfo{author}{T.~Bolemann}, \bibinfo{author}{H.~Frank},
  \bibinfo{author}{D.~Flad}, \bibinfo{author}{G.~Gassner},
  \bibinfo{author}{F.~Hindenlang}, \bibinfo{author}{M.~Hoffmann},
  \bibinfo{author}{T.~Kuhn}, \bibinfo{author}{M.~Sonntag},
  \bibinfo{author}{C.-D. Munz},
\newblock \bibinfo{title}{Flexi: A high order discontinuous galerkin framework
  for hyperbolic–parabolic conservation laws},
\newblock \bibinfo{journal}{Computers \& Mathematics with Applications}
  \bibinfo{volume}{81} (\bibinfo{year}{2021}) \bibinfo{pages}{186--219}.
  \bibinfo{note}{Development and Application of Open-source Software for
  Problems with Numerical PDEs}.
%Type = Misc
\bibitem[{Blind et~al.(2023)Blind, Gao, Kempf, Kopper, Kurz, Schwarz, and
  Beck}]{blind2023exascale-flexi}
\bibinfo{author}{M.~Blind}, \bibinfo{author}{M.~Gao},
  \bibinfo{author}{D.~Kempf}, \bibinfo{author}{P.~Kopper},
  \bibinfo{author}{M.~Kurz}, \bibinfo{author}{A.~Schwarz},
  \bibinfo{author}{A.~Beck}, \bibinfo{title}{Towards exascale cfd simulations
  using the discontinuous galerkin solver flexi}, \bibinfo{year}{2023}.
%Type = Article
\bibitem[{Melander et~al.(0)Melander, Strøm, Pind, Engsig-Karup, Jeong,
  Warburton, Chalmers, and Hesthaven}]{melander2023}
\bibinfo{author}{A.~Melander}, \bibinfo{author}{E.~Strøm},
  \bibinfo{author}{F.~Pind}, \bibinfo{author}{A.~P. Engsig-Karup},
  \bibinfo{author}{C.-H. Jeong}, \bibinfo{author}{T.~Warburton},
  \bibinfo{author}{N.~Chalmers}, \bibinfo{author}{J.~S. Hesthaven},
\newblock \bibinfo{title}{{Massively parallel nodal discontinous Galerkin
  finite element method simulator for room acoustics}},
\newblock \bibinfo{journal}{The International Journal of High Performance
  Computing Applications} \bibinfo{volume}{0} (\bibinfo{year}{0})
  \bibinfo{pages}{10943420231208948}.
%Type = Inproceedings
\bibitem[{Aditya and Donzis(2012)}]{konduri2012async}
\bibinfo{author}{K.~Aditya}, \bibinfo{author}{D.~A. Donzis},
\newblock \bibinfo{title}{{Poster: Asynchronous Computing for Partial
  Differential Equations at Extreme Scales}},
\newblock in: \bibinfo{booktitle}{Proceedings of the 2012 SC Companion: High
  Performance Computing, Networking Storage and Analysis}, SCC '12,
  \bibinfo{publisher}{IEEE Computer Society}, \bibinfo{address}{Washington, DC,
  USA}, \bibinfo{year}{2012}, p. \bibinfo{pages}{1444}.
%Type = Article
\bibitem[{Donzis and Aditya(2014)}]{donzis2014}
\bibinfo{author}{D.~A. Donzis}, \bibinfo{author}{K.~Aditya},
\newblock \bibinfo{title}{{Asynchronous finite-difference schemes for partial
  differential equations}},
\newblock \bibinfo{journal}{Journal of Computational Physics}
  \bibinfo{volume}{274} (\bibinfo{year}{2014}) \bibinfo{pages}{370--392}.
%Type = Article
\bibitem[{Aditya and Donzis(2017)}]{konduri2017at}
\bibinfo{author}{K.~Aditya}, \bibinfo{author}{D.~A. Donzis},
\newblock \bibinfo{title}{{High-order asynchrony-tolerant finite difference
  schemes for partial differential equations}},
\newblock \bibinfo{journal}{Journal of Computational Physics}
  \bibinfo{volume}{350} (\bibinfo{year}{2017}) \bibinfo{pages}{550--572}.
%Type = Article
\bibitem[{Kumari et~al.(2023)Kumari, Cleary, Desai, Donzis, Chen, and
  Aditya}]{komal2023reactions-at}
\bibinfo{author}{K.~Kumari}, \bibinfo{author}{E.~Cleary},
  \bibinfo{author}{S.~Desai}, \bibinfo{author}{D.~A. Donzis},
  \bibinfo{author}{J.~H. Chen}, \bibinfo{author}{K.~Aditya},
\newblock \bibinfo{title}{{Evaluation of finite difference based asynchronous
  partial differential equations solver for reacting flows}},
\newblock \bibinfo{journal}{Journal of Computational Physics}
  \bibinfo{volume}{477} (\bibinfo{year}{2023}) \bibinfo{pages}{111906}.
%Type = Misc
\bibitem[{Aditya et~al.(2019)Aditya, Gysi, Kwasniewski, Hoefler, Donzis, and
  Chen}]{aditya2019arXiv}
\bibinfo{author}{K.~Aditya}, \bibinfo{author}{T.~Gysi},
  \bibinfo{author}{G.~Kwasniewski}, \bibinfo{author}{T.~Hoefler},
  \bibinfo{author}{D.~A. Donzis}, \bibinfo{author}{J.~H. Chen},
  \bibinfo{title}{{A scalable weakly-synchronous algorithm for solving partial
  differential equations}}, \bibinfo{year}{preprint, arXiv: 1911.05769, 2019}.
%Type = Article
\bibitem[{Kumari and Donzis(2020)}]{komal2020dns-at}
\bibinfo{author}{K.~Kumari}, \bibinfo{author}{D.~A. Donzis},
\newblock \bibinfo{title}{{Direct numerical simulations of turbulent flows
  using high-order asynchrony-tolerant schemes: Accuracy and performance}},
\newblock \bibinfo{journal}{Journal of Computational Physics}
  \bibinfo{volume}{419} (\bibinfo{year}{2020}) \bibinfo{pages}{109626}.
%Type = Article
\bibitem[{Goswami et~al.(2023)Goswami, Matthew, and
  Aditya}]{goswami2023lserkat-jcp}
\bibinfo{author}{S.~K. Goswami}, \bibinfo{author}{V.~J. Matthew},
  \bibinfo{author}{K.~Aditya},
\newblock \bibinfo{title}{{Implementation of low-storage Runge-Kutta time
  integration schemes in scalable asynchronous partial differential equation
  solvers}},
\newblock \bibinfo{journal}{Journal of Computational Physics}
  \bibinfo{volume}{477} (\bibinfo{year}{2023}) \bibinfo{pages}{111922}.
%Type = Inproceedings
\bibitem[{Ghosh et~al.(2018)Ghosh, Saha, Gupta, and
  Tryggvason}]{ghosh2018event}
\bibinfo{author}{S.~Ghosh}, \bibinfo{author}{K.~K. Saha},
  \bibinfo{author}{V.~Gupta}, \bibinfo{author}{G.~Tryggvason},
\newblock \bibinfo{title}{{Event-Triggered Communication in Parallel
  Computing}},
\newblock in: \bibinfo{booktitle}{2018 IEEE/ACM 9th Workshop on Latest Advances
  in Scalable Algorithms for Large-Scale Systems (scalA)}, pp.
  \bibinfo{pages}{1--8}.
%Type = Inproceedings
\bibitem[{Ghosh et~al.(2019)Ghosh, Saha, Gupta, and
  Tryggvason}]{ghosh2019parallel}
\bibinfo{author}{S.~Ghosh}, \bibinfo{author}{K.~K. Saha},
  \bibinfo{author}{V.~Gupta}, \bibinfo{author}{G.~Tryggvason},
\newblock \bibinfo{title}{{Parallel Computation using Event-Triggered
  Communication}},
\newblock in: \bibinfo{booktitle}{2019 American Control Conference (ACC)}, pp.
  \bibinfo{pages}{4000--4005}.
%Type = Article
\bibitem[{Gravouil et~al.(2015)Gravouil, Combescure, and Brun}]{Gravouil2014}
\bibinfo{author}{A.~Gravouil}, \bibinfo{author}{A.~Combescure},
  \bibinfo{author}{M.~Brun},
\newblock \bibinfo{title}{{Heterogeneous asynchronous time integrators for
  computational structural dynamics}},
\newblock \bibinfo{journal}{International Journal for Numerical Methods in
  Engineering} \bibinfo{volume}{102} (\bibinfo{year}{2015})
  \bibinfo{pages}{202--232}.
%Type = Article
\bibitem[{{Mahjoubi} et~al.(2009){Mahjoubi}, {Gravouil}, and
  {Combescure}}]{mahjoubi2009}
\bibinfo{author}{N.~{Mahjoubi}}, \bibinfo{author}{A.~{Gravouil}},
  \bibinfo{author}{A.~{Combescure}},
\newblock \bibinfo{title}{{{Coupling subdomains with heterogeneous time
  integrators and incompatible time steps}}},
\newblock \bibinfo{journal}{Computational Mechanics} \bibinfo{volume}{44}
  (\bibinfo{year}{2009}) \bibinfo{pages}{825--843}.
%Type = Article
\bibitem[{{Fekak} et~al.(2017){Fekak}, {Brun}, {Gravouil}, and
  {Depale}}]{fekak2017}
\bibinfo{author}{F.-E. {Fekak}}, \bibinfo{author}{M.~{Brun}},
  \bibinfo{author}{A.~{Gravouil}}, \bibinfo{author}{B.~{Depale}},
\newblock \bibinfo{title}{{{A new heterogeneous asynchronous explicit-implicit
  time integrator for nonsmooth dynamics}}},
\newblock \bibinfo{journal}{Computational Mechanics} \bibinfo{volume}{60}
  (\bibinfo{year}{2017}) \bibinfo{pages}{1--21}.
%Type = Inproceedings
\bibitem[{Goswami and Aditya(????)}]{goswami2022asyncdg-aviation}
\bibinfo{author}{S.~K. Goswami}, \bibinfo{author}{K.~Aditya},
\newblock \bibinfo{title}{{An asynchronous discontinuous-Galerkin method for
  solving PDEs at extreme scales}},
\newblock in: \bibinfo{booktitle}{AIAA AVIATION 2022 Forum}.
%Type = Article
\bibitem[{Brus et~al.(2017)Brus, Wirasaet, Westerink, and
  Dawson}]{brus2017dgperformance}
\bibinfo{author}{S.~Brus}, \bibinfo{author}{D.~Wirasaet},
  \bibinfo{author}{J.~J. Westerink}, \bibinfo{author}{C.~Dawson},
\newblock \bibinfo{title}{Performance and scalability improvements for
  discontinuous galerkin solutions to conservation laws on unstructured grids},
\newblock \bibinfo{journal}{Journal of Scientific Computing}
  \bibinfo{volume}{70} (\bibinfo{year}{2017}) \bibinfo{pages}{210--242}.
%Type = Inproceedings
\bibitem[{Hoefler et~al.(2010)Hoefler, Schneider, and Lumsdaine}]{hoefler2010}
\bibinfo{author}{T.~Hoefler}, \bibinfo{author}{T.~Schneider},
  \bibinfo{author}{A.~Lumsdaine},
\newblock \bibinfo{title}{{Characterizing the Influence of System Noise on
  Large-Scale Applications by Simulation}},
\newblock in: \bibinfo{booktitle}{SC '10: Proceedings of the 2010 ACM/IEEE
  International Conference for High Performance Computing, Networking, Storage
  and Analysis}, pp. \bibinfo{pages}{1--11}.
%Type = Article
\bibitem[{VonNeumann and Richtmyer(1950)}]{neumann1950}
\bibinfo{author}{J.~VonNeumann}, \bibinfo{author}{R.~D. Richtmyer},
\newblock \bibinfo{title}{{{A Method for the Numerical Calculation of
  Hydrodynamic Shocks}}},
\newblock \bibinfo{journal}{Journal of Applied Physics} \bibinfo{volume}{21}
  (\bibinfo{year}{1950}) \bibinfo{pages}{232--237}.
%Type = Inbook
\bibitem[{Charney et~al.(1990)Charney, Fj{\"o}rtoft, and von
  Neumann}]{charney1990numerical}
\bibinfo{author}{J.~G. Charney}, \bibinfo{author}{R.~Fj{\"o}rtoft},
  \bibinfo{author}{J.~von Neumann}, \bibinfo{title}{{Numerical Integration of
  the Barotropic Vorticity Equation}}, \bibinfo{publisher}{American
  Meteorological Society}, \bibinfo{address}{Boston, MA}, pp.
  \bibinfo{pages}{267--284}.
%Type = Article
\bibitem[{Kumari and Donzis(2021)}]{komal2021stability}
\bibinfo{author}{K.~Kumari}, \bibinfo{author}{D.~A. Donzis},
\newblock \bibinfo{title}{{A generalized von Neumann analysis for multi-level
  schemes: Stability and spectral accuracy}},
\newblock \bibinfo{journal}{Journal of Computational Physics}
  \bibinfo{volume}{424} (\bibinfo{year}{2021}) \bibinfo{pages}{109868}.
%Type = Book
\bibitem[{Vichnevetsky and Bowles(1982)}]{vichnevetsky1982fourier}
\bibinfo{author}{R.~Vichnevetsky}, \bibinfo{author}{J.~B. Bowles},
  \bibinfo{title}{{Fourier Analysis of Numerical Approximations of Hyperbolic
  Equations}}, \bibinfo{publisher}{Society for Industrial and Applied
  Mathematics}, \bibinfo{year}{1982}.
%Type = Article
\bibitem[{Hu et~al.(1999)Hu, Hussaini, and Rasetarinera}]{fourierdg-fang1999}
\bibinfo{author}{F.~Q. Hu}, \bibinfo{author}{M.~Hussaini},
  \bibinfo{author}{P.~Rasetarinera},
\newblock \bibinfo{title}{{An Analysis of the Discontinuous Galerkin Method for
  Wave Propagation Problems}},
\newblock \bibinfo{journal}{Journal of Computational Physics}
  \bibinfo{volume}{151} (\bibinfo{year}{1999}) \bibinfo{pages}{921--946}.
%Type = Article
\bibitem[{Alhawwary and Wang(2018)}]{ALHAWWARY2018JCP}
\bibinfo{author}{M.~Alhawwary}, \bibinfo{author}{Z.~Wang},
\newblock \bibinfo{title}{{Fourier analysis and evaluation of DG, FD and
  compact difference methods for conservation laws}},
\newblock \bibinfo{journal}{Journal of Computational Physics}
  \bibinfo{volume}{373} (\bibinfo{year}{2018}) \bibinfo{pages}{835--862}.
%Type = Article
\bibitem[{Shu(2009)}]{shu2009}
\bibinfo{author}{C.-W. Shu},
\newblock \bibinfo{title}{{Discontinuous Galerkin methods: General approach and
  stability}},
\newblock \bibinfo{journal}{Numerical Solutions of Partial Differential
  Equations}  (\bibinfo{year}{2009}).
%Type = Article
\bibitem[{Zhang and Shu(2005)}]{shu2005}
\bibinfo{author}{Q.~Zhang}, \bibinfo{author}{C.-W. Shu},
\newblock \bibinfo{title}{{Error Estimates to Smooth Solutions of Runge-Kutta
  Discontinuous Galerkin Methods for Scalar Conservation Laws}},
\newblock \bibinfo{journal}{SIAM Journal on Numerical Analysis}
  \bibinfo{volume}{42} (\bibinfo{year}{2005}) \bibinfo{pages}{641--666}.
%Type = Article
\bibitem[{Sod(1978)}]{SOD1978JCP}
\bibinfo{author}{G.~A. Sod},
\newblock \bibinfo{title}{A survey of several finite difference methods for
  systems of nonlinear hyperbolic conservation laws},
\newblock \bibinfo{journal}{Journal of Computational Physics}
  \bibinfo{volume}{27} (\bibinfo{year}{1978}) \bibinfo{pages}{1--31}.
%Type = Book
\bibitem[{Toro(2009)}]{toro2009}
\bibinfo{author}{E.~F. Toro}, \bibinfo{title}{{Riemann Solvers and Numerical
  Methods for Fluid Dynamics: A Practical Introduction}},
  \bibinfo{publisher}{Springer Verlag}, \bibinfo{year}{2009}.
%Type = Article
\bibitem[{Cockburn and Shu(1998)}]{cockburn1998ldg}
\bibinfo{author}{B.~Cockburn}, \bibinfo{author}{C.-W. Shu},
\newblock \bibinfo{title}{{The Local Discontinuous Galerkin Method for
  Time-Dependent Convection-Diffusion Systems}},
\newblock \bibinfo{journal}{SIAM Journal on Numerical Analysis}
  \bibinfo{volume}{35} (\bibinfo{year}{1998}) \bibinfo{pages}{2440--2463}.
%Type = Article
\bibitem[{Williamson(1980)}]{WILLIAMSON1980}
\bibinfo{author}{J.~Williamson},
\newblock \bibinfo{title}{Low-storage runge-kutta schemes},
\newblock \bibinfo{journal}{Journal of Computational Physics}
  \bibinfo{volume}{35} (\bibinfo{year}{1980}) \bibinfo{pages}{48--56}.
%Type = Article
\bibitem[{Kennedy et~al.(2000)Kennedy, Carpenter, and Lewis}]{KENNEDY2000}
\bibinfo{author}{C.~A. Kennedy}, \bibinfo{author}{M.~H. Carpenter},
  \bibinfo{author}{R.~Lewis},
\newblock \bibinfo{title}{{Low-storage, explicit Runge–Kutta schemes for the
  compressible Navier–Stokes equations}},
\newblock \bibinfo{journal}{Applied Numerical Mathematics} \bibinfo{volume}{35}
  (\bibinfo{year}{2000}) \bibinfo{pages}{177--219}.
%Type = Article
\bibitem[{Zhang et~al.(2011)Zhang, Xi-Jun, and
  Guo-Zhong}]{ldg-burgers-zhang2011}
\bibinfo{author}{R.~Zhang}, \bibinfo{author}{Y.~Xi-Jun},
  \bibinfo{author}{Z.~Guo-Zhong},
\newblock \bibinfo{title}{{Local discontinuous Galerkin method for solving
  Burgers and coupled Burgers equations}},
\newblock \bibinfo{journal}{Chinese Physics B} \bibinfo{volume}{20}
  (\bibinfo{year}{2011}) \bibinfo{pages}{110205}.
%Type = Book
\bibitem[{Li(2010)}]{dg-book-li2006}
\bibinfo{author}{B.~Li}, \bibinfo{title}{{Discontinuous Finite Elements in
  Fluid Dynamics and Heat Transfer}}, Computational Fluid and Solid Mechanics,
  \bibinfo{publisher}{Springer London}, \bibinfo{year}{2010}.
%Type = Article
\bibitem[{Bar-Sinai et~al.(2019)Bar-Sinai, Hoyer, Hickey, and
  Brenner}]{data-driven-pnas2019}
\bibinfo{author}{Y.~Bar-Sinai}, \bibinfo{author}{S.~Hoyer},
  \bibinfo{author}{J.~Hickey}, \bibinfo{author}{M.~P. Brenner},
\newblock \bibinfo{title}{Learning data-driven discretizations for partial
  differential equations},
\newblock \bibinfo{journal}{Proceedings of the National Academy of Sciences}
  \bibinfo{volume}{116} (\bibinfo{year}{2019}) \bibinfo{pages}{15344--15349}.

\end{thebibliography}

% Authors are advised to submit their bibtex database files. They are
% requested to list a bibtex style file in the manuscript if they do
% not want to use model1-num-names.bst.

% References without bibTeX database:


% \bibitem must have the following form:
%   \bibitem{key}...
%


% \end{thebibliography}

\end{document}